\documentclass[aps,twocolumn,amsmath,amssymb,showpacs,showkeys,floatfix,a4]{revtex4}

\usepackage{epsfig}
\usepackage{amssymb}
\usepackage[ansinew]{inputenc}

\usepackage{soul,color}
\usepackage{graphicx}
\usepackage{dcolumn}
\usepackage{bm}

\usepackage{float}
\usepackage{mathrsfs}
\usepackage{tabularx}
\usepackage{midfloat}

\newcommand{\be}{\begin{equation}}
\newcommand{\ee}{\end{equation}}
\newcommand{\bea}{\begin{eqnarray}}
\newcommand{\eea}{\end{eqnarray}}
\newcommand{\ba}{\begin{array}}
\newcommand{\ea}{\end{array}}
\newcommand{\bi}{\begin{itemize}}
\newcommand{\ei}{\end{itemize}}
\newcommand{\lan}{\langle}
\newcommand{\ran}{\rangle}

\begin{document}

\title{Theory of Single Charge Exchange Heavy Ion Reactions}

\author{Horst Lenske$^{1}${\footnote{Electronic address: horst.lenske@theo.physik.uni-giessen.de}}}
\author{Jessica I.Bellone$^{2,3}$ {\footnote{Electronic address: jessica.bellone@ct.infn.it}}}
\author{Maria Colonna$^{2}$ {\footnote{Electronic address: colonna@lns.infn.it}}}
\author{Jos\'e-Antonio Lay$^{2,4}$ {\footnote{Electronic address: lay@us.es}}}

\affiliation{
$^1$Institut f\"{u}r Theoretische Physik, Justus-Liebig-Universit\"{a}t Giessen, D-35392 Giessen, Germany
\\
$^2$INFN-LNS, I-95123 Catania, Italy
\\
$^3$Dipartimento di Fisica e Astronomia, Universit\'a degli studi di Catania, 
Italy
\\
$^4$Departamento de FAMN, Universidad de Sevilla, Apartado 1065, E-41080 Sevilla, Spain\\}


\begin{abstract}
\begin{center}
(NUMEN Collaboration)
\end{center}

The theory of heavy ion single charge exchange reactions is reformulated. In momentum space the reaction amplitude factorizes into a product of projectile and target transition form factors, folded with the nucleon-nucleon isovector interaction and a distortion coefficient which accounts for initial and final state ion-ion elastic interactions. The multipole structure of the transition form factors is studied in detail for Fermi-type non-spin flip and Gamow-Teller-type spin flip transitions, also serving to establish the connection to nuclear beta decay. The reaction kernel is evaluated for central and rank-2 tensor interactions.  Initial and final state elastic ion-ion interaction are shown to be dominated by the imaginary part of the optical potential allowing to evaluate the reaction coefficients in the strong absorption limit, realized by the black disk approximation. In that limit the distortion coefficient is evaluated in closed form, revealing the relation to the total reaction cross section and the geometry of the transition form factors. It is shown that at small momentum transfer  distortion effects reduce to a simple scaling factor, allowing to define reduced forward-angle cross section which is given by nuclear matrix elements of beta decay-type.  The response function formalism is used to describe nuclear charge changing transitions. Spectral distributions obtained by a self-consistent HFB and QRPA approach are discussed for $\tau_\pm$ excitations of $^{18}O$ and $^{40}Ca$, respectively, and compared to spectroscopic data. The interplay of nuclear structure and reaction dynamics is illustrated for the single charge exchange reaction $^{18}O+^{40}Ca \to ^{18}F+^{40}K$ at $T_{lab}=270$~MeV. 
\end{abstract}
\pacs {21.60-n,21.60Jz,21.10.Dr}
\keywords{charge exchange reactions, Fermi and Gamow-Teller transitions, beta decay}
\maketitle

\section{Introduction}\label{sec:Intro}

Already quite early the large research potential of heavy ion charge exchange reactions was recognized as a versatile tool to study simultaneously the two branches of charge changing excitations in nuclei. By an appropriate choice of projectile and selection of a suitable ejectile exit channel both $np^{-1}$- and $pn^{-1}$-type target transitions will be accessible under the same experimental conditions. On the theoretical side, light ion reactions have been considered in detail already in the early days of ($p,n$) charge exchange reactions, starting with the discovery of the giant Gamow-Teller resonance ($GTR$) by the pioneering experiments at IUCF, summarized e.g. in \cite{Goodman:1980}. Distorted Wave Born Approximation (DWBA) methods were used in conjunctions with folding approaches for the nucleon-nucleus optical potentials and form factors. Somewhat later, the work of Taddeucci \emph{et al.} \cite{Taddeucci:1987} has given a comprehensive theoretical framework, still widely used for the description of light ion charge exchange data.  The focus of Taddeucci \emph{et al.} is on the relation of single charge exchange (SCE) cross sections and beta-decay nuclear matrix elements. Initial and final state projectile-target interactions are being estimated somewhat schematically by eikonal methods. The latter were used also by Bertulani in an early study of heavy ion charge exchange reaction at medium \cite{Bertulani:1992qs} and later at ultra-high energies \cite{Bertulani:1999cq}, respectively. At lower energies, heavy ion SCE reactions were investigated by microscopic approaches as early as in the 1980ies in connection the first experiments at GSI \cite{Brendel:1988}, GANIL \cite{Berat:1989amx}, and Hahn-Meitner Institute \cite{Bohlen:1988ten}. In \cite{Brendel:1988,Lenske:1989zz}, for example, and later also in \cite{Cappuzzello:2004afa,Cappuzzello:2004vka,Cappuzzello:201lyi} direct charge exchange mediated by the projectile-target isovector nucleon-nucleon (NN) interactions and two-step transfer charge exchange by sequential proton and neutron exchange were described by full scale distorted wave methods and microscopic nuclear transition form factors.

{The direct, one-step single charge exchange process is of central interest for spectroscopic investigations because of giving immediate access to nuclear isovector transition matrix elements.} 
In this paper, we present an update of the theory of heavy ion SCE reactions with special emphasis on direct charge exchange processes. 
As far as the spin and isospin structure is concerned, the SCE operators are of the same type as those encountered in beta-decay. For strong and weak interaction processes a connection can be established on the level of multipole operators, albeit with quite different form factors: While at nuclear scales weak interactions, mediated by the \emph{W} and \emph{Z} gauge bosons, are well described by contact interactions, charge exchange by strong interactions is given essentially by pion and rho-meson exchange, characterized by finite interaction ranges and considerably larger coupling constants. A clear advantage of nuclear charge exchange reactions is the availability of a large variety of projectile-target systems allowing to study the processes under well defined dynamical conditions, thus giving access to the less well explored isovector sector of nuclear spectroscopy. However, the peculiarities of heavy ion reactions demand for advanced theoretical methods allowing to extract the wanted spectroscopic information from the data. In the ideal case, theory should be able to describe the full complexity of such a reaction, including one-step direct and two-step transfer charge exchange. The task is simplified considerably by the fact that the reactions of interest are peripheral reactions which are only weakly coupled to the bulk of ion-ion interactions. Thus, a perturbative approach is possible in terms of distorted waves and DWBA methods.

Single charge exchange reactions with complex nuclei are covering a broad range of multipolarities where the peculiarities of heavy ion reactions favor transitions of high multipolarity. Thus, the high selectivity of weak interactions to $L=0$ Fermi or Gamow-Teller transitions is relaxed, allowing to study also matrix elements of the so-called forbidden transitions. With a suitable choice of projectile, even a filter to specific types of transitions can be set. For example, as discussed in \cite{Lenske:1989zz}, the ($^{12}C(0^+,g.s.),^{12}B(1^+,g.s.)$) and ($^{12}C(0^+,g.s.),^{12}N(1^+,g.s.)$) reactions will select in the target Gamow-Teller-type spin-flip transitions while both Fermi- and Gamow-Teller-type transitions are induced in reactions involving odd-even projectiles like ($^{3}He(\frac{1}{2}^+,g.s.)$,$^{3}H(\frac{1}{2}^+,g.s.)$), corresponding to a $p\to n$ reaction, and $(^{7}Li(\frac{3}{2}^-)$,$^{7}Be(\frac{3}{2}^-)$), corresponding to a $n\to p$ reaction, respectively. By a suitable choice of projectile and target the contributions of the transfer charge exchange branch by sequential proton and neutron transfer processes can be minimized as, for example, in the $(^7Li(\frac{3}{2}^-,g.s.),^7Be(\frac{3}{2}^-,g.s.))$ reaction, as discussed in \cite{Cappuzzello:2004afa,Cappuzzello:2004vka,Cappuzzello:201lyi}. That is also the scenario adopted in this work: we consider heavy ion SCE reactions for which one-step direct charge exchange processes should dominate over competing processes.

Experimental and theoretical activities on light ion induced reactions have led to a wealth of accurate spectral information on single charge exchange reactions, from which nuclear matrix elements for single beta decay were obtained, as e.g. in \cite{Ichimura:2006mq,Thies:2012xg,Frekers:2015wga}. For light ion reactions at intermediate energies the close relationship between measured forward angle SCE cross sections and $1\nu 1\beta$ nuclear matrix elements is well established also on theoretical grounds \cite{Taddeucci:1987}. In fact, experimental light ion SCE data have become an important source for spectroscopic results. As a new experimental approach, the NUMEN project at LNS Catania \cite{Cappuzzello:2015ixp,Cappuzzello:2016mxt} is designed to promote heavy ion single and, in particular, also double charge exchange reactions to a new level of accuracy with the perspective to determine nuclear matrix elements for $\Delta Z=\pm 1$ and $\Delta Z=\pm 2$ nuclear excitations. For that purpose, a quantitative reaction theory is necessary which accounts with sufficient accuracy for the interplay of reaction and nuclear structure dynamics. In section \ref{sec:Reaction} we recapitulate and extend the DWBA description of heavy ion SCE reactions where the main focus is on the description of the reaction dynamics and transition form factors. Different to the light ion case, here we have to consider simultaneously excitations in projectile and target which leads to an enriched spectrum of multipoles. In section \ref{sec:SCE_NME} the SCE form factors are considered in detail. A convenient and efficient method of calculation is to use the momentum representation. 

The description of the intrinsic nuclear transition is the topic of section \ref{sec:Structure}. Utilizing nuclear many-body theory we introduce nuclear response functions which lead to a very appropriate formulation of energy-differential heavy ion SCE cross sections. Effects beyond mean-field are briefly addressed.  A major difference between beta decay and hadronic charge exchange reactions are clearly the strong, non-negligible elastic interactions among the reaction partners, reflecting the non-elementary nature of nucleons and nuclei. The handling of the ion-ion initial state ($ISI$) and final state interactions ($FSI$) is discussed in section \ref{sec:DisCoeff}. In the fully microscopic approach ISI/FSI effects are taken into account by double folding optical potentials obtained with nuclear ground state densities from Hartree-Fock-Bogolubov (HFB) calculations and the isoscalar and isovector parts of the NN T-Matrix. Together with QRPA or shell model results for the transition form factors, folded with the isovector parts of the NN T-matrix, we have a powerful toolbox at hand, leading to an almost self-consistent microscopic description of the SCE reaction amplitude. However, in order to understand the reaction mechanism of heavy-ion SCE reactions at comparable low energies and the relation of measured cross sections to nuclear matrix elements, a deeper theoretical analysis is necessary. For that purpose, we discuss in \ref{sec:DisCoeff} an approach which allows to separate in the SCE reaction amplitude the ISI and FSI contributions from the nuclear transition form factors. At the energies of interest, heavy ion reactions are strongly absorbing systems. Under such conditions the \emph{black disk} approximation is shown to account for the essential part of the ISI and FSI distortion effects. Approximating the SCE transition potentials by form factors of Gaussian shape, the distortion coefficients can even be evaluated analytically, as shown in section \ref{sec:GaussFF}. The surprising and important result is that at forward angles heavy ion SCE cross sections can indeed be related by a simple scaling law to nuclear matrix elements, thus essentially matching the light ion case.
This is shown in section \ref{sec:LowMom}.

In section \ref{sec:Applications} the application of the theoretical tools to concrete case of physical interest is illustrated for the $^{18}O+^{40}Ca \to ^{18}F +^{40}K$ SCE reaction recently investigated by the NUMEN group at LNS Catania. Spectroscopic results for charge changing excitations in $^{18}O$ and $^{40}Ca$ obtained by the response function technique are discussed first. Broad space is given to the detailed comparison of full DWBA and plane wave cross sections and the relation to beta-decay transition probabilities. In section \ref{sec:Eikonal} we discuss the mass and energy dependence of distortion effects in the strong absorption limit by exploiting the fact that heavy ion reactions are accompanied by short wave lengths. In section \ref{sec:SumLook} the paper closes with a summary and an outlook. Mathematical details are discussed in a couple of appendices.

\section{Theory of Heavy Ion Single Charge Exchange Reactions}\label{sec:Reaction}

\subsection{Kinematics and Interactions of Single Charge Exchange Reactions}\label{ssec:KinDyn}
Here, we consider ion-ion SCE reactions according to the scheme
\be\label{eq:reaction}
^a_za+^A_ZA\to ^a_{z\pm 1}b+^A_{Z\mp 1}B
\ee
which retain the distribution of masses but change the charge partition by a balanced redistribution of protons and neutrons. For a reaction with a center-of-mass energy $s=(p_a+p_A)^2$, given by the projectile and target four-momenta $p_a$ and $p_A$, respectively, the Lorentz-invariant kinematical transformation into the center-of-momentum frame with the conserved total four-momentum $P=p_a+p_A$ and relative momentum $q$ is defined by
\be
p_a=q+x_aP\quad ;\quad p_A=-q+x_AP
\ee
where $x_a+x_A=1$ and
\be
x_a=\frac{s-m^2_A+m^2_a}{2s}
\ee
which is manifestly of Lorentz-invariant form. At low energies, we recover the well known relation $x_a\to m_a/(m_a+m_A)$ as the limiting result. The relative momentum $q$ is a space-like four-vector, $q^2<0$. In the center-of-momentum frame we have
\be
P_{\alpha,\beta}=\left(\sqrt{s},\mathbf{0}  \right)^T\quad ; \quad q_{\alpha,\beta}=\left(0,\mathbf{k}_{\alpha,\beta}  \right)^T
\ee
for the channels $\alpha=(a,A)$ and $\beta=(b,B)$. While the total 4-momentum $P_\alpha=P_\beta$ is conserved, the relative three-momenta $k_{\alpha,\beta}$ depend on the mass partition,
\be
k^2_{\alpha,\beta}=\frac{1}{4s}\left(s-(m_{a,b}+m_{A,B})^2 \right)\left(s-(m_{a,b}-m_{A,B})^2 \right).
\ee

In distorted wave approximation, the  direct charge exchange reaction amplitude is given by the expression
\be
M_{\alpha\beta}(\mathbf{k}_\beta,\mathbf{k}_\alpha)=\lan\chi^{(-)}_\beta, bB|\hat{T}_{NN}P_x|aA,\chi^{(+)}_\alpha\ran.
\ee
Incoming and outgoing distorted waves are denoted  by $\chi^{(\pm)}_{\alpha,\beta}$, taking care of the proper boundary conditions of asymptotically outgoing and incoming spherical waves, respectively. They depend on the respective channel momenta $\mathbf{k}_{\alpha,\beta}$ and the optical potentials, thus accounting for initial state (ISI) and final state (FSI) interactions.

The charge-changing process is described by the NN T-matrix $\hat{T}_{NN}$. Anti-symmetrization between target and projectile nucleons is taken care of by the operator
\be
P_x=1-P_\sigma P_\tau
\ee
where the spin and isospin projectors are defined as
\be
P_\sigma=\frac{1}{2}\left(1+\bm{\sigma}_a \cdot \bm{\sigma}_A \right) \quad ; \quad P_\tau=\frac{1}{2}\left(1+\bm{\tau}_a \cdot \bm{\tau}_A \right)
\ee
and $\bm{\sigma}_{a,A}$ and $\bm{\tau}_{a,A}$ are spin and isospin Pauli-matrices, acting in the projectile and target nucleus, respectively. We follow the widely used practice and contract $P_x$ and the T-matrix, resulting on the so-called anti-symmetrized T-matrix
\be
T_{NN}=\hat{T}_{NN}P_x
\ee
which corresponds to a (non-relativistic) Fierz transformation. In practice, anti-symmetrization is accomplished by means of a local momentum-dependent pseudo-potential, see e.g. \cite{Satchler:1983}, simplifying the coordinate structure by a localization procedure \cite{Love:1981gb,Franey:1985ye,Hofmann:1998}.

For the present purpose we consider the $\tau_\pm$ rank-1 isovector operators. In  non-relativistic momentum representation, the relevant isovector projectile-target interaction has the structure
\bea
&&T_{NN}(\mathbf{p})=
\sum_{S=0,1,T=1} \big\{ V^{(C)}_{ST}(p^2)\left[\bm{\sigma_a\cdot\sigma_B}\right]^S
\nonumber \\
&&+\delta_{S1} V^{(Tn)}_{T}(p^2)S_{12}(\mathbf{p})   \big\}
\left[ \bm{\tau}_a\cdot\bm{\tau}_A\right]^T
\eea
including isovector central spin-independent ($S=0$) and spin-dependent ($S=1$) interactions with form factors $V^{(C)}_{ST}(p^2)$, respectively, and rank-2 tensor interactions with form factors $V^{(Tn)}_{T}(p^2)$. The form factors are complex-valued scalar functions. Denoting the nucleon isospinors by $|p\ran$ and $|n\ran$, respectively, we use the convention $\lan p|\tau_0|p\ran=+1$ which implies $\tau_{-}|p\ran=|n\ran$.
The standard definition of the rank-2 tensor operator is
\be
S_{12}(\mathbf{p})=\frac{1}{p^2}\left(3\bm{\sigma}_a\cdot \mathbf{p}~\bm{\sigma}_A\cdot \mathbf{p}-\bm{\sigma}_a\cdot \bm{\sigma}_A p^2  \right).
\ee
but for applications to nuclear reactions an equivalent, more suitable representation is used, given by the scalar product of two rank-2 tensors, namely the spherical harmonic $Y_{2M}(\mathbf{\hat{p})}$ and the rank-2 spin operator
\be
S_{2M}=\left[\bm{\sigma_1}\otimes \bm{\sigma_2} \right]_{2M}=\sum_{m_1m_2}{\left(1m_11m_2|2M  \right)\sigma_{1m_1}\sigma_{2m_2}}
\ee
such that
\be
S_{12}=\sqrt{\frac{24\pi}{5}}Y_2\cdot S_2=\sqrt{\frac{24\pi}{5}}\sum_M{Y^*_{2M}(\mathbf{\hat{p}})S_{2M}}
\ee
where $Y^*_{2M}=(-)^M Y_{2-M}$. For the present discussion we neglect two-body spin-orbit interactions in order not to overload the presentation.

Following \cite{Taddeucci:1987} an elegant representation of the T-matrix is obtained in terms of the spin-isospin operators
\be\label{eq:OST}
O_{ST}(i)=\left(\bm{\sigma}_i\right)^S\left(\bm{\tau}_i\right)^T
\ee
which describe the operator structure of both the central and tensor interactions. 
 The operators $O_{ST}$ lead to the rather compact representation
\be
\begin{split}
&T_{NN}(p)=\sum_{S,T}  \big[ V^{(C)}_{ST}(p^2)O_{ST}(1)\cdot O_{ST}(2)\\
&+\delta_{S1}V^{(Tn)}_T(p^2)\sqrt{\frac{24\pi}{5}}Y^*_{2} (\mathbf{\hat{p}})\cdot \left[O_{ST}(1)\otimes O_{ST}(2)\right]_{2} \big]
\end{split}
\ee 
where scalar products are indicated as a dot-product and the rank-2 tensorial coupling affects of course only the spin degrees of freedom. Below, we shall consider only the subset of isovector operators, corresponding to  Fermi-type $S=0,T=1$ and Gamow-Teller-type $S=1$, $T=1$ operators.

\begin{figure}
\begin{center}
\epsfig{file=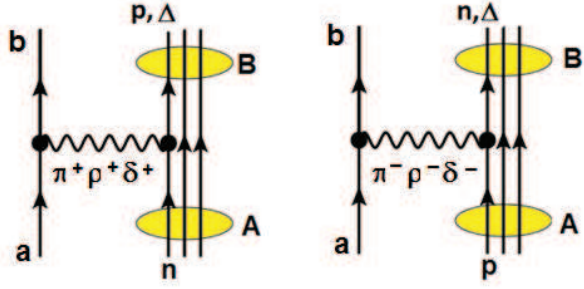, width=8cm}
\caption{Graphical representation of a single charge exchange heavy ion reaction by hadronic interactions corresponding to $\nu\beta$ processes. Both (n,p)-type (left) and (p,n)-type (right) reactions, as seen in the $A\to B$ transition in target system, are displayed, indicating also the exchanged meson.}
\label{fig:SCX}
\end{center}
\end{figure}

Charge changing reactions by strong interactions are off-shell processes mediated by the exchange of virtual particles. They require two reaction partners, which are acting mutually as the source or sink of the charge-changing virtual meson fields, as depicted in Fig.\ref{fig:SCX}. By experimental reasons, the projectile-like ejectile should be preferentially in a particle-stable state (see, however ($d$,$^2He$) reactions \cite{Bugg:1987zr}), thus simplifying the detection. If the ejectile has only a single bound state below the particle emission threshold, the calculations and the interpretation of the spectroscopic data are especially simple.

The matrix element of a single charge exchange reaction, Eq.(\ref{eq:reaction}), can be written in slightly different form as:
\be\label{eq:DWBA_SCX}
M_{\beta\alpha}(\mathbf{k}_\alpha,\mathbf{k}_\beta)=\lan \chi^{(-)}_\beta|\mathcal{U}_{\beta\alpha}|\chi^{(+)}_\alpha\ran
\ee
where $\alpha=\{J_aM_a,J_AM_A\cdots\}$ and $\beta=\{J_bM_b,J_BM_B\cdots\}$ account for the full set of (intrinsic) quantum numbers specifying the initial and final channel states. The nuclear structure information on multipolarities, transition strength and interactions are contained in the (anti-symmetrized) transition potential
\be\label{eq:Uab_SCX}
\mathcal{U}_{\alpha\beta}(\mathbf{r}_\beta,\mathbf{r}_\alpha)=
\lan J_bM_bJ_BM_B|T^{(C)}_{NN}+T^{(Tn)}_{NN}...|J_aM_aJ_AM_A\ran 
\ee
depending on the channel coordinates $\mathbf{r}_{\alpha,\beta}$. If recoil effects due to the change of the mass partitions can be neglected and anti-symmetrization is taken into account by an equivalent effective local interaction, 
one can just consider the local
transition potential
$\mathcal{U}_{\alpha\beta}(\mathbf{r})$ where $\mathbf{r}=\mathbf{r}_\alpha=\mathbf{r}_\beta$. 
Obviously,
by means of Eq.(\ref{eq:Uab_SCX}) the reaction amplitude, Eq.(\ref{eq:DWBA_SCX}) can be rewritten in terms of a sum of reaction amplitudes defined by the tensorial rank $\mathfrak{r}$ of the $NN$-interaction,
\bea
&&M_{\beta\alpha}(\mathbf{k}_\alpha,\mathbf{k}_\beta)=M^{(C)}_{\beta\alpha}(\mathbf{k}_\alpha,\mathbf{k}_\beta)+
M^{(Tn)}_{\beta\alpha}(\mathbf{k}_\alpha,\mathbf{k}_\beta)+...
\nonumber \\
&&=\sum_{\mathfrak{r}=C,Tn...}{M^{(\mathfrak{r})}_{\beta\alpha}(\mathbf{k}_\alpha,\mathbf{k}_\beta)}.
\eea
The differential SCE cross section is defined as
\bea\label{eq:xsec_gen}
&&d\sigma_{\alpha\beta}=\frac{m_\alpha m_\beta}{(2\pi\hbar^2)^2}\frac{k_\beta}{k_\alpha}\frac{1}{(2J_a+1)(2J_A+1)} \times
\nonumber \\
&&\sum_{M_a,M_A\in \alpha;M_b,M_B\in \beta}{\left|\sum_{\mathfrak{r}}{M^{(\mathfrak{r})}_{\alpha\beta}(\mathbf{k}_\alpha,\mathbf{k}_\beta)}\right|^2}d\Omega.
\eea
Reduced masses in the incident and exit channel, respectively, are denoted by $m_{\alpha,\beta}$. In relativistic notation we have
\be
\frac{1}{m_\alpha}=\frac{1}{E_a(k_\alpha)}+\frac{1}{E_A(k_\alpha)}= \frac{1}{m_a}+\frac{1}{m_A}+\mathcal{O}(\frac{k^2_\alpha}{(m_a+m_A)})
\ee
where $E_{a,A}(k_\alpha)=\sqrt{m^2_{a,A}+k^2_\alpha}$ is the relativistic energy in the center-of-momentum frame. $m_\beta$ is defined accordingly.

\subsection{Momentum Representation}\label{ssec:MomRep}
In order to obtain a deeper insight into the interplay of nuclear structure dynamics and beta decay matrix elements on the one side and heavy ion reaction dynamics on the other side, a more detailed study of the process is necessary. A convenient approach is to consider the reaction amplitude in momentum representation. A considerable advantage of that representation is that the transition potential becomes separable into target and projectile transition form factors. They are defined by matrix elements of one-body operators:
\bea\label{eq:FF_SCE}
F^{(ab)}_{ST}(\mathbf{p})&=&\frac{1}{4\pi}\lan J_bM_b| e^{+i\mathbf{p}\cdot\mathbf{r_a}}O_{ST}|J_aM_a\ran \label{eq:Fproj}\\
F^{(AB)}_{ST}(\mathbf{p})&=&\frac{1}{4\pi}\lan J_BM_B| e^{+i\mathbf{p}\cdot\mathbf{r_A}}O_{ST}|J_AM_A\ran \label{eq:Ftarg}
\eea
where $\mathbf{r}_{a,A}$ indicate the intrinsic nuclear coordinates of projectile and target, respectively. For convenience, we have introduced a normalization to the surface volume of the unit sphere. The transitions are determined by the reaction kernel
\be\begin{split}
&K^{(ST)}_{\alpha\beta}(\mathbf{p})=
(4\pi)^2( V^{(C)}_{ST}(p^2)F^{ab\dag}_{ST}(\mathbf{p})\cdot F^{AB}_{ST}(\mathbf{p})\\
&+\delta_{S1}\sqrt{\frac{24\pi}{5}}V^{(Tn)}_{ST}(p^2)
Y^*_{2}(\mathbf{\hat{p}})\cdot \left[F^{ab\dag}_{ST}(\mathbf{p})\otimes F^{AB}_{ST}(\mathbf{p})\right]_{2})
\end{split}
\label{eq:kern}
\ee
where, as before, the rank-2 tensorial coupling relates to the spin degrees of freedom only. Through the form factors $F^{ab,AB}_{ST}$, the kernels contain the spectroscopic information on the nuclear transitions, and the dynamics by the interaction form factors $V^{C,Tn}_{ST}$. In the central interaction part, the scalar product indicates the contraction of the projectile and target form factor with respect to the spin and isospin degrees of freedom. The isospin degrees of freedom are of course projected by the nuclear transitions to the proper combination of $\tau_\pm$ operators. In terms of the reaction kernels, the Fourier transform of the transition potential is found as
\be\label{eq:TransPot}
\mathcal{U}_{\alpha\beta}(\mathbf{p})=\sum_{ST}{K^{(ST)}_{\alpha\beta}(\mathbf{p}) }.
\ee
The transition potential in coordinate space is obtained by the inverse Fourier transform
\be
\mathcal{U}_{\alpha\beta}(\mathbf{r})=\int{\frac{d^3p}{(2\pi)^3}e^{-i\mathbf{p}\cdot \mathbf{r}}}\mathcal{U}_{\alpha\beta}(\mathbf{p}),
\ee
to be used in standard DWBA (or coupled channels) calculations, as e.g. in \cite{Lenske:1989zz}.

Here, however, we continue to use the momentum space approach by reasons which are becoming obvious below. Within that formulation, it remains to evaluate the integration over the relative motion degrees of freedom which leads to the distortion coefficients
\be\label{eq:Nab}
N_{\alpha\beta}(\mathbf{k}_\alpha,\mathbf{k}_\beta,\mathbf{p})=\frac{1}{(2\pi)^3}\lan \chi^{(-)}_\beta|e^{-i\mathbf{p}\cdot \mathbf{r}}|\chi^{(+)}_\alpha\ran ,
\ee
which will be discussed in detail below in sect. \ref{sec:DisCoeff}. Finally, by folding the kernels with the distortion coefficients  we obtain the full reaction amplitudes,
\be\label{eq:ME_SCE}
M_{\alpha\beta}(\mathbf{k}_\alpha,\mathbf{k}_\beta)=\sum_{ST}{\int{d^3p K^{(ST)}_{\alpha\beta}(\mathbf{p})N_{\alpha\beta}(\mathbf{k}_\alpha,\mathbf{k}_\beta,\mathbf{p})  }},
\ee

now dressing the reaction amplitude by initial and final state ion-ion interactions. Formally, the above relation is fully equivalent to the corresponding DWBA amplitude. The momentum representation, however, has the important advantage that
the intrinsic nuclear transition dynamics and the reaction dynamics are separated, although to the expense of an additional momentum integration, Eq.(\ref{eq:ME_SCE}). The latter, however, does not pose a special problem in our case. As will be shown below, for heavy ion scattering the distortion coefficient can be evaluated in closed form under realistic assumptions and the nuclear transition form factors typically decrease rapidly beyond $p\sim 300$~MeV/c, thus facilitating numerical evaluations.

\section{SCE Form Factors and Nuclear Matrix Elements}\label{sec:SCE_NME}

\subsection{General Features of SCE Form Factors}
The projectile and target transition form factors, Eq.(\ref{eq:FF_SCE}), (\ref{eq:Ftarg}), are of a very general structure accounting for the complete set of multipoles as contained in the plane waves. The integration over the nuclear intrinsic coordinates, however, will project on a subset of multipoles according to the multipolarity of the transitions $a\to b$  and $A\to B$, respectively.

A special feature is encountered in the rank-2 tensor amplitudes $M^{(Tn)}_{\alpha\beta}$. By evaluating the integrals explicitly one finds that the presence of the spherical harmonics of order 2 induces a corresponding rank-2 tensorial coupling of the nuclear transition multipoles, Eq.(\ref{eq:kern}). This has important consequences for ion-induced SCE reactions.
Indeed, except for transitions involving only s-wave proton and neutron orbitals, Gamow-Teller like excitations are typically a mixture of a leading multipolarity $L_1$ and a sub-leading one with $L_2=|L_1\pm 2|$. Beta decay strongly favors the multipolarity with the lower value of $L_{1,2}$. That selectivity is missing in strong interactions. Since in heavy ion charge exchange reactions especially processes with large angular momentum transfer are favored, the whole spectrum of multipolarities becomes visible. The rank-2 tensor interactions are mixing the orbital angular momenta $L=|J\pm 1|$ for a given total angular momentum transfer $J$, thus additionally enhancing and giving access to the \emph{beta decay-forbidden} components.

The nuclear transitions in either target or projectile are induced by one-body operators of the type
\be
\mathcal{R}_{ST}(\mathbf{p},\mathbf{r})=\frac{1}{4\pi}e^{i\mathbf{p}\cdot \mathbf{r}}O_{ST}
\ee
In a broader formal context, these operators are in fact to be considered as vertex operators as part of a Lagrangian interaction density in the sense of a (non-relativistic) field theory. Such considerations pave the way to second quantization. In second quantization the charge-changing transition operator is given by
\bea
&&\mathcal{R}_{ST}(\mathbf{p},\mathbf{r})\mapsto \mathcal{R}_{ST}(\mathbf{p},a^\dag a) \nonumber \\
&&= \sum_{j_p,m_p,j_n,m_n}\lan j_pm_p|\mathcal{R}_{ST}|j_nm_n\ran a^\dag_{j_pm_p} a_{j_nm_n} \nonumber \\
&&+\lan j_nm_n|\mathcal{R}_{ST}|j_pm_p\ran a^\dag_{j_nm_n} a_{j_pm_p}
\eea
where the summation extends over a complete set of proton and neutron single particle states and ($j_{p,n},m_{p,n}$) represent the full set of quantum numbers specifying the orbitals. Thus, the transition operator has been expressed in terms of the non-diagonal elements of the one-body density matrix  $\sim a^\dag_i a_{j|_{j\neq i}}$. A partial wave expansion of the plane wave leads to the multipole tensor representation
\bea
&&\mathcal{R}_{ST}(\mathbf{p},a^\dag a)\nonumber \\
&&= \sum_{LM_L}Y^*_{LM_L}(\mathbf{\hat{p}})\sum_{j_pm_p,j_nm_n}\big\{ U^{STLM_L}_{j_pm_pj_nm_n}(p^2) a^\dag_{j_pm_p} a_{j_nm_n}\nonumber \\
&&+U^{STLM_L}_{j_nm_nj_pm_p}(p^2)a^\dag_{j_nm_n} a_{j_pm_p}\big\}
\eea
with particle-hole type one-body transition matrix elements
\be
\begin{split}
&U^{STLM_L}_{j_pm_pj_nm_n}(p^2)=\lan j_pm_p|j_L(pr)i^L Y_{LM_L}(\mathbf{\hat{r}})O_{ST}|j_nm_n\ran \\
&U^{STLM_L}_{j_nm_nj_pm_p}(p^2)=\lan j_nm_n|j_L(pr)i^L Y_{LM_L}(\mathbf{\hat{r}})O_{ST}|j_pm_p\ran
\end{split}
\ee
describing $n\to p$ and $p\to n$ transitions, respectively. We introduce the irreducible tensor operators (for $S=0,1$)
\be\label{eq:RSLJ}
\begin{split}
&R_{LSJM}(\mathbf{r},\bm{\sigma})=\sum_{M_L,M_S}{\left(LM_L SM_S|JM \right)i^L}Y_{LM_L}(\mathbf{\hat{r}})\left(\sigma_{M_S}  \right)^S \\
&=\left[i^LY_{L}(\mathbf{\hat{r}})\otimes \left(\bm{\sigma}\right)^S \right]_{JM}
\end{split}
\ee
by which the one-body transition matrix elements become
\begin{widetext}
\bea
U^{STLM_L}_{j_pm_pj_nm_n}(p^2)=\sum_{JM}{\left(LM_LSM_S|JM \right)\lan j_pm_p|j_L(pr)R_{LSJM}(\mathbf{r},\bm{\sigma})\tau_+|j_nm_n\ran }\\
U^{STLM_L}_{j_nm_nj_pm_p}(p^2)=\sum_{JM}{\left(LM_LSM_S|JM \right)\lan j_nm_n|j_L(pr)R_{LSJM}(\mathbf{r},\bm{\sigma})\tau_-|j_pm_p\ran }
\eea
\end{widetext}
Applying the Wigner-Eckart theorem \cite{Edmonds:1957} the matrix elements separate into a Clebsch-Gordan coefficient and a reduced matrix element. This allows to perform the summation over the proton and neutron magnetic quantum numbers leading to the one-body transition density operators
\be
A^\dag_{JM}(j_pj_n)=\sum_{m_pm_n}{\left(j_pm_pj_nm_n|JM  \right)a^\dag_{j_pm_p}\tilde{a}_{j_nm_n}  }.
\ee
where $\tilde{a}_{jm}=(-)^{j+m}a_{j-m}$ denotes the conjugated operator. The proton-neutron and the neutron-proton particle-hole operators are related by Hermitian conjugation,
\be
A^\dag_{JM}(j_nj_p)=(-)^{J+M}A_{J-M}(j_pj_n)
\ee
reflecting charge-conjugation symmetry. The reduced isovector matrix elements are
\bea
\bar{U}^{LSJ}_{j_pj_n}(p^2)=\frac{\sqrt{2}}{\hat{J}}\lan \ell_p s_p j_p||j_L(pr)R_{LSJ} ||\ell_n s_nj_n\ran\\
\bar{U}^{LSJ}_{j_nj_p}(p^2)=\frac{\sqrt{2}}{\hat{J}}\lan \ell_ns_n j_n||j_L(pr)R_{LSJ} ||\ell_p s_pj_p\ran
\eea
where $\hat{J}=\sqrt{2J+1}$ and $s_p=s_n=\frac{1}{2}$. The factor $\sqrt{2}$ results from the isospin structure of the isovector nucleon-meson vertices. These steps lead to the representation of the transition operators in terms of irreducible tensor components of conserved total angular momentum 
{$J$}
\begin{widetext}
\bea
\sum_{m_pm_n}{U^{STLM_L}_{j_pm_pj_nm_n}(p^2)a^\dag_{j_pm_p} a_{j_nm_n}}=\sum_{JM}{\left(LM_LSM_S|JM \right)\bar{U}^{LSJ}_{j_pj_n}(p^2)  A^\dag_{JM}(j_pj_n) }\\
\sum_{m_pm_n}{U^{STLM_L}_{j_nm_nj_pm_p}(p^2)a^\dag_{j_nm_n} a_{j_pm_p}}=\sum_{JM}{\left(LM_LSM_S|JM \right)\bar{U}^{LSJ}_{j_nj_p}(p^2)  A^\dag_{JM}(j_nj_p) }
\eea
\end{widetext}
Thus, the transition operator becomes
\bea
\mathcal{R}_{ST}(\mathbf{p},a^\dag a)=\sum_{LM_LJM}Y^*_{LM_L}(\mathbf{\hat{p}})\left(LM_LSM_S|JM \right)\nonumber \\
\sum_{j_p,j_n}{ \left\{\bar{U}^{LSJ}_{j_pj_n}(p^2)A^\dag_{JM}(j_pj_n)+\bar{U}^{LSJ}_{j_nj_p}(p^2)A^\dag_{JM}(j_nj_p)  \right\} }
\eea
The transition form factors are now given as
\bea
&&F^{(ab)}_{SM_S}(\mathbf{p})=\sum_{\lambda_a\mu_aI_aN_a}Y^*_{\lambda_a\mu_a}(\mathbf{\hat{p}})\left(\lambda_a\mu_aSM_S|I_aN_a \right) \nonumber \\
&&\bigg\{\sum_{j_p,j_n}{
\bar{U}^{\lambda_aSI_a}_{j_pj_n}(p^2)\lan J_bM_b|A^\dag_{I_aN_a}(j_pj_n)|J_aM_a\ran} \nonumber \\
&&+
\sum_{j_p,j_n}{\bar{U}^{\lambda_aSI_a}_{j_nj_p}(p^2)\lan J_bM_b|A^\dag_{I_aN_a}(j_nj_p)|J_aM_a\ran }\bigg\}
\eea
and correspondingly
\bea
&&F^{(AB)}_{SM_S}(\mathbf{p})=\sum_{\lambda_A\mu_AI_AN_A}Y^*_{\lambda_A\mu_A}(\mathbf{\hat{p}})\left(\lambda_A\mu_ASM_S|I_AN_A\right) \nonumber \\
&&\bigg\{\sum_{j_p,j_n}{
\bar{U}^{\lambda_ASI_A}_{j_pj_n}(p^2)\lan J_BM_B|A^\dag_{I_AN_A}(j_pj_n)|J_AM_A\ran}  \nonumber \\
&&+ \sum_{j_p,j_n}{\bar{U}^{\lambda_ASI_A}_{j_nj_p}(p^2)\lan J_BM_B|A^\dag_{I_AN_A}(j_nj_p)|J_AM_A\ran } \bigg\}
\eea
Of course, for a given reaction only one of the two
terms in Eqs.(40) and (41)
is effectively contributing to the transition: if e.g. a $p^{-1}n$-type transition is occurring in the projectile, only the parts containing operators of $a^\dag_n a_p$ structures give non-vanishing contributions while the complementary $a^\dag_p a_n$ operator-branch is only active in the target and \emph{vice versa}.

The spectroscopy of the charge exchange process is now contained in one-body transition density matrix elements defined as
\be
D^{JM}_{j_cj_d}(J_fM_f,J_iM_i)=\lan J_fM_f|A^\dag_{JM}(j_cj_d)|J_iM_i\ran
\ee
The Wigner-Eckart theorem leads to
\be
\begin{split}
&D^{JM}_{j_cj_d}(J_fM_f,J_iM_i)\\ 
&=(-)^{J_f-M_f}\left(J_fM_fJ_i-M_i|J-M \right)\bar{D}^J_{j_cj_d}(J_f,J_i)\\
\end{split}
\ee
with the reduced one-body transition density
\be
\bar{D}^J_{j_cj_d}(J_f,J_i)=\frac{1}{\hat{J}}\lan J_f||A^\dag_J(j_cj_d)||J_i\ran
\ee

If the parent state has $J_i=0$, the result simplifies to
\be
D^{JM}_{j_cj_d}(J_fM_f,00)=\bar{D}^J_{j_cj_d}\delta_{JJ_f}\delta_{MM_f}
\ee
where
\be
\bar{D}^J_{j_cj_d}=\frac{1}{\hat{J}}\lan J||A^\dag_J(j_cj_d)||0^+\ran
\ee
The same simplification is obtained for the case $J_f=0$.

Obviously, the one-body transition densities are the elements of central importance for the spectroscopy of the charge exchange process. They are providing access to the many-body structure of the underlying nuclear wave functions. The evaluation of the one-body transition densities requires knowledge of the structure of the initial and final nuclear states which is a demanding task for nuclear theory.

\subsection{ Multipole Structure of the Reaction Kernel}
After discussion of the transition operators, now we can investigate the multipole content of the nuclear transition form factors, defined in Eq.(\ref{eq:FF_SCE}) and Eq.(\ref{eq:Ftarg}) for projectile and target excitations, respectively. By standard angular momentum coupling techniques, we obtain
\begin{widetext}
\bea
F^{(ab)}_{ST}(\mathbf{p})&=&\sum_{L,M_L,J_1,M_1}{\left(J_aM_aJ_bM_b|J_1M_1 \right)\left(LM_{L}SM_S|J_1M_1 \right)f^{(ab)}_{LSJ_1}(p^2)i^{L}Y_{LM_{L}}(\bm{\hat{p}})}  \label{eq:FprojM}\\
F^{(AB)}_{ST}(\mathbf{p})&=&\sum_{L,M_{L},J_2,M_2}{\left(J_AM_AJ_BM_B|J_2M_2 \right)\left(LM_{L}SM_S|J_2M_2 \right)f^{(AB)}_{LSJ_2}(p^2)i^{L}Y_{LM_{L}}(\bm{\hat{p}})}
\label{eq:FtargM}
\eea
\end{widetext}
The total angular momentum transfer in the projectile and target system are given by $J_{1,2}$, defining the set of multipole components which are contributing to a given reaction leading from initial states $J_{a,A}$ to final states $J_{b,B}$. These relations are expressed by the first Clebsch-Gordan coefficient in the above equations. In accordance with the investigations of the previous section, these multipoles carry substructures given by the coupling of orbital ($L_{1,2}$) and spin ($S_{1,2}$) angular momentum transfers, as expressed by the second Clebsch-Gordan coefficients in Eq.(\ref{eq:FprojM}) and Eq.(\ref{eq:FtargM}), respectively.

The recoupling procedure follows standard rules \cite{Edmonds:1957} and is discussed in Appendix \ref{app:Irreps}. Anticipating the results, the reaction kernels become
\bea
&&K^{(ST)}_{\alpha\beta}(\mathbf{p})=
\sum_{J_1M_1,J_2M_2,LM}\left(J_aM_aJ_bM_b|J_1M_1 \right)\nonumber \\
&&\left(J_AM_AJ_BM_B|J_2M_2 \right)\left(J_1M_1J_2M_2|LM \right)
i^LY_{LM}(\bm{\hat{p}})\nonumber \\
&&\left(V^{(C)}_{ST}(p^2)F^{J_1J_2}_{LS}(p^2) +\delta_{S1}V^{(Tn)}_T(p^2) H^{J_1J_2}_{LS}(p^2) \right),
\eea
including central and rank-2 tensor interactions. Correspondingly, for the reaction amplitude we obtain the expression
\be
\begin{split}
&M_{\alpha\beta}(\mathbf{k}_\alpha,\mathbf{k}_\beta)=\sum_{J_1M_1,J_2M_2,LM}\left(J_aM_aJ_bM_b|J_1M_1 \right) \\
&\left(J_AM_AJ_BM_B|J_2M_2 \right)\left(J_1M_1J_2M_2|LM \right) \\
&\int d^3p N_{\alpha\beta}(\mathbf{\mathbf{k}}_\alpha,\mathbf{k}_\beta,\mathbf{p}) i^LY_{LM}(\bm{\hat{p}}) M^{(J_aJ_A,J_bJ_B)}_{LJ_1J_2}(p^2)
\end{split}
\ee
where
\be
\begin{split}
&M^{(J_aJ_A,J_bJ_B)}_{LJ_1J_2}(p^2)=\sum_{S,T}\delta_{T1}\\
&\left(V^{(C)}_{ST}(p^2)F^{J_1J_2}_{LS}(p^2) +\delta_{S1}V^{(Tn)}_T(p^2) H^{J_1J_2}_{LS}(p^2)  \right)
\end{split}
\ee
Exploiting the completeness and orthogonality relations of Clebsch-Gordan coefficients, the double-differential cross section becomes
\begin{widetext}
\be
\frac{d^2\sigma_{\alpha\beta}}{d\Omega dE_x}=
\sum_{bB}\frac{m_\alpha m_\beta}{(2\pi\hbar^2)^2}\frac{k_\beta}{k_\alpha}\frac{1}{(2J_a+1)(2J_A+1)}
\sum_{LM,J_1J_2}
\left|\int{d^3p N_{\alpha\beta}(\mathbf{k}_\alpha,\mathbf{k}_\beta,\mathbf{p})
Y_{LM}(\mathbf{\hat{p}})M^{(J_aJ_A,J_bJ_B)}_{LJ_1J_2}(p^2)}   \right|^2\delta(E^*_b+E^*_B-E_x)
\ee
\end{widetext}
where by practical reasons we may impose the constraint that the ejectile $b$ of excitation energy $E^*_b$ should be in a bound state. The Dirac delta-function projects the sum of excitation energies onto the total effective energy loss $E_x$.

A substantial simplification is found for $J_a=0^+=J_A$. In that case, $J_1=J_b$ and $J_2=J_B$ and
\bea
&&F^{(ab)}_{ST}(\mathbf{p}) = \sum_{L,M}\left(LMSM_S|J_bM_b \right)
\nonumber \\
&&\times f^{(ab)}_{LSJ_b}(p^2)i^{L}Y_{LM}(\bm{\hat{p}})\\
&&F^{(AB)}_{ST}(\mathbf{p}) = \sum_{L,M}\left(LMSM_S|J_BM_B \right)
\nonumber \\
&&\times f^{(AB)}_{LSJ_B}(p^2)i^{L}Y_{LM}(\bm{\hat{p}})
\eea
Besides the triangle rule of angular momentum coupling, the allowed values of the orbital angular momentum transfer $L$ are constrained further by parity selection rules. For $0^+\to J^{\pi_J}$ transitions $\pi_J=(-)^{J}$ and $\pi_J=(-)^{J+1}$ for natural and unnatural parity transitions, respectively, and $\pi_J=(-)^L$ must be fulfilled. For natural parity transitions with $L=J$, non-spin-flip $S=0$ and spin-flip $S=1$ transitions are allowed while for unnatural parity transitions with $L=|J\pm 1|$ only transitions with $S=1$  will contribute. Finally, an important feature of the rank-2 tensor interaction is that transition form factors differing by $\Delta L=2$ in total orbital angular momentum transfer are coupled by the rank-2 spherical harmonic in a parity-conserving manner, see e.g. Ref.\cite{Lenske:1989zz}.

\section{Response Function Theory of Charge Changing Nuclear Excitations}\label{sec:Structure}

\subsection{Survey of the Response Function Method}
\begin{figure}
\begin{center}
\epsfig{file=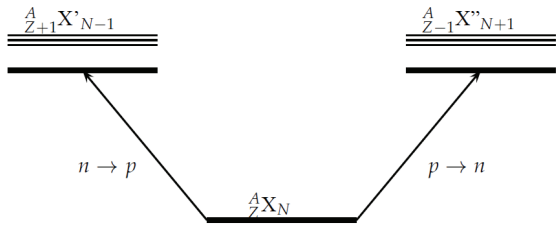, width=8cm}
\caption{Illustration of the two branches of a nuclear charge changing excitation, including the $np^{-1}$ (right) and the complementary $pn^{-1}$ (left) branch, respectively.}
\label{fig:nppn}
\end{center}
\end{figure}

The one-body operators acting in SCE reactions couple directly to the one particle-one hole components ($NN^{-1}$) of the nuclear states. If a $\beta^+$-type $np^{-1}$ branch is exited in the target, the projectile undergoes the complementary $pn^{-1}$ transition and \emph{vice versa}. In Fig. \ref{fig:nppn} the two branches probed in either a $p\to n-$ or a $n\to p-$type SCE reactions are illustrated. Thus, in particular we must consider the nuclear response in the $NN^{-1}$ or two-quasiparticle (2QP) excitation channel. A formally and practically elegant method to describe the response of a nucleus for an external perturbation is the Green's function method and the related polarization propagator, well known in the theory of interacting quantum many-body systems \cite{FW:1971}. In our case, the perturbation is given by the effective one-body fields provided by the projectile-target interaction. In \cite{Baker:1997} the theoretical background and the application of that approach to light-ion reaction data has been discussed in due detail. In \cite{Cappuzzello:2004afa,Cappuzzello:2004vka,Cappuzzello:201lyi} previous applications to heavy ion SCE reactions are found. Here, we only sketch the essential steps of importance for charge exchange reactions.

The key quantity of the response function formalism is the polarization propagator $\Pi_{\kappa\lambda}$, defined as the ground state expectation value of (external) one-body fields $T_{\lambda}$ and $T_{\kappa}$ with the interacting 4-point function $G$.
\be\label{eq:PolProp1}
\Pi_{\kappa\lambda}(\mathbf{q}',\mathbf{q},\omega)=\lan 0|T^\dag_\lambda(\mathbf{q}')G(\omega)T_\kappa(\mathbf{q})|0\ran
\ee
where $\mathbf{q}$ and $\omega$ are the momentum and energy transfer, respectively. The 4-point function $G$ describes the propagation of interacting 2QP states. $G$ is defined in terms of the non-interacting 4-point function $G_0$ and the residual 2QP interaction $V_{QQ}$  and obeys the Dyson equation
\be\label{eq:Dyson}
G(\omega) = G_0(\omega)+G_0(\omega)V_{QQ}G(\omega)
\ee

A particularly simple approach is obtained by expressing $V_{QQ}$ in separable form. This amounts to expand the 2QP residual interaction into a series of bilinears of one-body multipole operators
\be
V_{QQ}(1,2)=\sum_{\gamma\in \{LSJM\}}{\kappa_{\gamma}U^\dag_{\gamma}(1)U_{\gamma}(2)}
\ee
where $S=0,1$ and only the $T=1$ isovector components will contribute to charge changing excitations. The one-body multipole components  $U_\gamma$, where $\gamma={LSJM}$, are given by the previously introduced one-body multipole tensor operators $R_{LSJM}$, Eq.(\ref{eq:RSLJ}), and a scalar (radial) form factor. This technique allows to solve the Dyson equation algebraically, thus obtaining the
polarization propagator $\Pi_{\kappa\lambda}$ of the interacting system. The spectroscopic response functions are then defined by
\be
R_{\kappa\lambda}(\mathbf{q}',\mathbf{q},\omega)=-\frac{1}{\pi}Im\left[ \Pi_{\kappa\lambda}(\mathbf{q}',\mathbf{q},\omega) \right]
\ee
Explicitly,
\be
R_{\kappa\lambda}(\mathbf{q}',\mathbf{q},\omega)=\sum_c{\lan 0|T^\dag_\lambda(\mathbf{q}')|c\ran\delta(E_c-\omega)\lan c|T_\kappa(\mathbf{q})|0\ran}
\ee
which shows that the response functions contain the spectral distribution of states $|c\ran$ and the transition strength due to the coupling to the external fields $T_{\kappa,\lambda}$ 
with a structure given by the operators $O_{ST}$.

Extending the description to higher order dissipative phonon self-energies, the Dirac delta-function is changed into a (shifted and fragmented) Lorentz-type energy distribution with a finite width which is given by the imaginary part of the particle-hole self-energy.

\subsection{Response Functions and Double Differential Cross Sections}
In a heavy ion SCE reaction both ions may be excited. Thus, depending on the detection method, experiments may record the spectral distribution in both the projectile and the target, in only one of the nuclei, or in a fully inclusive measurement only the unresolved full yield of a collision.  The most involved case is the differential measurement of the outgoing nuclei in coincidence with the decay products of excited states allowing to identify the spectral state of each nucleus. A less demanding, semi-inclusive approach is to identify the ejectile by mass and charge, thus excluding the projectile excitations to unbound particle states. Such measurements, however, are recording essentially the total energy loss and momentum transfer and any excited state of the projectile with energy below the particle emission threshold is contributing to such a cross section, differential in momentum transfer (i.e. scattering angle) and energy loss. The response function formalism accounts appropriately for such conditions. The double differential SCE cross section at total excitation energy (or energy loss) $E_x$ is given as:
\begin{widetext}
\bea
\frac{d^2\sigma_{\alpha\beta}}{dE_xd\Omega}\sim tr_{SS'}&&\int dE_B \int d^3p_1 \int d^3p_2
N_{\alpha\beta}(\mathbf{p}_1,\mathbf{q}_{\alpha\beta})N^*_{\alpha\beta}(\mathbf{p}_2,\mathbf{q}_{\alpha\beta}) V^{(C)}_{ST}(p^2_1)V^{(C)*}_{S'T}(p^2_2)
\nonumber \\&&
 R^{(a)}_{SS'}(\mathbf{p}_1,\mathbf{p}_2,E_x-E_B)R^{(A)}_{SS'}(\mathbf{p}_1,\mathbf{p}_2,E_B)
\eea
\end{widetext}
where the reaction amplitudes have been expressed in the momentum representation and ISI and FSI effects are contained in the distortion coefficients $ N_{\alpha\beta}$. For simplicity, only  central interactions have been considered. The momentum transfer is indicated by $\mathbf{q}_{\alpha\beta}=\mathbf{k}_\alpha-\mathbf{k}_\beta$. $R^{(a)}_{SS'}$ denotes the isovector SCE projectile response function for spin transfer $S$ and $S'$, respectively, and a corresponding notation is used for the target contributions. As indicated by the traces over the spin projections of $S$ and $S'$, the target and projectile response functions are contracted such that, in total, a scalar function in spin and all other intrinsic and kinematical degrees of freedom is obtained. By an expansion of the response functions into multipoles the detailed spectroscopic structure of projectile and target is accessible. This is achieved by essentially the same techniques as applied in sect.\ref{sec:SCE_NME} and in Appendix \ref{app:Irreps}.

\section{Initial and Final State Interactions}\label{sec:DisCoeff}

\subsection{Distorted Waves and Distortion Coefficient}

For heavy ion reactions the elastic interactions in the initial and the final channel are playing a key role for a quantitative description of cross sections. In a microscopic description, the optical potentials are obtained in a double-folding approach \cite{Satchler:1983}. In the many cases where elastic scattering data are not available the folding approach is in fact the only way to obtain information on elastic ion-ion interactions. The double-folding potential is defined in terms of the NN T-matrix and the ground state densities of the interacting nuclei. Thus, specific contributions e.g. due to the coupling to break-up and transfer channels or rotational and vibrational excitations are not included. Experience, however, shows that at kinetic energies above the Coulomb-barrier the double folding potential are accounting surprisingly well for the elastic interactions. The reason is that most of the interaction effects are already covered by the multiple scattering series inherent to an elastic amplitude iterated to all orders, as in the case of the solutions of a Schroedinger-type wave equation. 
A commonly used approach is the impulse approximation, amounting to consider the isoscalar and isovector parts of the free space NN T-matrix. Since there are no heavy ion polarization data available, spin-dependent interactions are neglected. Coulomb-interactions, of course, must be included as well. They are treated by folding the two-body projectile-target nucleon Coulomb-interaction with the nuclear charge densities. Thus, we use
\be
U_{opt}(\mathbf{r})=V(\mathbf{r})-iW(\mathbf{r})+U_c(\mathbf{r})
\ee
where the imaginary part must in total correspond to an absorptive potential, guaranteeing a positive reaction cross section. The distorted waves are then defined by wave equations with the  generic structure
\be
\left(-\frac{\hbar^2}{2m_{\gamma}}\bm{\nabla}^2+U_\gamma(\mathbf{r})-E^{(rel)}_\gamma \right)\chi^{(\pm)}_\gamma(\mathbf{r},\pm \mathbf{k})=0
\ee
for $\gamma\in \{\alpha,\beta\}$ and $E^{rel}_{\alpha,\beta}=\sqrt{s}-M_{A,B}-M_{a,b}$ denotes the kinetic energy available in the projectile-target rest frame.

\subsection{Separation Approach to the Distortion Coefficient}
From Eq.(\ref{eq:Nab}) the limiting case of a system without ISI and FSI interactions is immediately found by replacing the distorted waves by plane waves ($PW$). Then, the distortion coefficient reduces to
\be
N^{(PW)}_{\alpha\beta}(\mathbf{k}_\alpha,\mathbf{k}_\beta,\mathbf{p})=\delta(\mathbf{k}_\alpha-\mathbf{k}_\beta-\mathbf{p})
\ee
and we retrieve the reaction amplitude, Eq.(\ref{eq:ME_SCE}), in lowest order Born approximation as
\be
M^{(B)}_{\alpha\beta}(\mathbf{k}_\alpha,\mathbf{k}_\beta)\equiv \mathcal{U}_{\alpha\beta}(\mathbf{q}_{\alpha\beta})
\label{Born_a}
\ee
In order to establish the connection of the full distorted wave (DW) amplitudes to those of the PW limit we need to consider the distortion coefficient in more detail. For that purpose, we separate the distorted waves $|\chi^{(\pm)}_{\alpha,\beta}\ran$ into plane waves $|\mathbf{k}_{\alpha,\beta}\ran$ and a residual distortion amplitude $u^{(\pm)}_{\alpha,\beta}(\mathbf{k}_{\alpha,\beta},\mathbf{r})$. On very general grounds, such a separation is justified by the representation of an interacting wave in terms of the  M{\o}ller-wave operator acting on a plane wave \cite{WuOhmura:1962}. Using
\be\label{eq:DisAmp}
\eta_{\alpha\beta}=u^{(-)\dag}_{\beta}u^{(+)}_\alpha
\ee
and assuming that $u^{(\pm)}_{\alpha,\beta}$ and $\mathcal{U}_{\alpha\beta}$ commute, the DWBA matrix element, Eq.(\ref{eq:DWBA_SCX}), becomes a matrix element of formal PW-structure
\be\label{eq:M_Born}
M_{\alpha\beta}(\mathbf{k}_\alpha,\mathbf{k}_\beta)=
\lan \mathbf{k}_\beta|\eta_{\alpha\beta}\mathcal{U}_{\alpha\beta}|\mathbf{k}_\alpha\ran 
\ee
but with a kernel modified by the ion-ion ISI and FSI effects as contained in the distortion amplitude $\eta_{\alpha\beta}$. Then, from Eq.(\ref{eq:Nab}) we find
\be
N_{\alpha\beta}(\mathbf{k}_\alpha,\mathbf{k}_\beta,\mathbf{p})=\eta_{\alpha\beta}(\mathbf{q}_{\alpha\beta}-\mathbf{p})
\ee
Since for a non-interacting system $\eta_{\alpha\beta}(\mathbf{r})\to \eta^{(PW)}_{\alpha\beta}(\mathbf{r})=1$ it is useful to consider $\Delta_{\alpha\beta}(\mathbf{r})=1-\eta_{\alpha\beta}(\mathbf{r})$. This allows to split the distortion coefficient as follows
\be
N_{\alpha\beta}(\mathbf{k}_\alpha,\mathbf{k}_\beta,\mathbf{p})=
N^{(PW)}_{\alpha\beta}(\mathbf{k}_\alpha,\mathbf{k}_\beta,\mathbf{p})-\Delta_{\alpha\beta}(\mathbf{q}_{\alpha\beta}-\mathbf{p})
\ee
where now the ISI and FSI effects are fully contained in the Fourier transform of $\Delta_{\alpha\beta}$. Correspondingly, the reaction amplitude becomes
\bea
&&M_{\alpha\beta}(\mathbf{k}_\alpha,\mathbf{k}_\beta)= \nonumber \\
&&M^{(B)}_{\alpha\beta}(\mathbf{q}_{\alpha\beta})
-\int d^3p \Delta_{\alpha\beta}(\mathbf{q}_{\alpha\beta}
-\mathbf{p})M^{(B)}_{\alpha\beta}(\mathbf{p})=\nonumber \\
&&M^{(B)}_{\alpha\beta}(\mathbf{q}_{\alpha\beta})
-\int{d^3q}\Delta_{\alpha\beta}(\mathbf{q})M^{(B)}_{\alpha\beta}(\mathbf{q}_{\alpha\beta}-\mathbf{q})
\label{reac_amp}
\eea
Assuming that $\Delta_{\alpha\beta}$ is spherical symmetric, we obtain
\be\label{M_1}
\begin{split}
&M_{\alpha\beta}(\mathbf{k}_\alpha,\mathbf{k}_\beta)=
M^{(B)}_{\alpha\beta}(\mathbf{q}_{\alpha\beta})\\
&- 4\pi 
\int^{\infty}_0dq q^2\Delta_{\alpha\beta}(q)\bar{M}^{(B)}_{\alpha\beta}(\mathbf{q}_{\alpha\beta},q)\\
\end{split}
\ee
where
\be
\bar{M}^{(B)}_{\alpha\beta}(\mathbf{q}_{\alpha\beta},q)=\frac{1}{4\pi}\int{d\Omega_q M^{(B)}_{\alpha\beta}(\mathbf{q}_{\alpha\beta}-\mathbf{q})}
\label{M_2}
\ee
denotes the Born-amplitude averaged over the orientations of $\mathbf{q}$. Referring to the definition of the Born amplitude, Eq.(\ref{Born_a}), the angle integral can be performed analytically and we obtain
\be
\bar{M}^{(B)}_{\alpha\beta}(\mathbf{q}_{\alpha\beta},q)=\int{d^3r e^{i\mathbf{q}_{\alpha\beta}\cdot \mathbf{r}}\mathcal{U}_{\alpha\beta}( \mathbf{r})j_0(qr)}
\label{M_3}
\ee

The above relations involve in fact different scales which allow a separation \emph{ansatz}: The distribution of the momenta $q$ is controlled by the optical model quantity $\Delta_{\alpha\beta}$ with a typical momentum spread of the order of the potential radius, i.e. $\Delta q_{reac}\sim \frac{1}{R_{opt}}\leq 50$~MeV/c. The momentum structure of the Born-amplitude is determined by the charge-changing nuclear form factors $F^{(ab),(AB)}$. Their overall momentum dependence is closely related to the Fermi-momenta of protons and neutrons, thus $\Delta q_{nucl}\sim k_F \sim 300$~MeV/c. Therefore, we introduce the separation \emph{ansatz}
\be
\bar{M}^{(B)}_{\alpha\beta}(\mathbf{q}_{\alpha\beta},q)\simeq M^{(B)}_{\alpha\beta}(\mathbf{q}_{\alpha\beta})h_{\alpha\beta}(q)
\label{M_4}
\ee
where the separation function $h_{\alpha\beta}(q)$ is determined by the variation of the Born-amplitude off the physical 3-momentum shell $\mathbf{q}_{\alpha\beta}$.

Now, we perform the remaining integral and define the absorption index
\be
n_{\alpha\beta}={4\pi}\int^\infty_0{dq q^2 \Delta_{\alpha\beta}(q) h_{\alpha\beta}(q)}
\label{eta}
\ee
The full reaction amplitude obtains a considerably simplified structure
\be\label{eq:Mab_sep}
M_{\alpha\beta}(\mathbf{k}_\alpha,\mathbf{k}_\beta)=
M^{(B)}_{\alpha\beta}(\mathbf{q}_{\alpha\beta})\left(1-n_{\alpha\beta} \right)
\ee
given in leading order by the Born-amplitude, scaled by a distortion coefficient which should depend only weakly on the momentum transfer for a meaningful factorization of $M_{\alpha\beta}$.

\section{Separation Function for Gaussian Form Factors}\label{sec:GaussFF}
\subsection{Transition Potential in Gaussian Approximation}\label{ssec:GaussTr}
The separation \textit{ansatz} discussed above can be checked, on an analytical basis,
if one adopts a Gaussian shape, $U_G$, for the transition potential $\mathcal{U}_{\alpha\beta}(\mathbf{p})$. Indeed nuclear
SCE transitions are well modeled by surface form factors for which the Gaussian shape
is a quite convenient and realistic choice. For the present purpose, it is sufficient to consider a transition potential with a single Gaussian form factor:
\be\label{eq:UGauss}
\mathcal{U}_G(\mathbf{r},\mathbf{R})=\frac{1}{4\pi}U_0e^{-\frac{(\mathbf{r}-\mathbf{R})^2}{2\sigma^2}}
\ee
which can be adjusted to microscopically derived shapes by an appropriate choice of the centroid parameter $R$ and the width parameter $\sigma$. Considered as a classical quantities, $R$ and $\sigma$ are determined, in principle, by the radii and surface thicknesses of the colliding ions.   
$\mathcal{U}_G$ contains a rich multipole structure
\be\label{eq:UG_MultiP}
\mathcal{U}_G(\mathbf{r},\mathbf{R})=\sum_{LM}{Y^*_{LM}(\mathbf{\hat{R}})U_{LM}(\mathbf{r},R)}
\ee
with the multipole form factors
\bea\label{eq:ULM}
U_{LM}(\mathbf{r},R)&=&\int{d\hat{R}Y_{LM}(\mathbf{\hat{R}})U_G(\mathbf{r},\mathbf{R})}\\
&=&U_0e^{-\frac{r^2+R^2}{2\sigma^2}}i_L(rR/\sigma^2)Y_{LM}(\mathbf{\hat{r}})
\eea
where $i_L(x)=i^{L}j_L(ix)$ is a modified spherical Bessel function. As discussed in Appendix \ref{app:GFF}, the connection to the microscopic structure of the intrinsic nuclear transitions involved in projectile and target is recovered by imposing on $Y_{LM}(\hat{\mathbf{R}})$ a quantization condition in terms of the projectile and target state operators, similar to the collective model of nuclear excitations. There, it is also shown that within the Gaussian approximation $R$ and $\sigma$ are determined by the corresponding projectile and target quantities. The strength parameter $U_0$ is related to the volume integral of the NN T-matrix. However, for the following those details are of minor relevance because state-independent, universal properties of distortion effects in non-elastic ion-ion reactions are investigated. Thus, for simplicity we neglect the state dependence, choose $U_0=1$ and leave the determination of $R$ and $\sigma$ for later.

The Fourier-Bessel transform is derived analytically:
\be\label{Ug_gaus}
\mathcal{U}_G(\mathbf{p},\mathbf{R})=\sqrt{\frac{\pi}{2}}\sigma ^3 e^{i\mathbf{p}\cdot \mathbf{R}}e^{-\frac{1}{2}\sigma^2p^2}
\ee
and the momentum space multipoles are obtained as above by projecting on $Y_{LM}(\mathbf{\hat{R}})$. This amounts to expand the plane wave into partial waves resulting in:
\be\label{Ulm}
U_{LM}(\mathbf{p},R)=4\pi \sqrt{\frac{\pi}{2}}\sigma ^3 e^{-\frac{1}{2}\sigma^2p^2}j_L(pR)i^L Y_{LM}(\mathbf{\hat{p}})
\ee
According to Eq.(\ref{reac_amp}), we need to evaluate $U_G(\mathbf{p})$  at $\mathbf{p}=\mathbf{q}_{\alpha\beta}-\mathbf{q}$.
 This leads to
\be
\mathcal{U}_G(\mathbf{q}_{\alpha\beta}-\mathbf{q},\mathbf{R})=\mathcal{U}_G(\mathbf{q}_{\alpha\beta},\mathbf{R})\mathcal{H}_{\alpha\beta}(\mathbf{q},\bm{\rho})
\label{UG_eq}
\ee
describing the (partial) separation of the dependencies on the physical momentum transfer $\mathbf{q}_{\alpha\beta}$ and the momentum shift $\mathbf{q}$ due to the ISI/FSI interactions by means of
\be
\mathcal{H}_{\alpha\beta}(\mathbf{q},\bm{\rho})=e^{-\frac{1}{2}\sigma^2 q^2}e^{-i\mathbf{q}\cdot \bm{\rho}}
\ee
with the pseudo-radius
\be
\bm{\rho}=\mathbf{R}+i\sigma^2 \mathbf{q}_{\alpha\beta}
\ee
which is shifted into the complex plane by an amount controlled by the width parameter $\sigma$. We use $\rho=\sqrt{\rho^2}$ where
\be
\rho^2=R^2-\sigma^4 q^2_{\alpha\beta}+2i\sigma^2 \mathbf{q}_{\alpha\beta}\cdot \mathbf{R}
\ee
Since $\rho$ also depends on the on-shell momentum transfer $\mathbf{q}_{\alpha\beta}$, the separation of variables is not yet fully achieved.
The function $h_{\alpha\beta}$ of Eq.(\ref{M_4}) is given as:
\be
h_{\alpha\beta}(q,\rho)=\frac{1}{4\pi}\int{d\mathbf{\hat{q}}\mathcal{\mathcal{H}}_{\alpha\beta}(\mathbf{q},\bm{\rho})}= e^{-\frac{1}{2}\sigma^2 q^2}j_0(q\rho)
\label{mono_sep}
\ee
and the distortion coefficient $(1-n_{\alpha\beta})$ is found according to Eq.(\ref{eta}). Further insight into the modification introduced by the ion-ion ISI and FSI interactions is obtained by using the addition theorem for Bessel functions \cite{Watson:1966}
\be
j_0(q\rho)=\sum_\lambda{(2\lambda+1)}P_\lambda(\cos{\gamma})j_\lambda(qR)i^\lambda i_\lambda(qq_{\alpha\beta}\sigma^2)
\ee
where $\gamma$ denotes the angle between $\mathbf{R}$ and $\mathbf{q}_{\alpha\beta}$. Furthermore, using the addition theorem of spherical harmonics we find
\be
\begin{split}
&h_{\alpha\beta}(q,\rho)
=(4\pi) e^{-\frac{1}{2}\sigma^2 q^2}\\
&\sum_{\lambda\mu}{i^\lambda Y_{\lambda\mu}(\mathbf{\hat{q}}_{\alpha\beta})Y^*_{\lambda\mu}(\mathbf{\hat{R}})j_\lambda(qR)i_\lambda(qq_{\alpha\beta}\sigma^2)}
\end{split}
\ee
For momentum transfers in the range $q_{\alpha\beta}\ll 1/\sigma$, which amounts to about the order of $100$~MeV/c, the sum is well approximated by the monopole term,
\be
h_{\alpha\beta}(q)= e^{-\frac{1}{2}\sigma^2 q^2}j_0(qR)i_0(qq_{\alpha\beta}\sigma^2)
\label{mono}
\ee
indicating a remaining dependence on the reaction momentum transfer. This derivation, based on the Gaussian form factor, allows one to understand the range of validity of the separation ansatz, Eq.(\ref{M_4}). Indeed, for transferred momenta approaching zero, one recovers the complete factorization discussed above, i.e.
\be
h_{\alpha\beta}(q) \mapsto e^{-\frac{1}{2}\sigma^2 q^2}j_0(qR)
\label{comp_fact}
\ee

\subsection{Distortion Coefficient in Black Disk Approximation}\label{ssec:BDA}
In the derivation of Eq.(\ref{eq:Mab_sep}) the critical step is clearly the treatment of the distortion effects which we consider next. For strongly absorbing systems like ion-ion scattering, the distorted waves are almost completely suppressed in the overlap region, thus reflecting the large amount of channel coupling which leads to a redirection of the incoming elastic probability flux into a multitude of non-elastic reaction channels. Such systems are described by optical potentials with a strong imaginary part of a strength comparable in magnitude to the real, diffractive part. Under such conditions, the distortion amplitude introduced before resembles in coordinate space a step function, $\eta_{\alpha\beta}(\mathbf{r})\sim e^{i\phi(\mathbf{r})}\Theta(r-R_{abs})$. In the following, we neglect the phase factor given by $\phi$. This picture coincides with the \emph{black disk assumption} (BD) where one assumes that a major part of the incoming flux is consumed by a (spherical) absorber of radius $R_{abs}$ resulting in the total absorption cross section
\be
\sigma^{(BD)}_{abs}(\sqrt{s})=\pi R^2_{abs}(\sqrt{s})
\ee
and by equating $\sigma^{(BD)}_{abs}$ and the quantum mechanical reaction cross sections $\sigma^{(\alpha,\beta)}_{abs}$ the absorption radius $R_{abs}$ is obtained. 
Considering that $\sigma_{abs}\sim 1...3$~barn as a representative range of values for ion-ion reaction cross sections at energies of a few $10AMeV$ we find $R_{abs}\sim 5...10 fm$. These values are implying a variation of the function $j_0(qR_{abs})$ on a momentum scale $\Delta q_{reac}\sim 1/R_{abs} \sim 20...40$~MeV/c, thus complying perfectly well with the previous estimates.

In the BD-limit we can evaluate the distortion coefficient in closed form. We find
\be
\Delta^{(BD)}_{\alpha\beta}(q)=\frac{1}{2\pi^2}\frac{R_{abs}}{q}\left(-\frac{\partial}{\partial q} \right) j_0(qR_{abs})
\ee
and the scaling function is given by
\be
n^{(BD)}_{\alpha\beta}(R_{abs})=\frac{2R_{abs}}{\pi}\int^\infty_0{dq j_0(qR_{abs})\frac{\partial}{\partial q}\left(q h_{\alpha\beta}(q) \right)}
\label{scal}
\ee
which corresponds to a Fourier-Bessel transform of $h_{\alpha\beta}(q)$, mapping the dependence on the variable $q$ to the complementary variable $R_{abs}$. As discussed in Appendix \ref{app:GaussBD}, for $h(q)$ given by Eq.(\ref{mono_sep}) the black disk distortion coefficient can be calculated in closed form, resulting in a superposition of error integrals and Gaussians.

In  $h_{\alpha\beta}$, see Eq.(\ref{mono_sep}), the parameter $\sigma$ controls the slope of the momentum distribution around the physical momentum transfer $q_{\alpha\beta}$. By the arguments given above, we expect $\sigma\sim \mathcal{O}(1/k_F)$, thus being related to the binding properties of nuclei. Hence, the width of the Gaussian form factor is determined by the surface diffuseness of nuclear density distributions. The (off-shell) diffraction structure of the transition form factors, which is described by $R$, is more directly affected by the nuclear geometry, which to a large extent is a mean-field effect, thus related to the radii of the nuclear densities, $R_{a,A}$. Taking into account the modifications by the folding with the NN-interaction, we estimate therefore $R \sim \mathcal{O}(R_{pot}) $ where $R_{pot}$ is the radius of the ion-ion potential.

\section{Nuclear Response at Low Momentum Transfer}\label{sec:LowMom}

\subsection{Form Factors in the Low-Momentum Transfer Limit and $1\nu 1\beta$ Nuclear Matrix Elements}
The reduced form factors $f^{XY}_{LSJ}$, introduced in Eq.(\ref{eq:FprojM}) and Eq.(\ref{eq:FtargM}), are the quantities of central interest for charge changing reactions and beta decay studies. They contain the complete information on the nuclear configurations which are contributing to the transitions. In that sense, they are the fingerprints characterizing a nuclear species.  The form factors are related to the corresponding reduced radial transition densities by a Fourier-Bessel transform,
\be
f^{(XY)}_{LSJ}(p^2)=\int^\infty_0{drr^2\rho^{(XY)}_{LSJ}(r)j_L(pr)}
\label{eq:fredXY}
\ee
For small momenta $p\to 0$, we find
\be
f^{(XY)}_{LSJ}(p^2)\sim \frac{p^{L}}{(2L+1)!!}\left(\int^\infty_0{drr^{2+L}\rho^{(XY)}_{LSJ}(r)}+\mathcal{O}(p^2)\right)
\label{eq:tr1}
\ee
and the transition densities are normalized such that the matrix element
\be
b^{(XY)}_{LSJ}=\int^\infty_0{drr^{2+L}\rho^{(XY)}_{LSJ}(r)}
\label{eq:tr2}
\ee
is the reduced transition amplitude belonging to the multipole operator
\be
\mathcal{B}_{(LST)JM}(\mathbf{r})=r^L[Y_L\otimes \left(\bm{\sigma}\right)^S]_{JM}\left(\bm{\tau}\right)^T
\ee
which is of the same functional structure as the beta-decay transition operators.

The excitation probability is given by
\be
B^{(XY)}_{LSJ}=\frac{1}{2J+1}\left|\lan J_Y||\mathcal{B}_{LSTJ}||J_X\ran \right|^2=\left|b^{(XY)}_{LSJ}\right|^2
\ee
\subsection{Cross Sections at Low-momentum transfer}
In the limit of low momentum transfer, the cross section simplifies considerably because the separation approach can be applied. Further simplification is gained when considering transitions from
$0^+$ ground states.
In this case there will be for $J>0$ in general two contributing multipole form factors, namely those of the $S=0,1$ transitions of the same $L=J$ for natural parity and those of the $L=J\pm 1$ transitions with fixed $S=1$ for unnatural parity. For natural parity transitions the superposition will not modify the low-momentum behaviour of the cross sections but has to be taken into account for the extraction of the corresponding transition strengths. At forward angles the cross section describing {\it natural parity}
transition in both nuclei will be of the type (see Eqs.(\ref{eq:tr1}),(\ref{eq:tr2}),(\ref{eq:Mab_sep}))
\be
\begin{split}\label{cross_1}
&\frac{d\sigma^{FF}}{d\Omega}\sim \frac{q^{2(J_a+J_A)}}{\left[(2J_a+1)!!(2J_A+1)!!\right]^2} |1-n_{\alpha\beta}|^2 \\
&\big|V^{(C)}_{01}(0)b^{AB}_{J_A0J_A}b^{ab}_{J_a0J_a}
+e^{i\phi_{aA}}V^{(C)}_{11}(0)b^{(AB)}_{J_A1J_A}b^{(AB)}_{J_A1J_A}  \big|^2
\end{split}
\ee
where $q=|k_\alpha-k_\beta|$ denotes the momentum transfer at forward direction and $\phi_{aA}$ accounts for possible relative phase factors of the target and projectile matrix elements. If one of the nuclei undergoes a $J=0^+$ monopole excitation, i.e. a $0^+_{g.s.}\to 0^+_{E_x}$ transition, the $S=1$ components will not contribute and irrespective of the multipolarity of the excitations in the second nucleus, only $S=0$ transitions will be observed.

For unnatural parity states the multipole mixtures lead to a modification of the momentum dependence because for $J\neq 0^-$ we have two angular momentum transfers, $L=J-1$ and $L=J+1$. The forward cross section for {\it unnatural parity}
transitions in both nuclei behaves as
\be
\begin{split}\label{cross_2}
&\frac{d\sigma^{GG}}{d\Omega}\sim \frac{q^{2(J_a+J_A-2)}}{\left[(2J_a-1)!!(2J_A-1)!!\right ]^2}|V^{(C)}_{11}(0)|^2 \\
&[~|b^{(AB)}_{J_A-11J_A}+\frac{q^2}{(2J_A+1)(2J_A+3)}b^{(AB)}_{J_A+11J_A}|^2 \\
&|b^{(ab)}_{J_a+11J_a}+\frac{q^2}{(2J_a+1)(2J_a+3)}b^{(ab)}_{J_a+11J_a}|^2 ]  |1-n_{\alpha\beta}|^2
\end{split}
\ee
The contributions from the rank-2 tensor interactions are not shown because they will be suppressed at small momentum transfer.
The multipole mixtures  will change with the effective momentum transfer at forward directions. If there is a $0^-$ transition in one of the two nuclei, the corresponding transition form factor reduces to a single contribution with $L=1$ and $S=1$.

In addition, there are mixed transitions, combining a natural parity spin-flip excitation in one nucleus with unnatural excitations in the other nucleus. The corresponding cross sections are obtained in a similar way and are easily deduced by an appropriate combination of the above results.

\section{Applications to Heavy Ion Induced SCE Reactions}\label{sec:Applications}

\subsection{Spectroscopy of Charge Changing Nuclear Excitations}\label{sec:CC_Spectra}
The theoretical methods developed in the previous sections are applied in the following to a case of practical interest, for the SCE reaction $^{18}O+^{40}Ca$ $\to$
$^{18}F+^{40}K$, at $T_{lab}= 15\, AMeV$ \cite{Mariangela:PhD}. 
Experimentally, this reaction has been recently investigated by the NUMEN collaboration \cite{Cappuzzello:2015ixp}. In this section we consider first charge changing nuclear excitations in a self-consistent approach utilizing nuclear Hartree-Fock-Bogolubov (HFB) mean-field theory for ground states and QRPA theory in the polarization propagator formulation. In combination, these two methods provide a versatile toolbox with appropriate instruments for the proper description of nuclear spectroscopy over most of the nuclear mass table, except for the lightest nuclei. The reaction theoretical aspects will be addressed afterwards. There, the focus will be in the first place to clarify and establish a couple of special aspects of heavy ion reactions at intermediate energies, rather than fitting data.

\subsubsection{HFB Mean-field description of the A=18 and A=40 Isobars}

For the practical calculations the quasiparticle spectrum and the single particle wave functions are obtained by density functional theory (DFT). An energy density functional (EDF) along the line of Refs. \cite{Hofmann:1998} and \cite{Tsoneva:2017kaj} is constructed, using a G-Matrix interaction, supplemented by three-body terms. First variation leads the to single particle wave equations with effective density dependent potentials and pairing interactions, solved self-consistently by HFB and BCS methods. In the particle-particle channel an effective density dependent contact pairing interaction is used. The strength is derived from the $nn$ and $pp$ singlet-even Born matrix elements of the Bonn interaction in non-relativistic reduction found in \cite{Machleidt:1987pr}. Such an approach leads to state dependent pairing gaps which are determined self-consistently in parallel to the HFB iteration procedure.
In Tab. \ref{tab:HFB} HFB results for the ground states of mass-18 and mass-40 nuclei are listed. For the $A=18$ isobars the measured binding energies are reproduced by better than 4\%. As typical for a mean-field description with global parameter sets, the agreement improves with increasing mass. The binding energies of the $A=40$ isobars are described by better than about 1\%. A similar dependence will also be detected for the QRPA results discussed below. For the single-particle spectra entering into the QRPA calculations, proton and neutron continuum states are included up to single particle energies of 100~MeV. They are obtained by using the self-consistent HFB mean-field potentials, thud avoiding artificial, non-physical non-orthogonality effects. The single particle continua are described by a dense spectrum of discrete states. Enclosing the system into a spherical cavity of a size of up to 100~fm, an average energy spacing of about 20~keV is obtained.

\begin{table*}
\begin{center}
\begin{tabular}{|c|c|c|c|c|}
  \hline
  Nucleus & $B_{exp}(A)/A$ [MeV/A] & $B_{theo}(A)/A$ [MeV/A]  & $r_d$ [fm] & $r_{chrg}$ [fm] \\ \hline
  $^{18}N$ & 7.038  & 7.236  & 2.790  & 2.693    \\ \hline
  $^{18}O$ & 7.767  & 7.894  & 2.740  & 2.757  \\ \hline
  $^{18}F$ & 7.632 & 7.329   & 2.763  & 2.854  \\ \hline \hline
  $^{40}K$  & 8.538 & 8.620 & 3.369 & 3.449 \\ \hline
  $^{40}Ca$ & 8.551 & 8.618 & 3.373 & 3.486 \\ \hline
  $^{40}Sc$ & 8.174 & 8.269 & 3.381 & 3.524 \\
  \hline
\end{tabular}
\end{center}
\caption{Ground state properties of the $A=18$ and the $A=40$ isobars. The observed and the calculated binding energies per nucleon are shown in the second and third column. Density and charge root-mean-square radii are noted by $r_d$ and $r_{chrg}$, respectively. The data are taken from the AMDC mass evaluation \cite{Audi:2017asy}.  \label{tab:HFB}}
 \end{table*}

\begin{figure*}
\begin{center}
\includegraphics[width = 6.5cm]{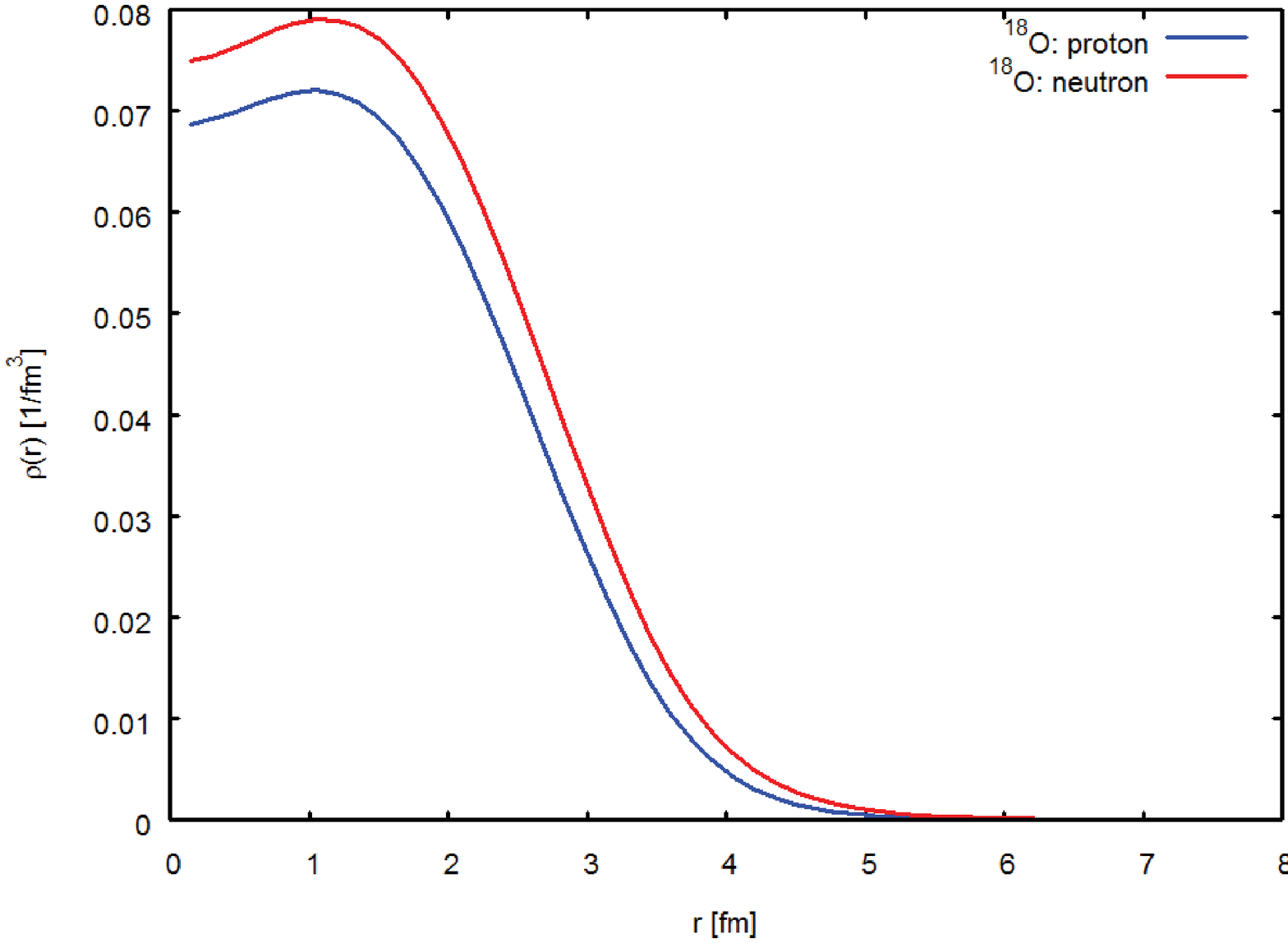}
\includegraphics[width = 6.5cm]{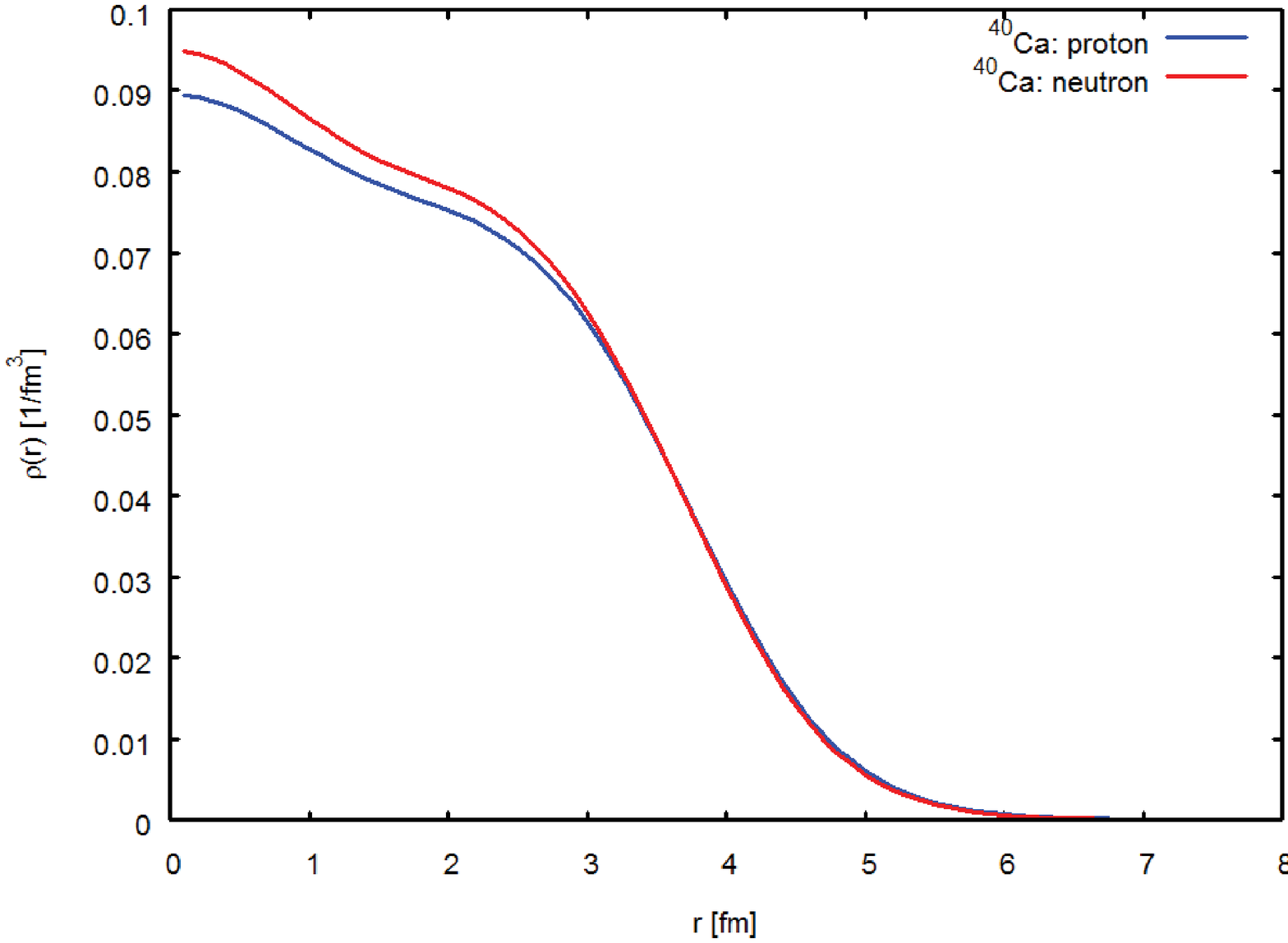}
\end{center}
\caption{(Color online) Proton and neutron HFB ground state densities for $^{18}O$ (left) and $^{40}Ca$ (right), respectively. In $^{18}O$ the onset of a neutron skin is visible.}
\label{fig:gsd}
\end{figure*}

The optical potentials discussed below are calculated with the $^{18}O$ and $^{40}Ca$ HFB ground state densities. They are displayed in Fig. \ref{fig:gsd}.

\subsubsection{SCE Response Functions in Oxygen and Calcium}
As illustrated in Fig. \ref{fig:nppn} nuclear charge changing excitations consist of two branches: The $np^{-1}$ branch probes the $\tau_+$-response and $pn^{-1}$ excitations probes the $\tau_-$-response, intimately related to the $\beta^\pm$ processes of weak interactions. The retarded propagators introduced above included both branches because the $np^{-1}$ branch is connected by time reversal to the $pn^{-1}$ branch and vice versa. This is true in particular for systems where pairing is non-negligible. In both $^{18}O$ and $^{40}Ca$, however, the mixing of the two branches is negligible.

Physically, the 2QP configurations will be coupled to 4QP and higher order many-body configurations. These couplings induce non-hermitian polarization self-energies $\Sigma_{QQ}=\Delta_c-\frac{i}{2}\Gamma_c$. The real part $\Delta_c(\omega)$ leads to additional state dependent energy shifts. Below the particle emission threshold the imaginary part $Im(\Sigma(\omega))=-\frac{1}{2}\Gamma^\downarrow_c(\omega)$ describes the damping effects due to the redistribution of the 2QP spectroscopic strength over the high order background states.  Above the particle emission threshold, a decay width $\Gamma^\uparrow_c(\omega)$ has to be added, leading the total width $\Gamma_c(\omega)$. Thus, the 2QP QRPA states  $|c\ran$ 
are in fact doorway states of finite life time $t_{1/2}\sim 1/\Gamma_c$ which eventually will decay into more complex configurations. The contributions of the dispersive self-energies are taken into account approximately by replacing in the propagators the bare 2QP energies by the polarized energies $E_{j_1}+E_{j_2}+\Sigma_{QQ}(\omega)$ where $\Sigma_{QQ}(\omega)$ is an averaged, global self-energy. With the self-energy insertions the propagators contain a finite imaginary part, thus shifting the poles far into the complex plane. The self-energies are described by a global energy dependent parametrization of the imaginary part according to the procedure discussed in \cite{Baker:1997,Mahaux:1982eig}. At the Fermi-edge the damping width vanishes and then increases to $\Gamma^\downarrow \simeq 3$~MeV in the giant resonance region. At large energies, the damping width decreases again. In order to preserve analyticity, also the real part must be included and it is derived in a self-consistent manner by dispersion theory.

The residual interactions are derived by second variation form of the same EDF as used in the HFB ground state calculations. The variational approach leads to density dependent Landau-Migdal parameters. For the present purpose, the isovector interactions are of primary interest. Because of the density dependence the Landau-Migdal interactions include rearrangement contributions describing an effective screening of vertices. In infinite nuclear matter, the spin-independent isovector interaction is the strongest at low densities and decreases rapidly towards the saturation point. Slightly above the saturation density the corresponding Landau-Migdal parameter $F'_0(\rho)$ changes sign and at much larger densities levels off at a value of about $F'_0\sim -0.85$.  We obtain a symmetry energy $E_{sym}=30.2$~MeV at $\rho=\rho_{sat}$. The Landau-Migdal parameter $G'_0$ describing the interaction strength in the $\bm{\sigma\tau}$ spin-isovector channel increases with density, reaching the value $G'_0(\rho_{sat})=0.77$.

Below, results of our nuclear structure calculations will be discussed for charge changing excitations of $^{18}O$ and $^{40}Ca$. As test operators we use the multipole operators
\be\label{eq:TLSJ}
T_{LSJM}=\left(\frac{r}{R_d} \right)^L\left[ \bm{\sigma}^S\otimes Y_L\right]_{JM}\tau_{\pm}
\ee
which are of a structure similar to the weak interaction operators of nuclear beta-decay. However, here we consider the full spectrum of spatial and spin multipoles, i.e we also include response functions for transitions which would be strongly suppressed in beta-decay.

In order to obtain spectral distribution of comparable magnitude the radial form factors are normalized to the half-density radius $R_d$ of the respective parent nucleus. By definition, the response functions include the complete combined spectroscopic information on energy levels and transition strengths for the operators of Eq.(\ref{eq:TLSJ}). In the following, all data on energy spectra were taken from the NNDC online compilation \cite{NNDC}.

\subsubsection{Charge Changing Response Functions for $^{18}O$}\label{sssec:O18}
The HFB ground state of $^{18}O$ is given by a semi-magic configuration: For the protons the perfect $Z=8$ shell closure as in $^{16}O$ is maintained but the two valence neutrons are in an open-shell configuration in the $d_{\frac{5}{2}}$ shell. Thus, the two charge exchange branches involve quite different configurations. The low-energy $np^{-1}$-excitations lead to negative parity $J^P=0^-,1^-,2^-,3^-$ ground state multiplet of states in $^{18}N$, as allowed by the transitions from the 1p-proton shell to the (2s,1d)-neutron shell. Experimentally, one finds $^{18}N(1^-,g.s.)$, followed by states at $E_x=115$~keV and $E_x=588$~keV, tentatively assigned as $J^P=2^-$, and a tentative $J^P=3^-$ state at $E_x=747$~keV but the $J^P=0^-$ state is missing.

\begin{figure*}
\begin{center}
\includegraphics[width = 6.5cm]{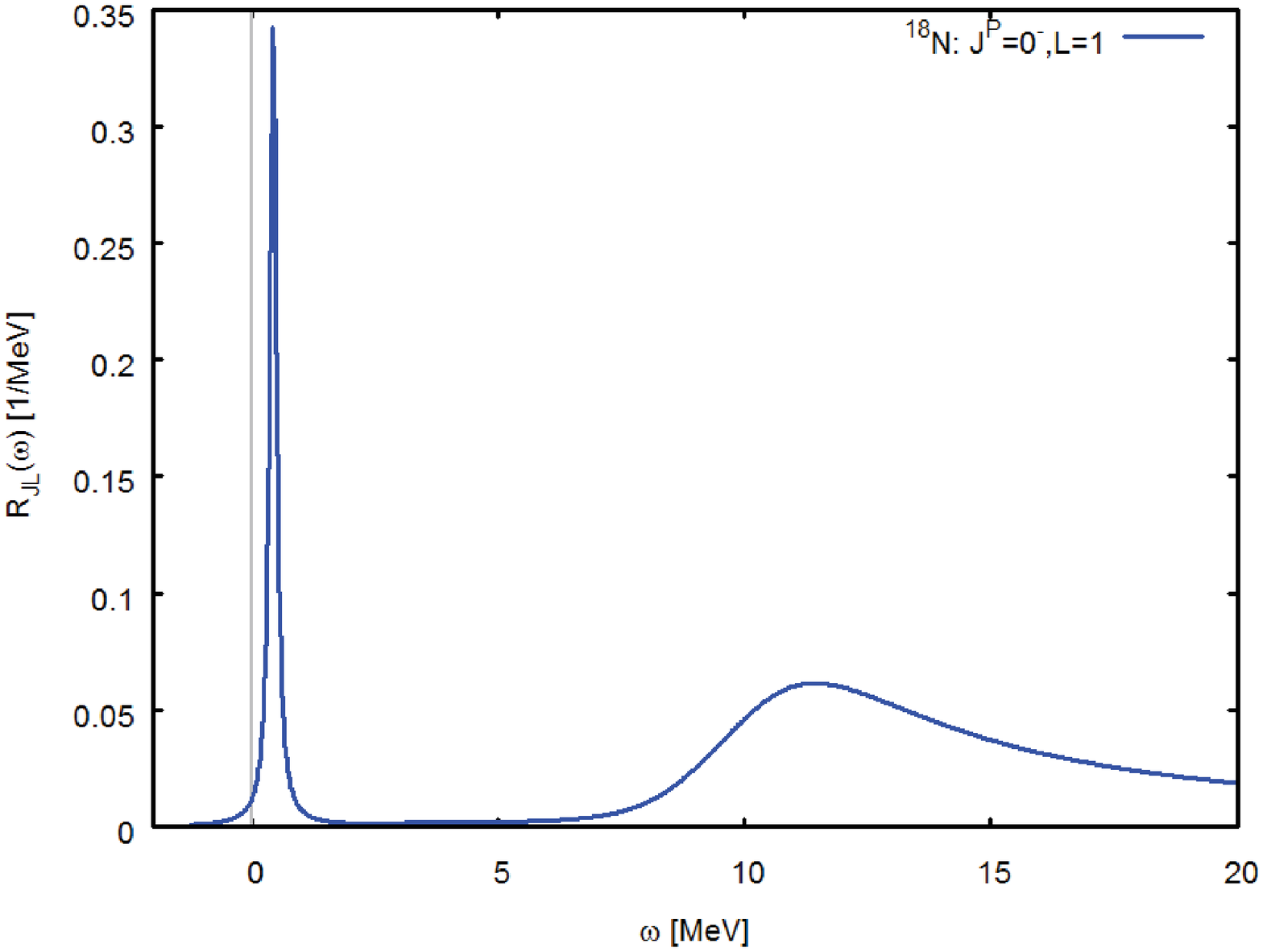}
\includegraphics[width = 6.5cm]{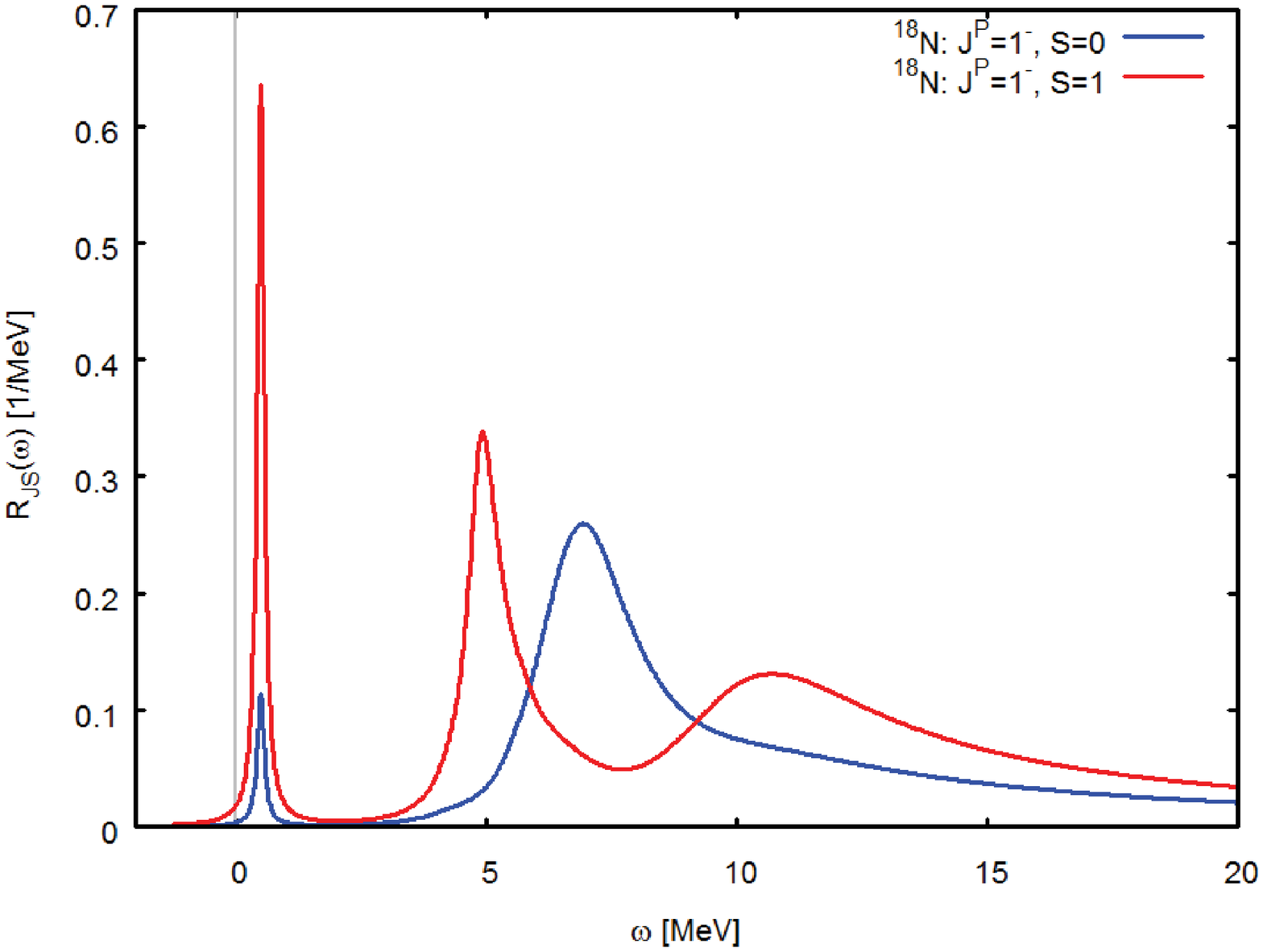}
\includegraphics[width = 6.5cm]{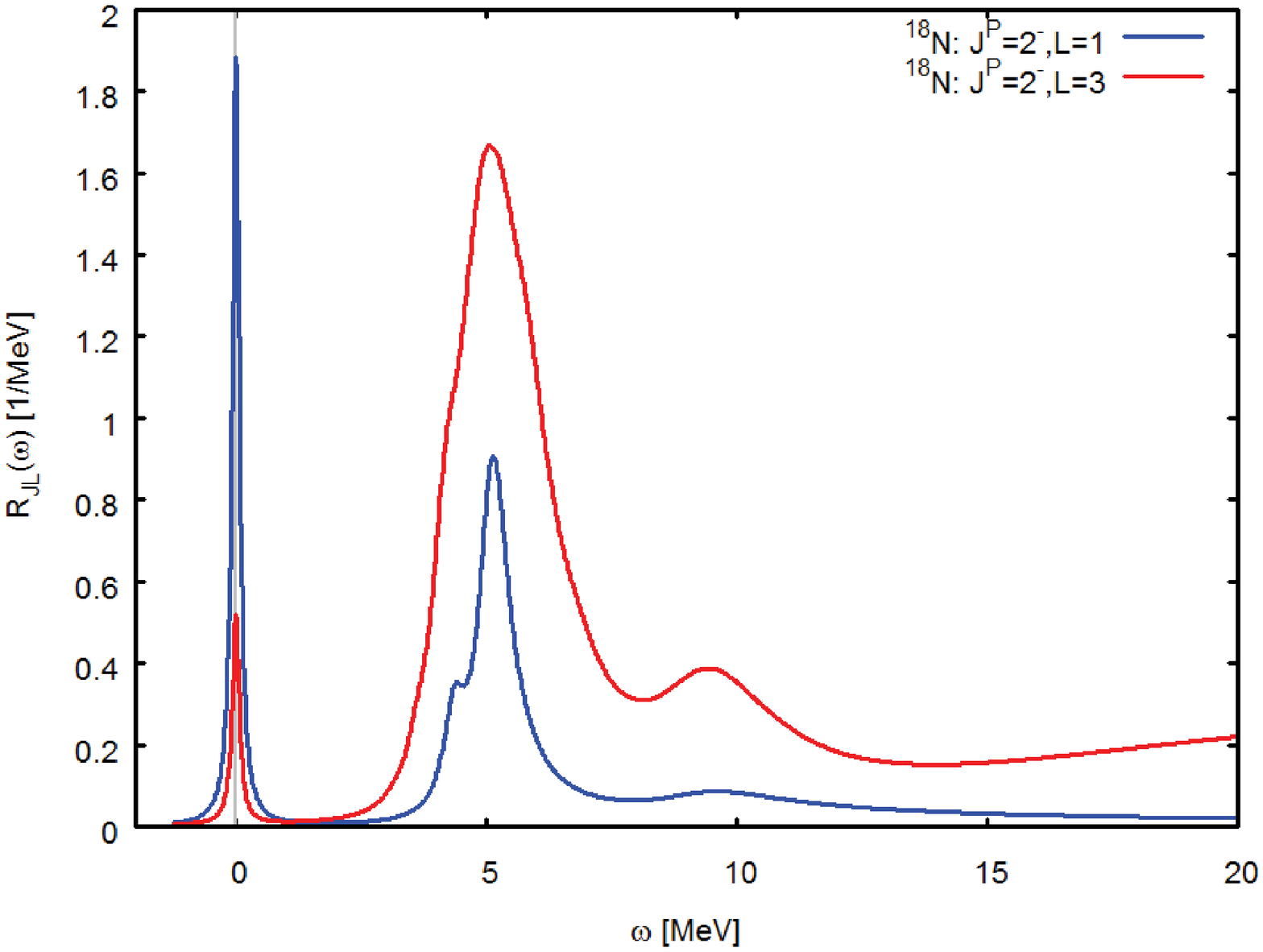}
\includegraphics[width = 6.5cm]{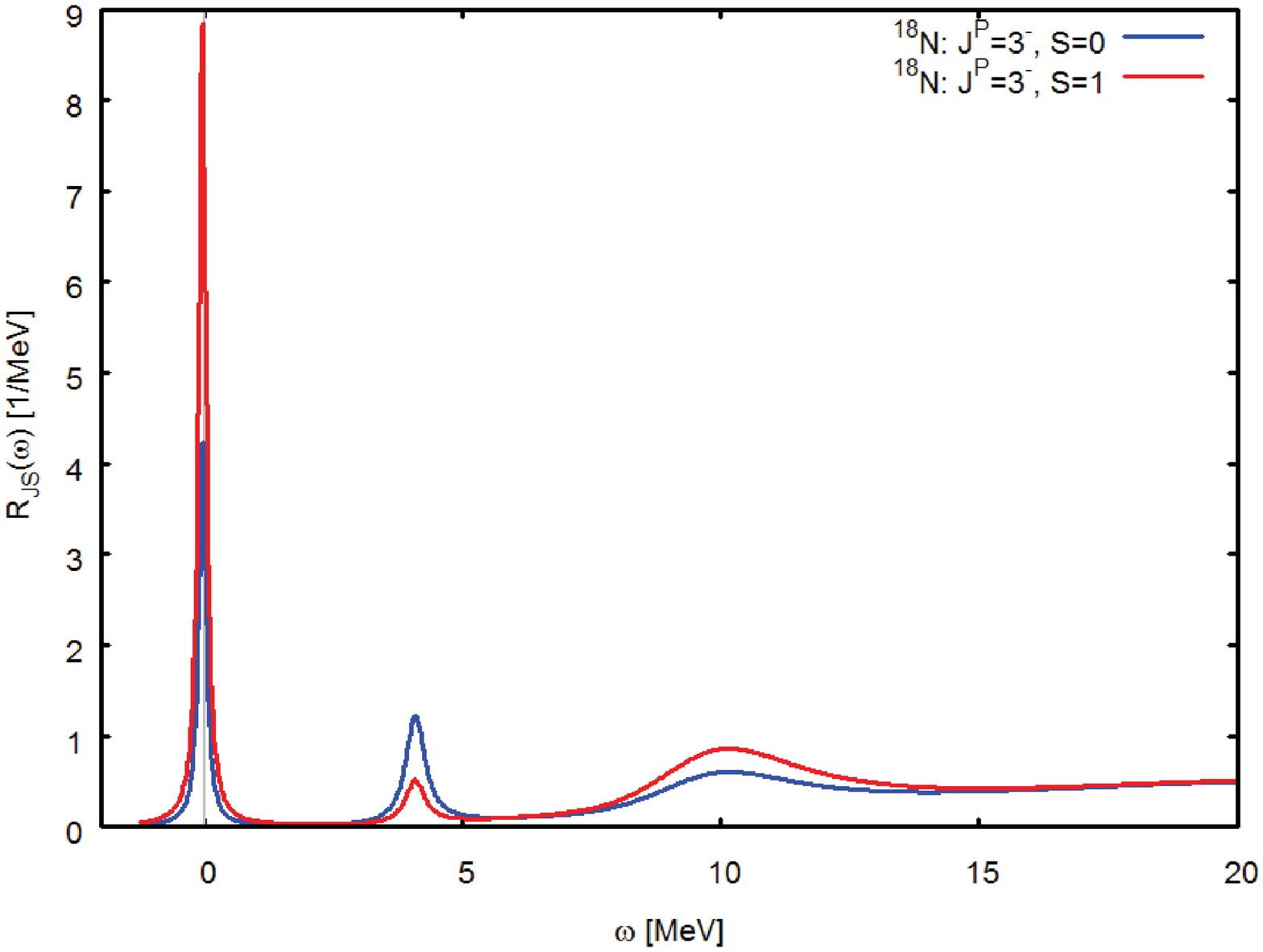}
\caption{(Color online) QRPA response functions for $^{18}O\to ^{18}N$ transitions. Results for the multipole transition operators $T_{LSJM}=\left(\frac{r}{R_d} \right)^L\left[ \bm{\sigma}^S\otimes Y_L\right]_{JM}\tau_{-}$ are shown where $R_d=2.74$~fm corresponds to the half-density radius of $^{18}O$.}
\label{fig:18N}
\end{center}
\end{figure*}

The results of our QRPA calculations are shown in Fig.\ref{fig:18N}. Similar to the data, the theoretical spectrum predicts the complete multiplet within 500~keV. The level ordering, however, is different: The $J^P=2^-,3^-$ doublet comes first, followed by $J^P=0^-$ at $E_x=411$~keV and $J^P=1^-$ at $E_x=484$~keV. Above $E_x\sim 5.5$~MeV, the neutron continuum is reached, allowing to populate unbound p- and f-wave neutron states, also giving rise to positive parity continuum configurations.

The low-energy spectrum of the complementary $pn^{-1}$ branch, populating states in $^{18}F$, is determined by configuration of $1d_{\frac{5}{2}}$ neutron hole states and proton states in the (2s,1d) shell. In principle, this allows a ground state sextet with $J^P=0^+,1^+,2^+,3^+,4^+,5^+$. Experimentally, a $^{18}F(1^+,g.s.)$ is found, followed by a $J^P=3^+$ state at $E_x=937$~keV, a $J^P=0^+$ state at $E_x=1042$~keV, and a $J^P=5^+$ state at $E_x=1121$~keV.  The first $J^P=2^+$ state is found at the much higher energy $E_x=2523$~keV. Thus, a much richer spectrum than in $^{18}N$ is observed. At $E_x=1181$~keV, a $J^P=0^-$ is observed and at $E_x=2101$~keV a $J^P=2^-$ state is seen. These negative-~parity intruder states indicate an imperfect closure of the proton 1p-shell.

\begin{figure*}
\begin{center}
\includegraphics[width = 6.5cm]{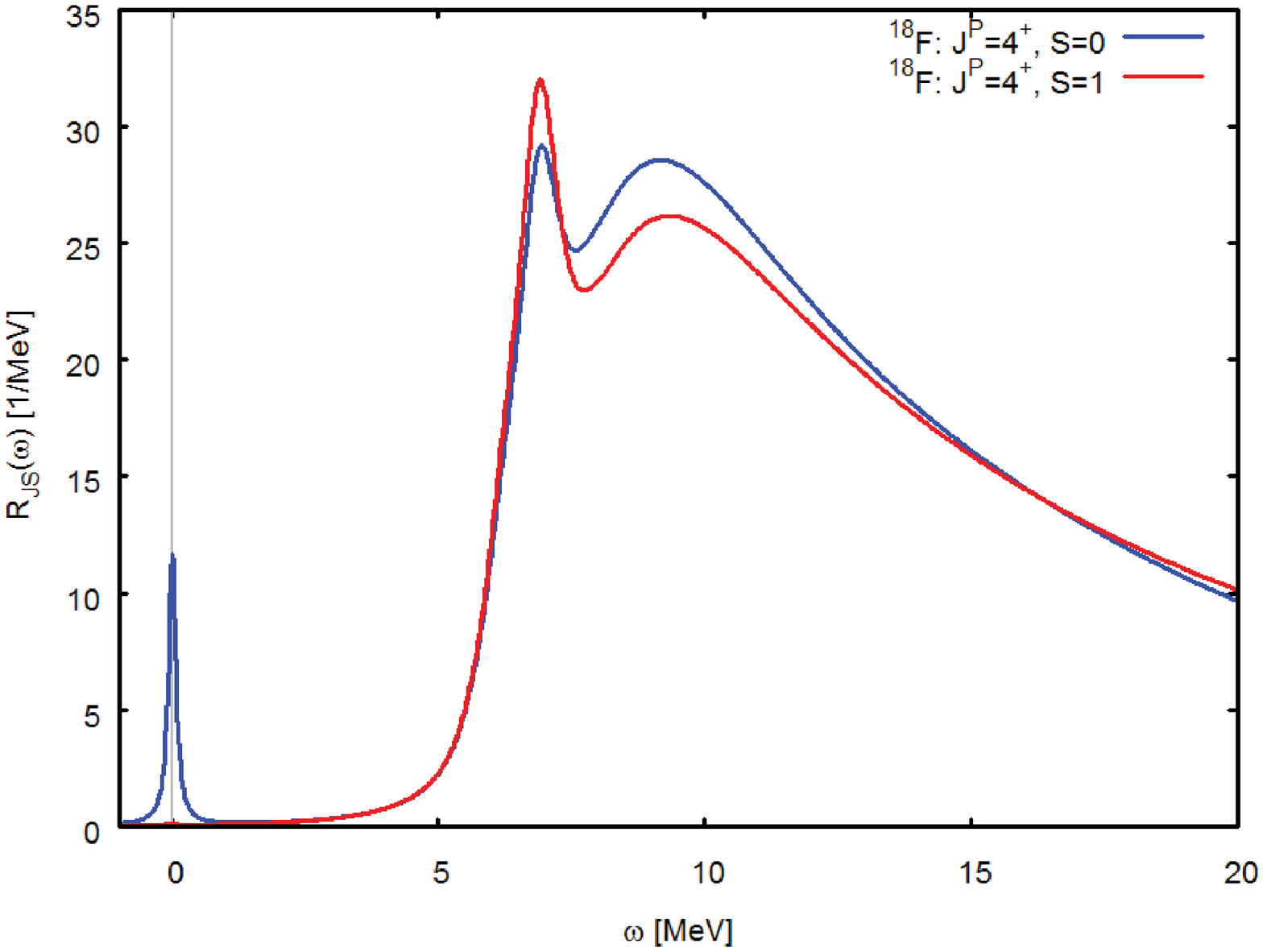}
\includegraphics[width = 6.5cm]{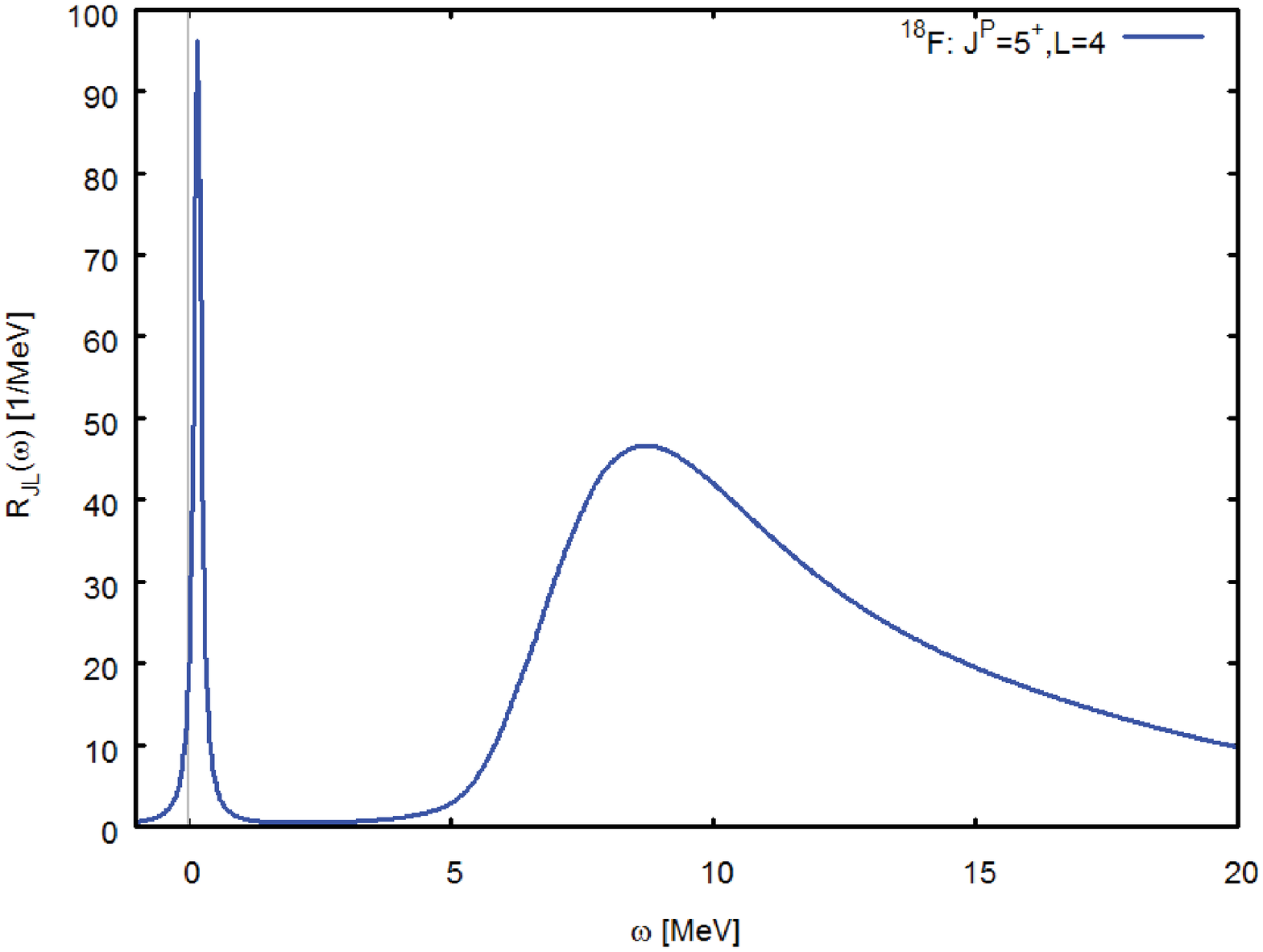}
\includegraphics[width = 6.5cm]{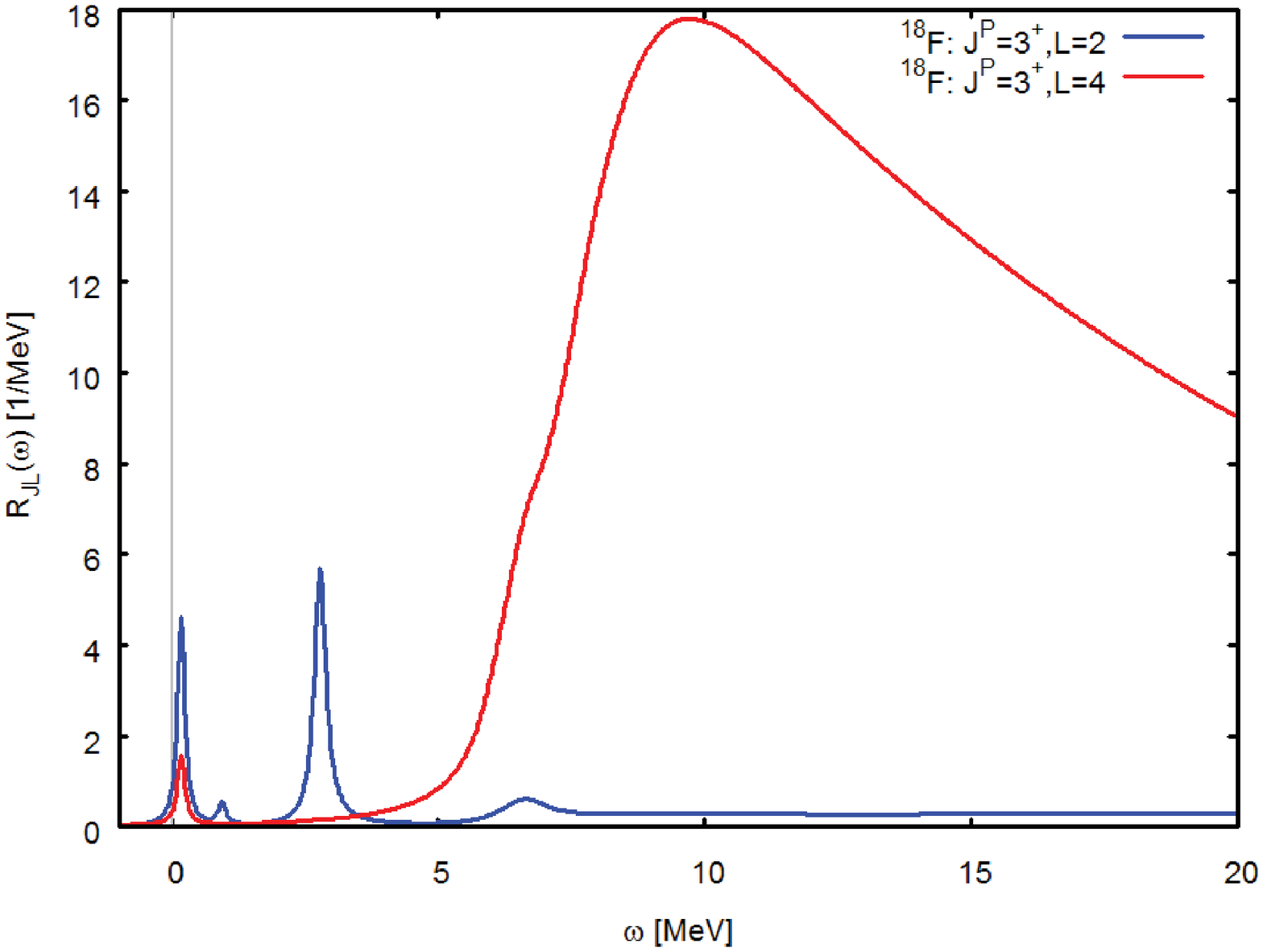}
\includegraphics[width = 6.5cm]{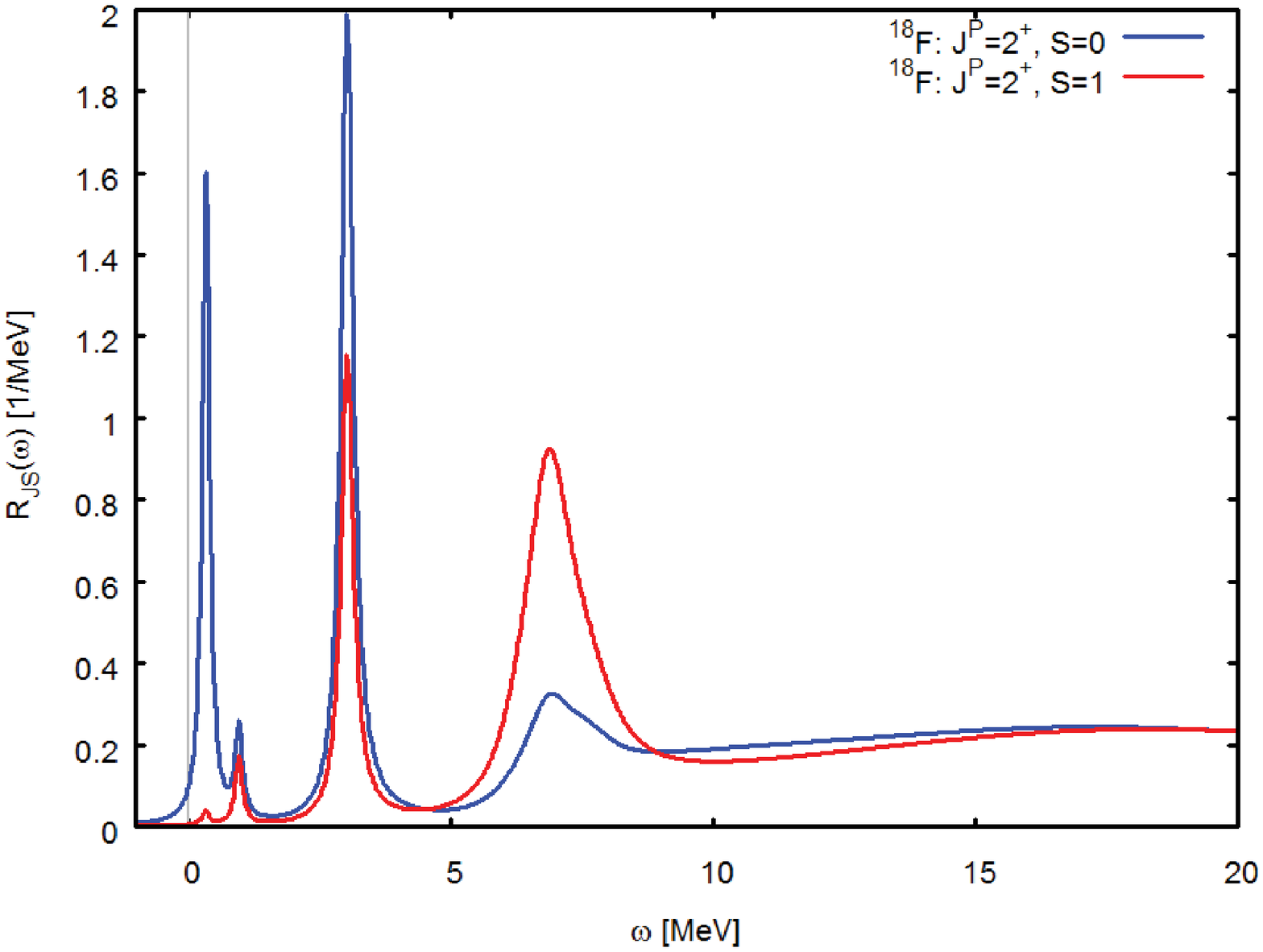}
\includegraphics[width = 6.5cm]{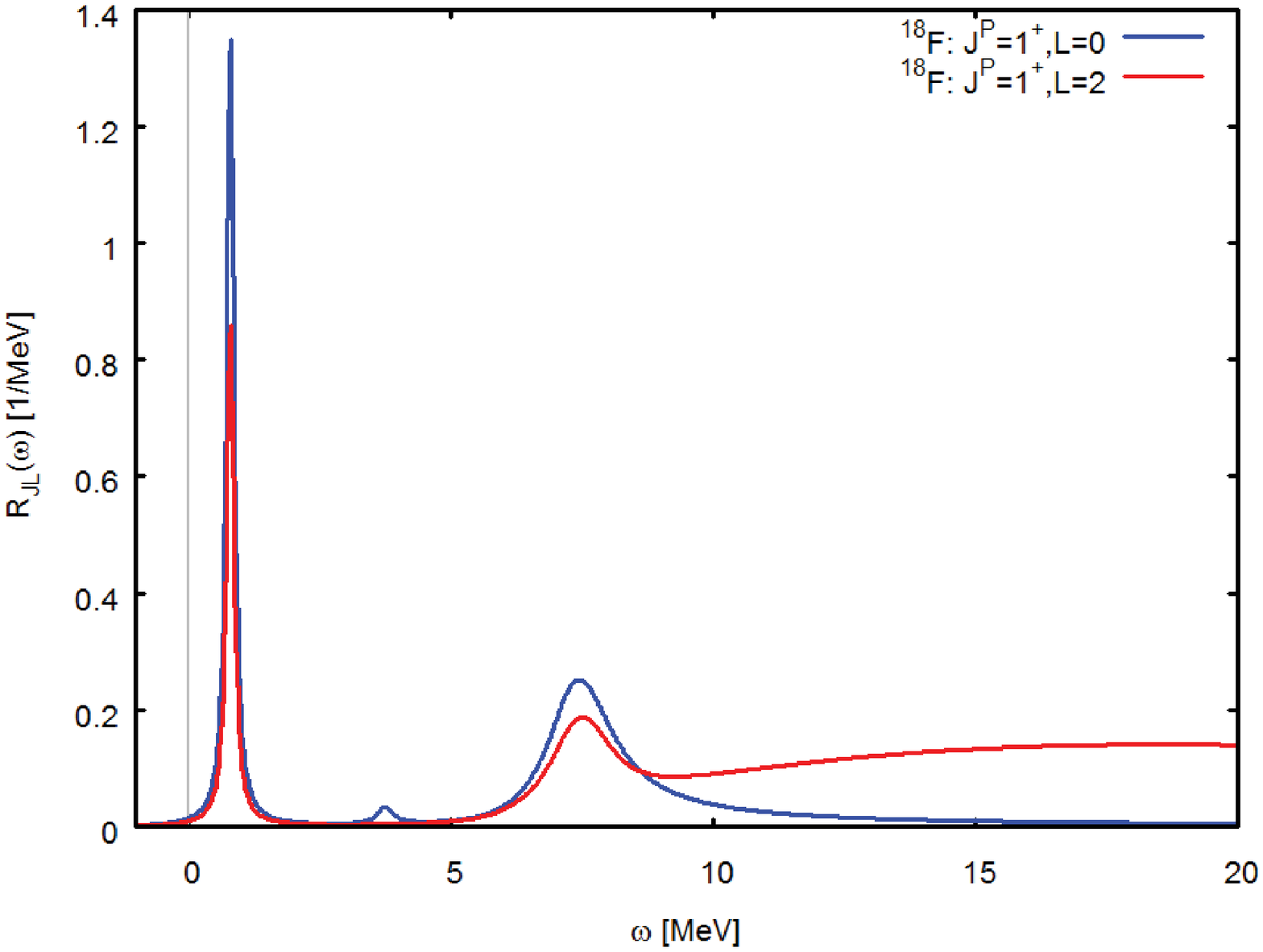}
\includegraphics[width = 6.5cm]{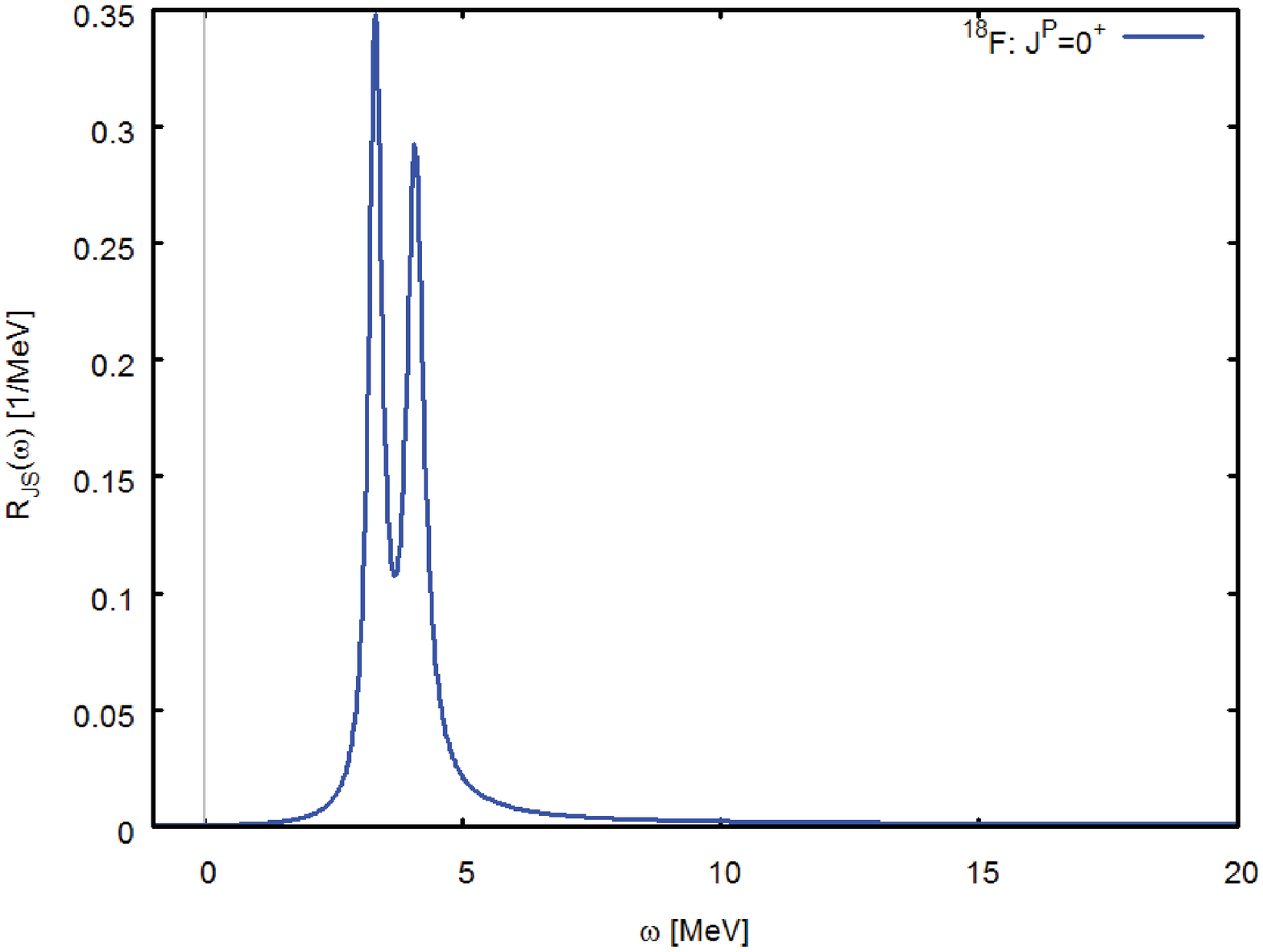}
\caption{(Color online) QRPA response functions for $^{18}O\to ^{18}F$ transitions. Results for the multipole transition operators $T_{LSJM}=\left(\frac{r}{R_d} \right)^L\left[ \bm{\sigma}^S\otimes Y_L\right]_{JM}\tau_{+}$ are shown where $R_d=2.74$~fm corresponds to the half-density radius of $^{18}O$.}
\label{fig:18F}
\end{center}
\end{figure*}

In contrast to the data, the QRPA calculations lead to a somewhat more spread out spectrum. Overall, however, the agreement is very satisfactory in view of the restriction to the 2QP-configuration space. The model calculations, shown in Fig.\ref{fig:18F}, predict a $J^P=4^+$ ground state, followed by a $J^P=5^+$ state at $E_x=172$~keV, a nearby $J^P=3^+$ state at $E_x=197$~keV, and a $J^P=2^+$ state at $E_x=305$~keV. Another $J^P=2^+$ state is obtained at $E_x=980$~keV. At $E_x=3298$~keV and $E_x=4049$~keV a $J^P=0^+$ doublet is predicted. The two states may be the theoretical counterparts of the two observed $J^P=0^+$ states at $E_x=1042$~keV and $E_x=4753$~MeV, respectively. Above $E_x\sim 5.6$~MeV the proton continuum is populated, thus leading to particle unstable states.

Overall, the rather complex spectra of the two odd-odd nuclei are described surprisingly well by the QRPA calculations which is especially worthwhile emphasizing since global model parameters were used without any attempt of fine-tuning.

\subsubsection{Charge Changing Response Functions for $^{40}Ca$ }\label{sssec:Ca40}

Since $^{40}Ca$ is a (double-magic) $N=Z$ nucleus, protons and neutrons are
occupying the same single particle orbitals. Therefore, also the odd-odd daughter nuclei $^{40}K$ and $^{40}Sc$ are of a mirror-like level structure, reflecting the almost conserved isospin symmetry. The low energy part of both excitation branches is determined by hole states in the (2s,1d)-shell and particle states in the (2p,1f)-shell. Thus, negative parity states with $J^P=0^-...5^-$ will prevail in the spectra. Experimentally, one finds for both daughter nuclei a $J^P=4^-$ ground state. In $^{40}K$, a triplet of $J^P=3^-,2^-,5^-$ states is seen at $E_x=29,800, 821$~keV. Another $J^P=2^-,3^-$ doublet is found at $E_x=2047,2070$~keV and the first $J^P=1^-$ occurs at $E_x=2104$~keV. At $E_x=2626$~keV a $J^P=0^-$ state is seen. However, there are also positive-parity intruder states which, similar to the $A=18$ systems, indicate the lack of perfect shell closures. Above $E_x\sim 2.5$~MeV a dense spectrum of positive and negative parity states is observed.  The spectrum of $^{40}Sc$ is less well known, but tentative assignments of spins and parity indicate at least for the ground state multiplet a very similar $J^P=4^-,2^-,3^-,5^-$ level sequence with a comparable spacing.

\begin{figure*}
\begin{center}
\includegraphics[width = 6.5cm]{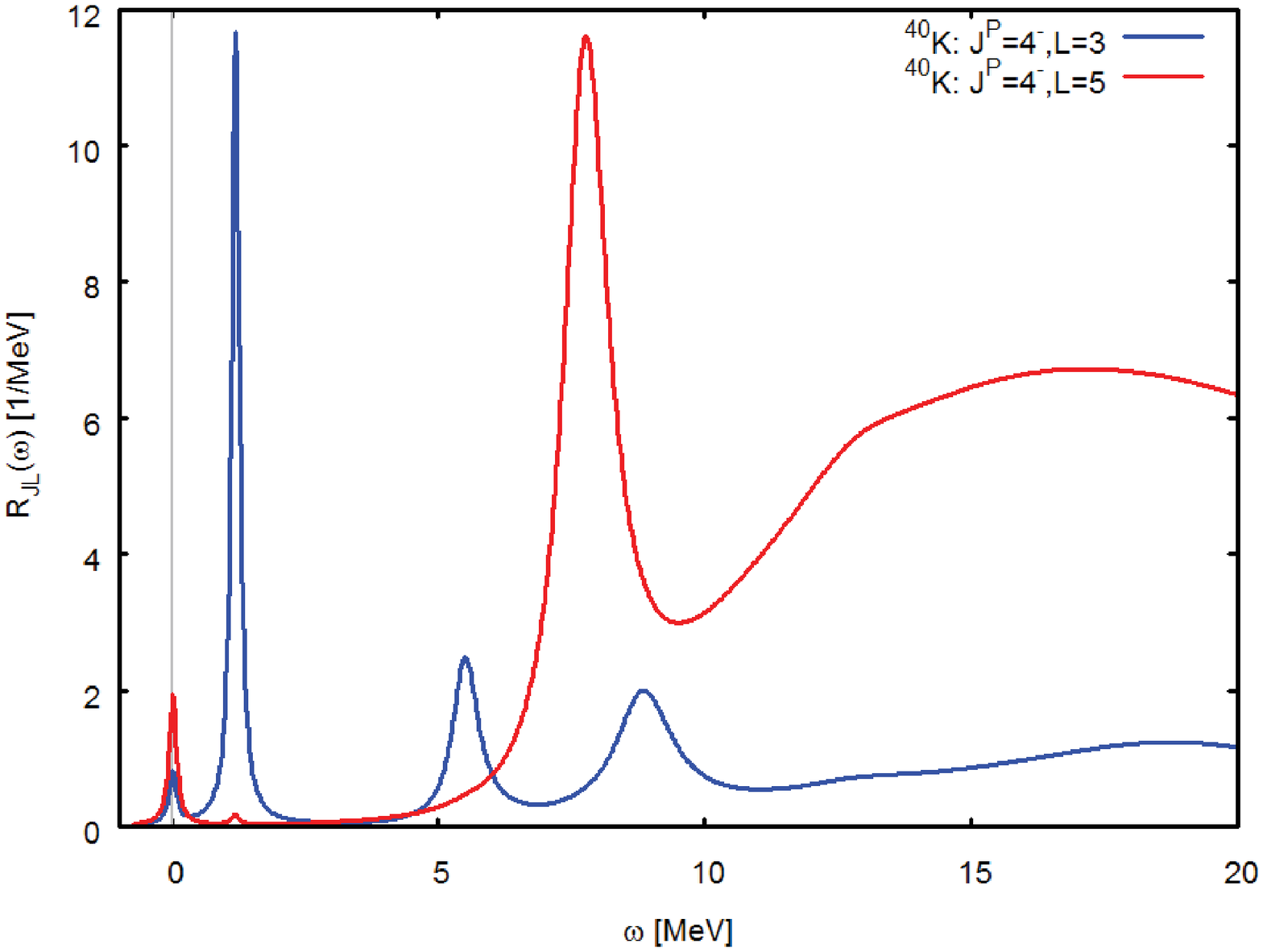}
\includegraphics[width = 6.5cm]{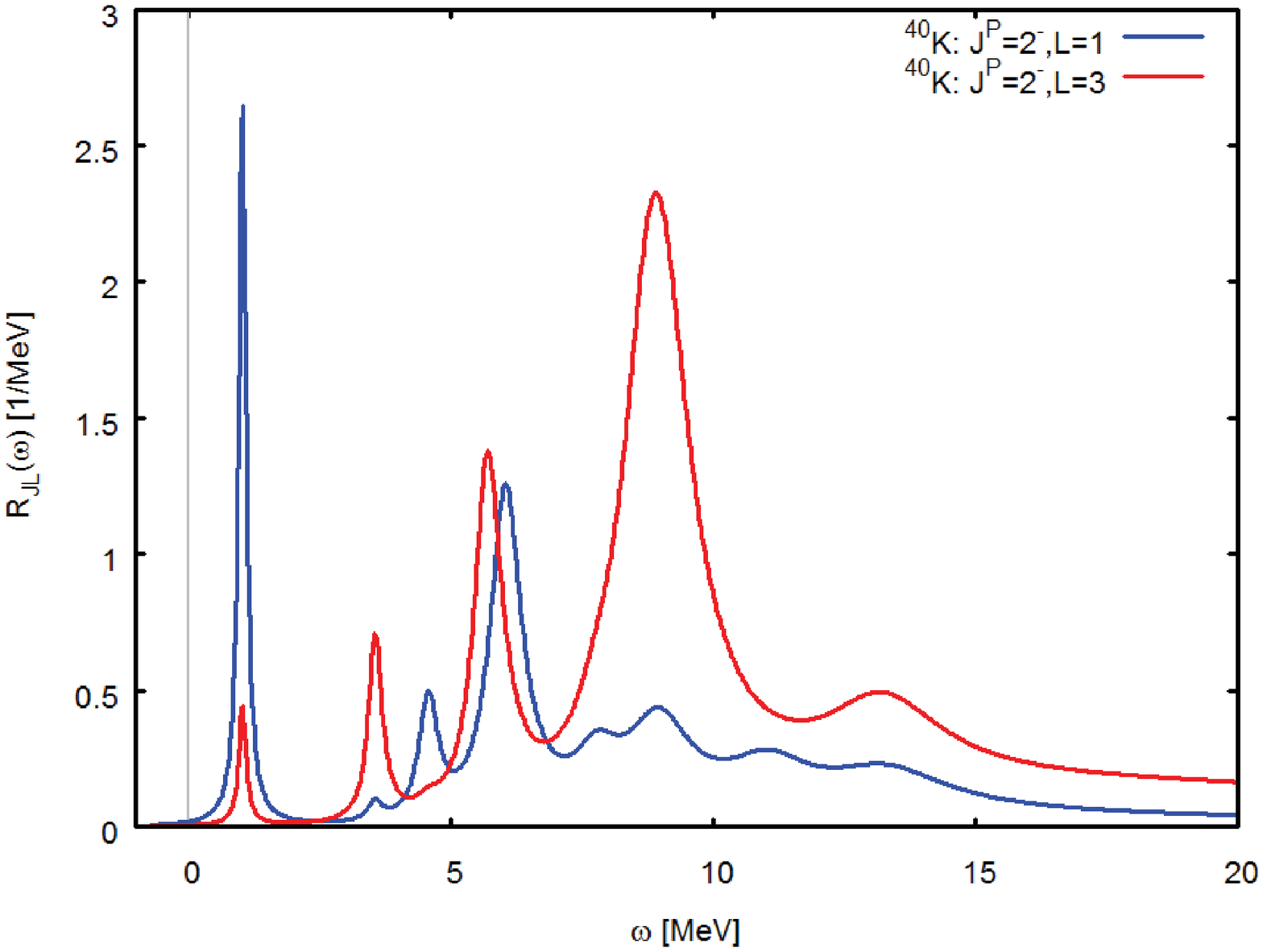}
\includegraphics[width = 6.5cm]{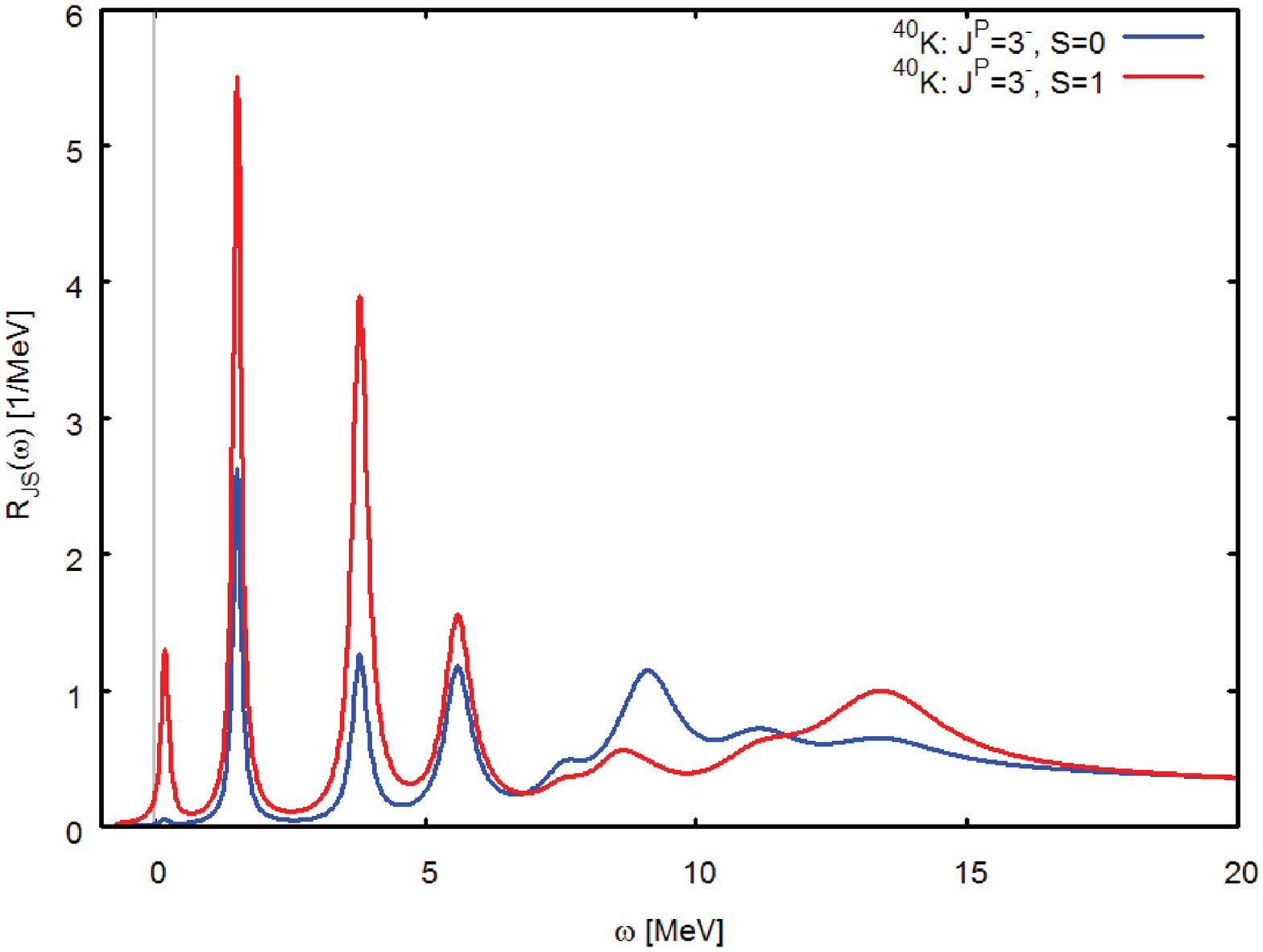}
\includegraphics[width = 6.5cm]{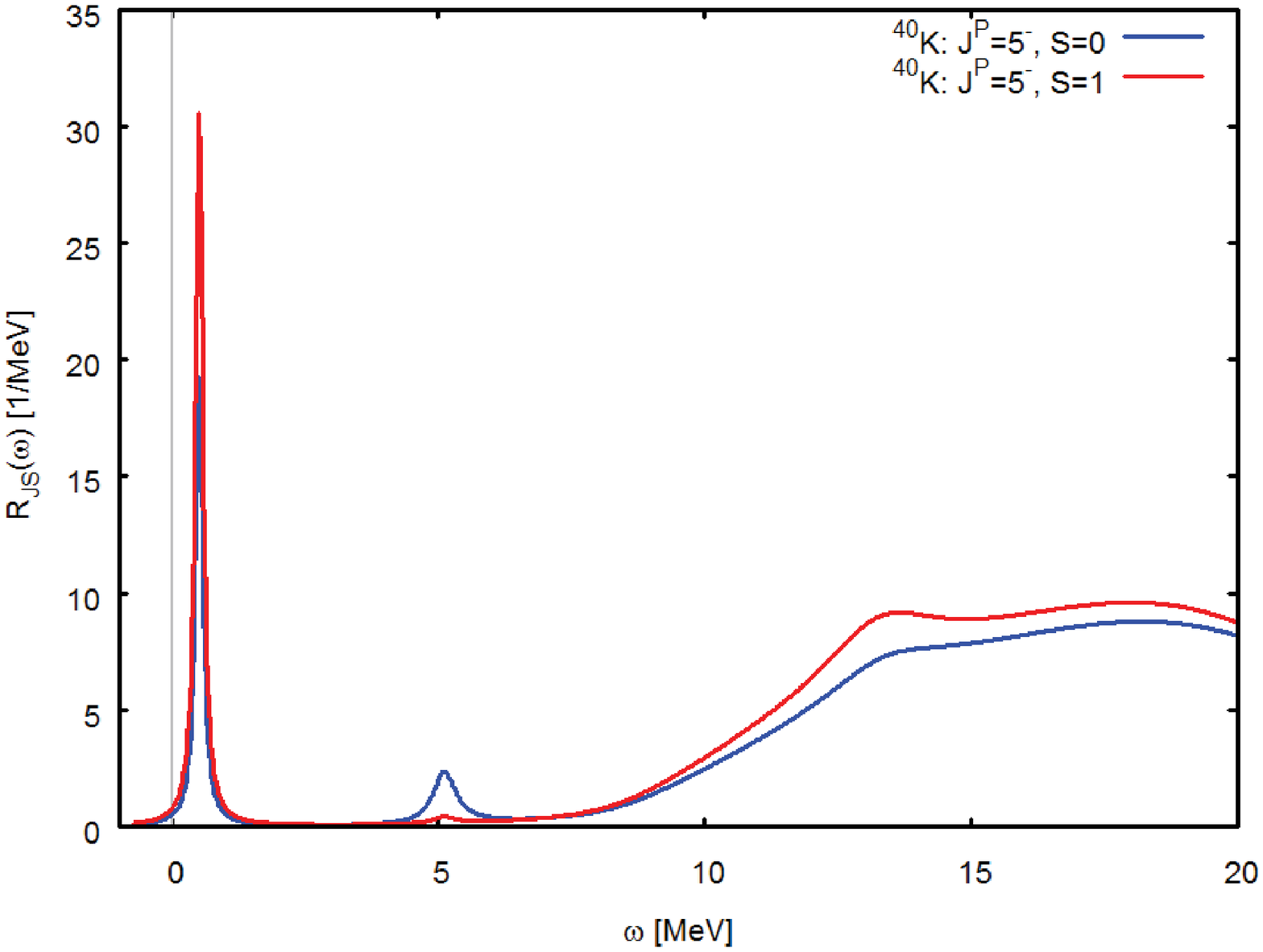}
\includegraphics[width = 6.5cm]{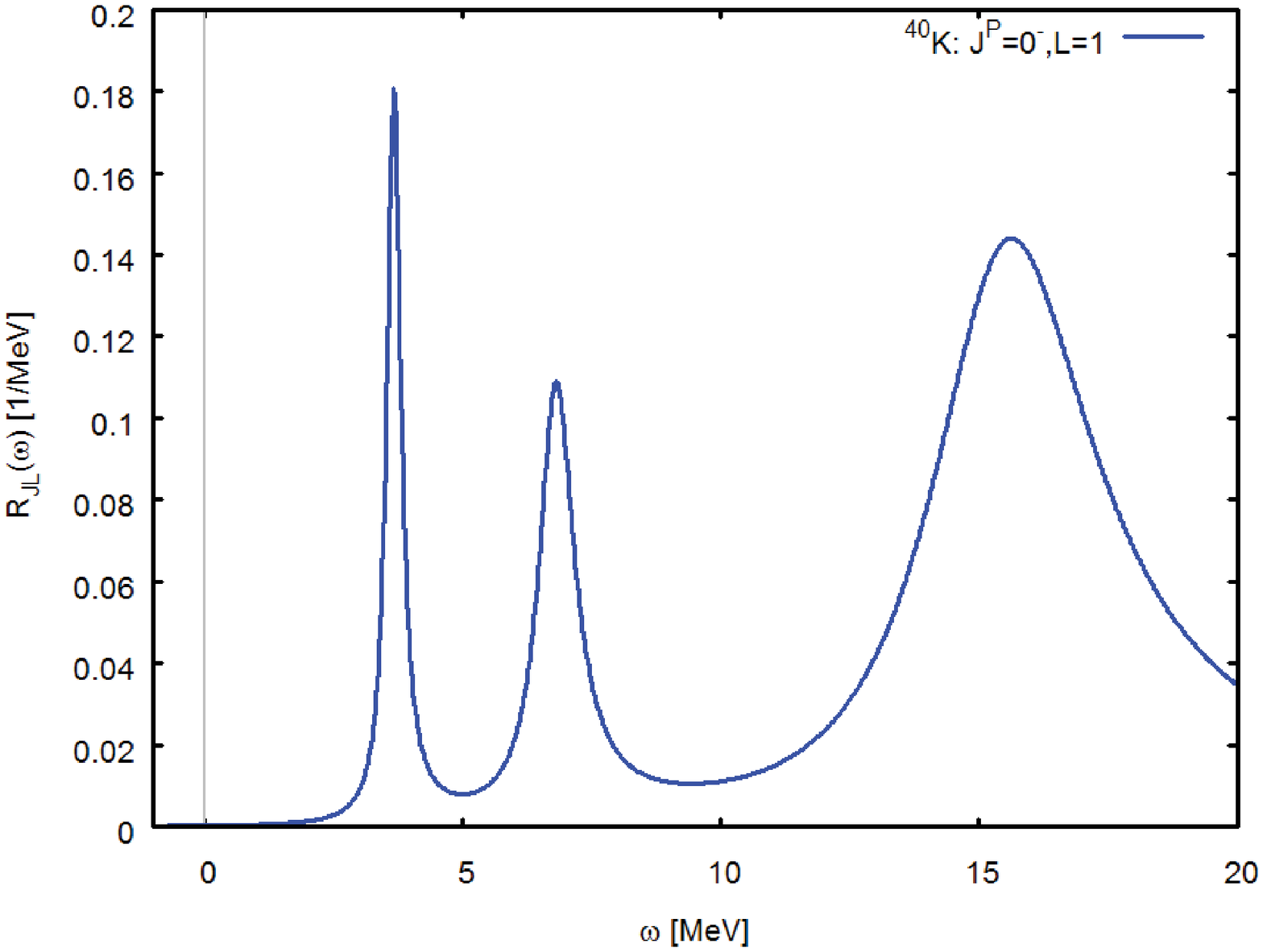}
\includegraphics[width = 6.5cm]{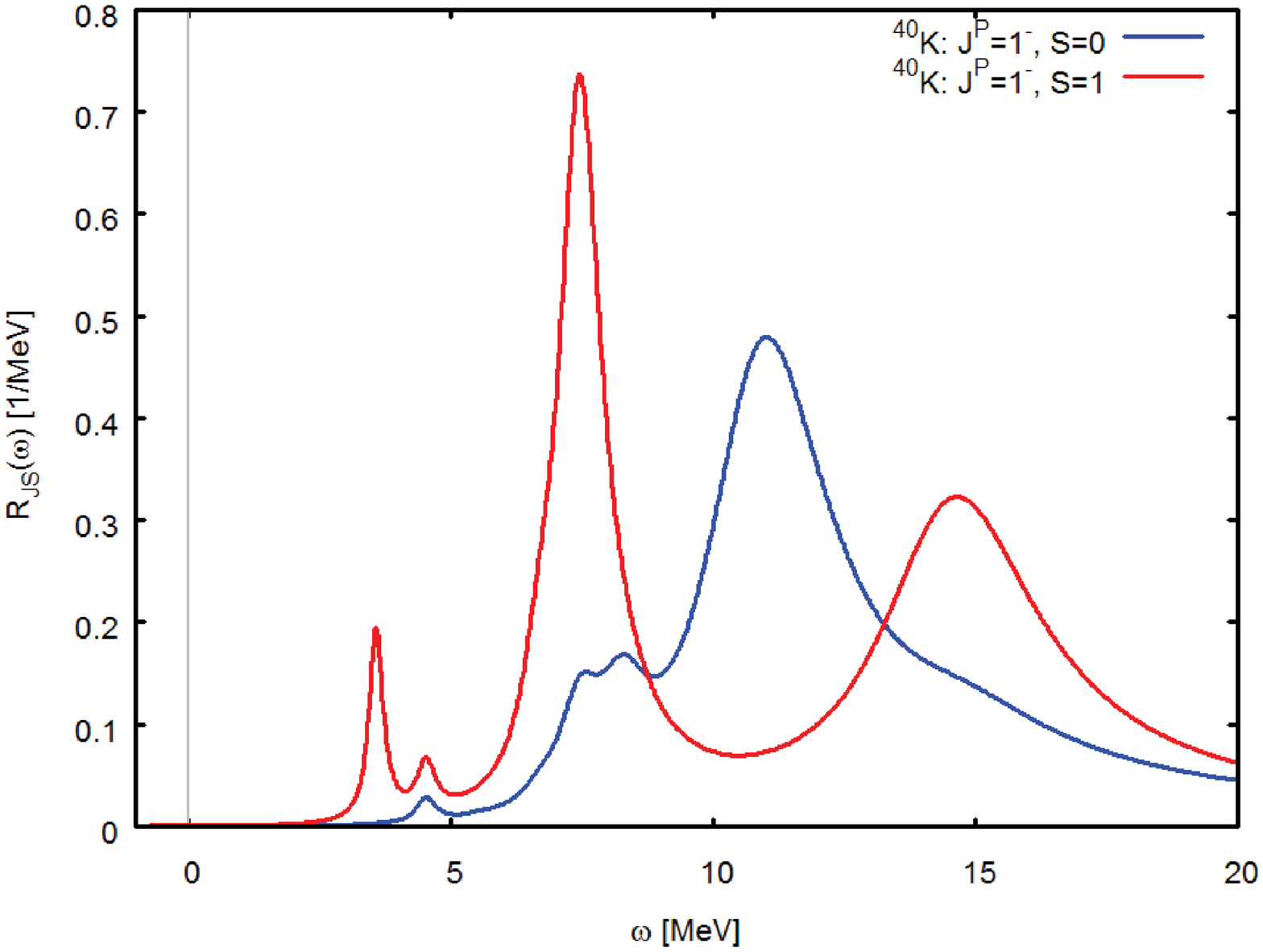}
\caption{(Color online) QRPA response functions for $^{40}Ca\to ^{40}K$ transitions. Results for the multipole transition operators $T_{LSJM}=\left(\frac{r}{R_d} \right)^L\left[ \bm{\sigma}^S\otimes Y_L\right]_{JM}\tau_{-}$ are shown where $R_d=3.72$~fm corresponds to the half-density radius of $^{40}Ca$.}
\label{fig:40K}
\end{center}
\end{figure*}

\begin{figure*}
\begin{center}
\includegraphics[width = 6.5cm]{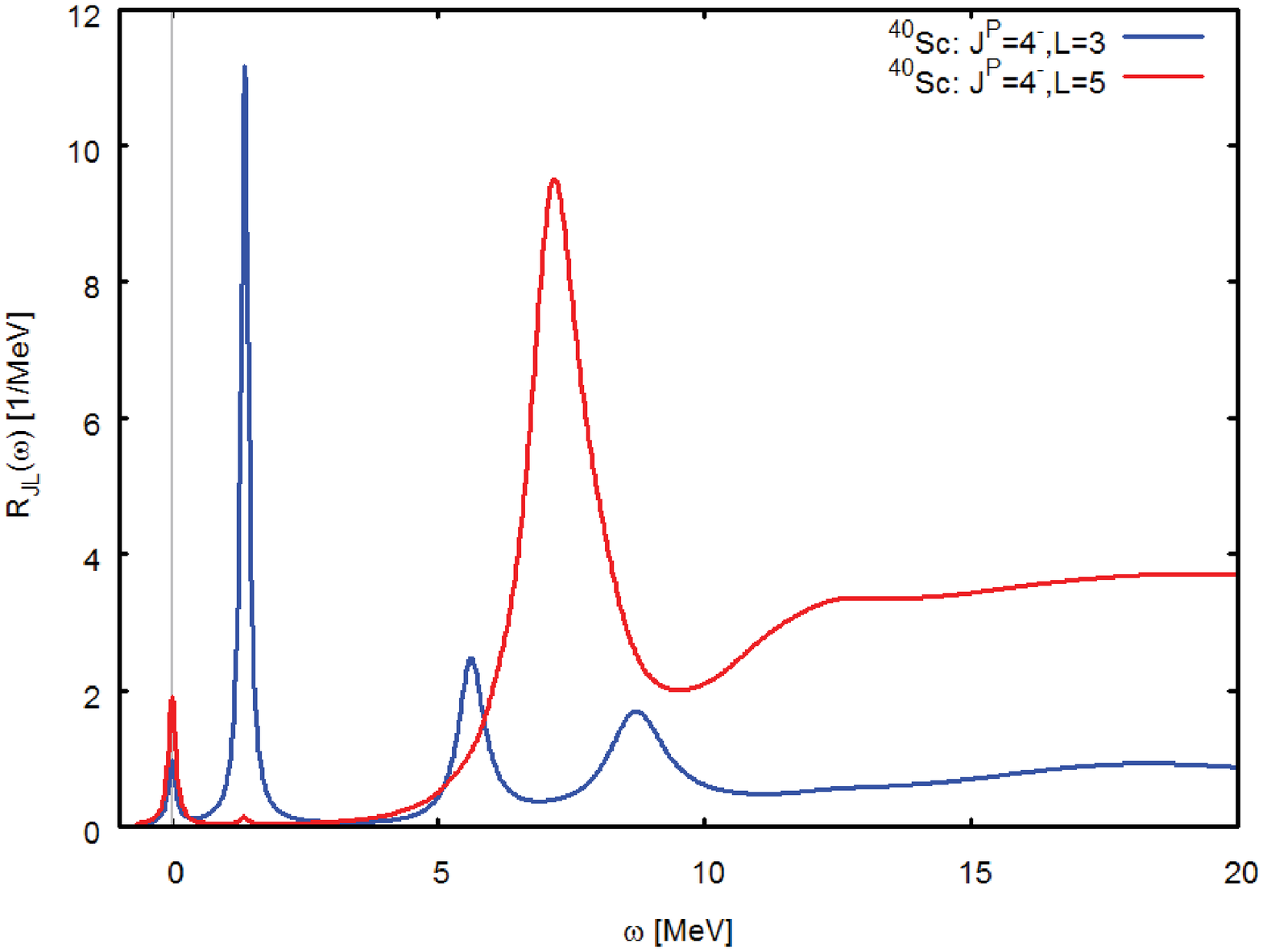}
\includegraphics[width = 6.5cm]{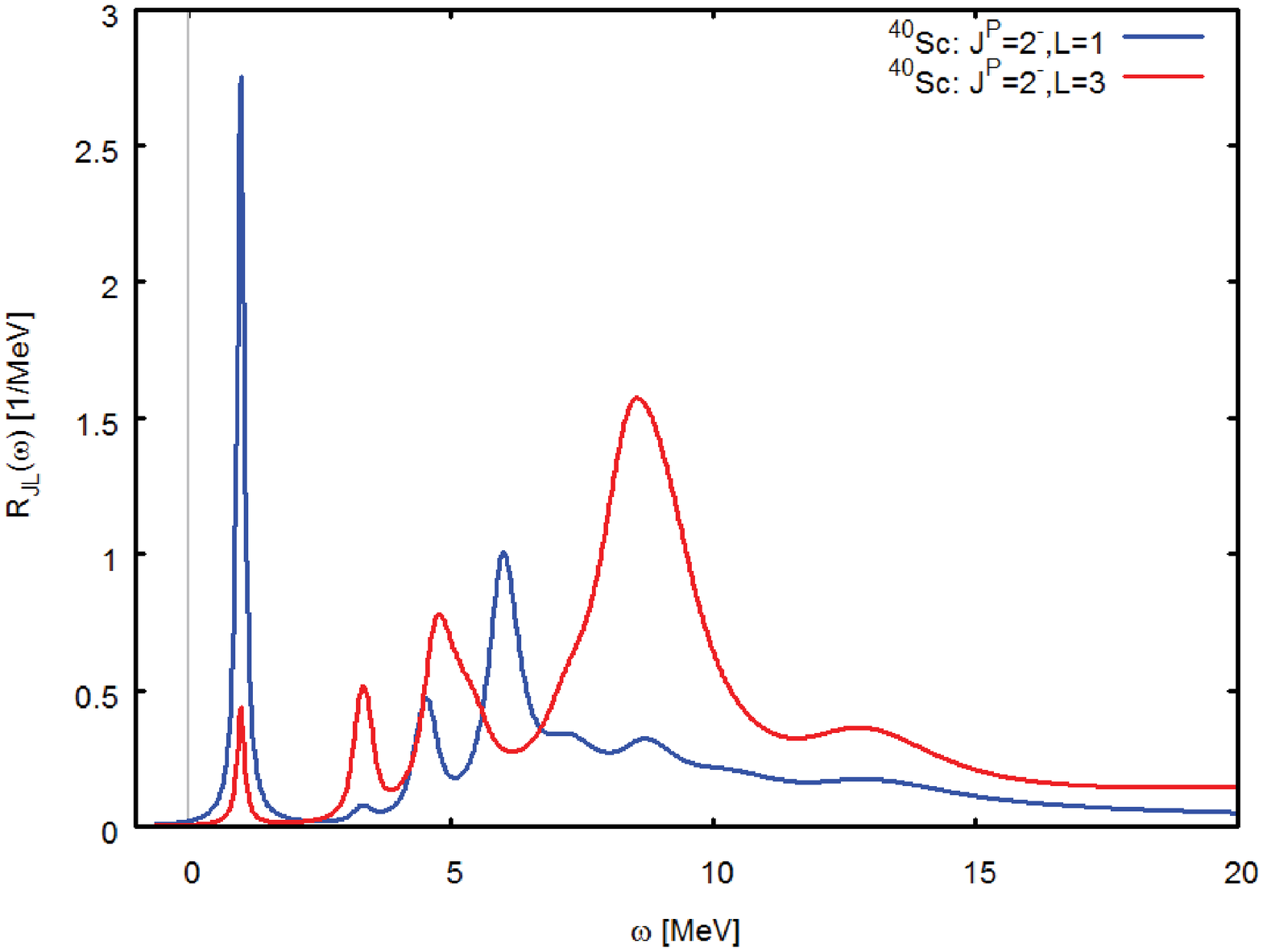}
\includegraphics[width = 6.5cm]{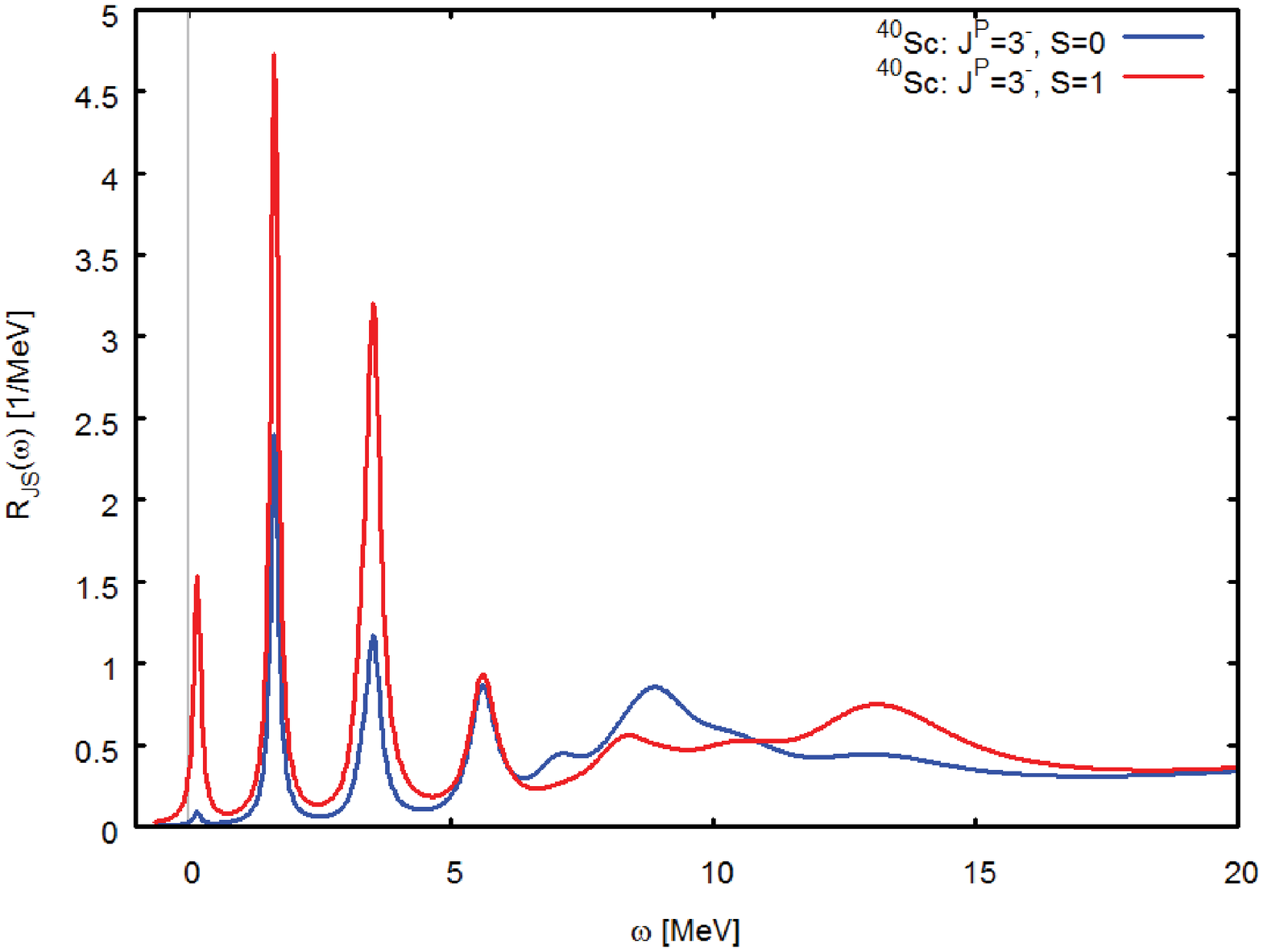}
\includegraphics[width = 6.5cm]{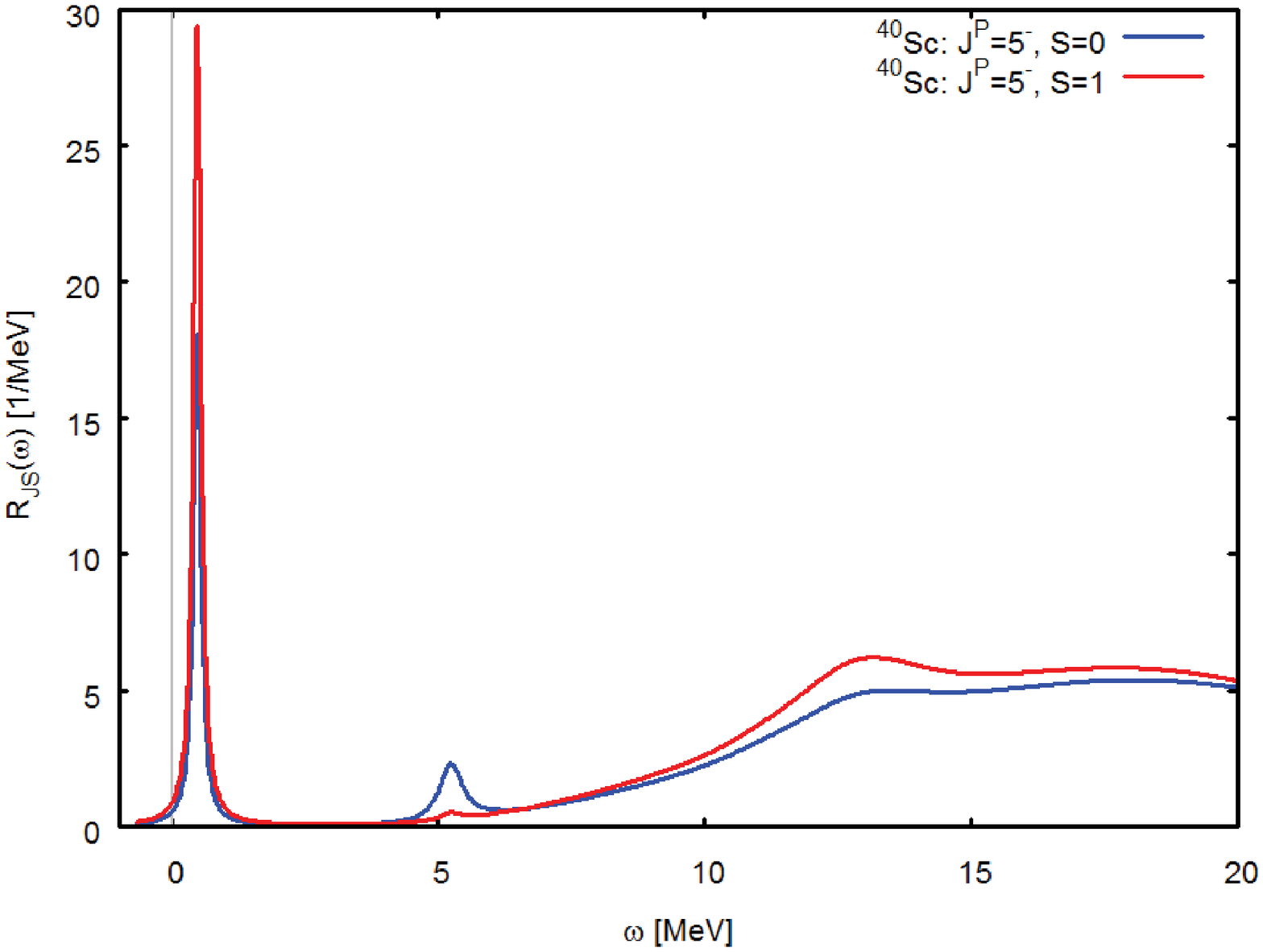}
\includegraphics[width = 6.5cm]{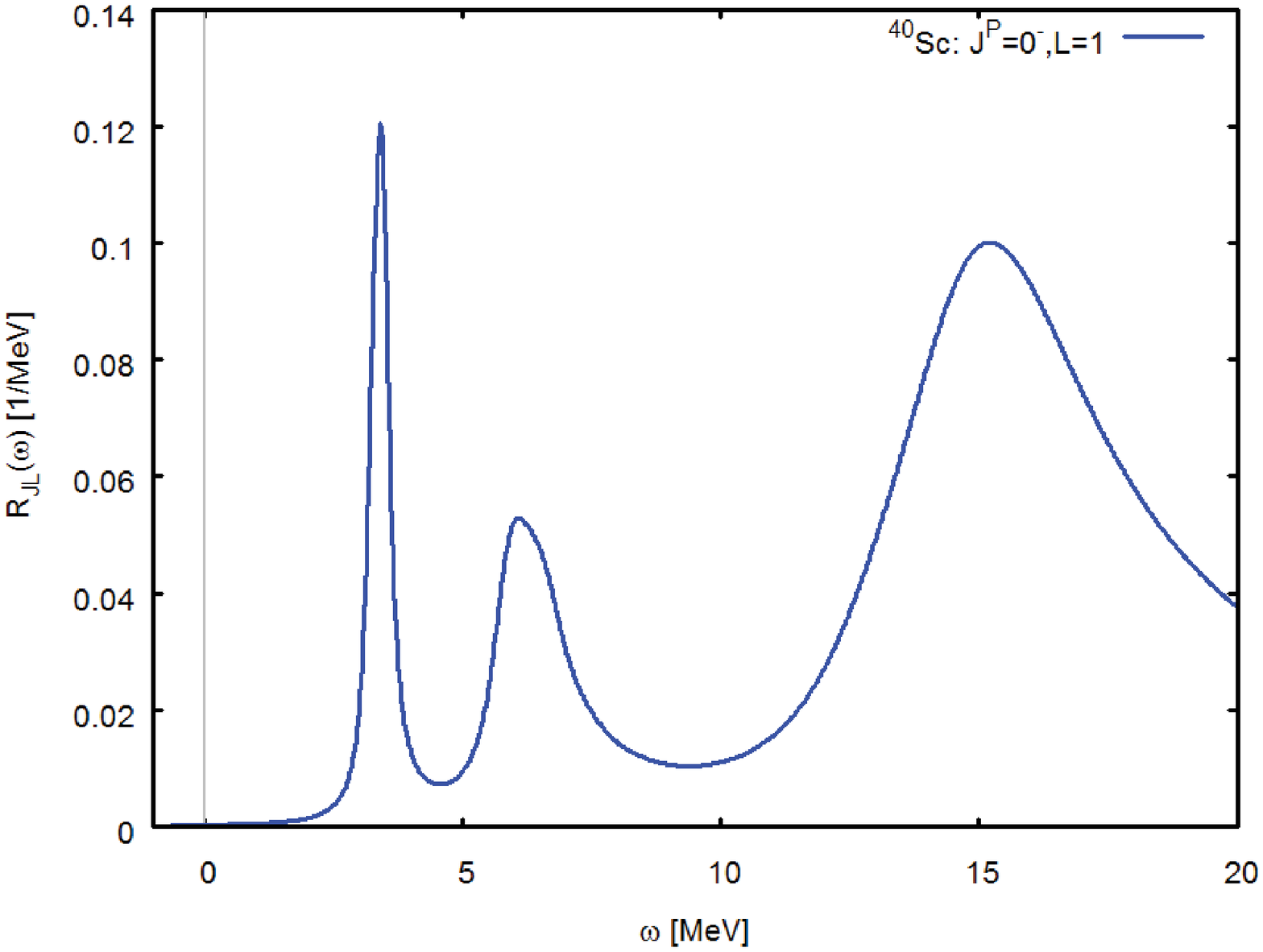}
\includegraphics[width = 6.5cm]{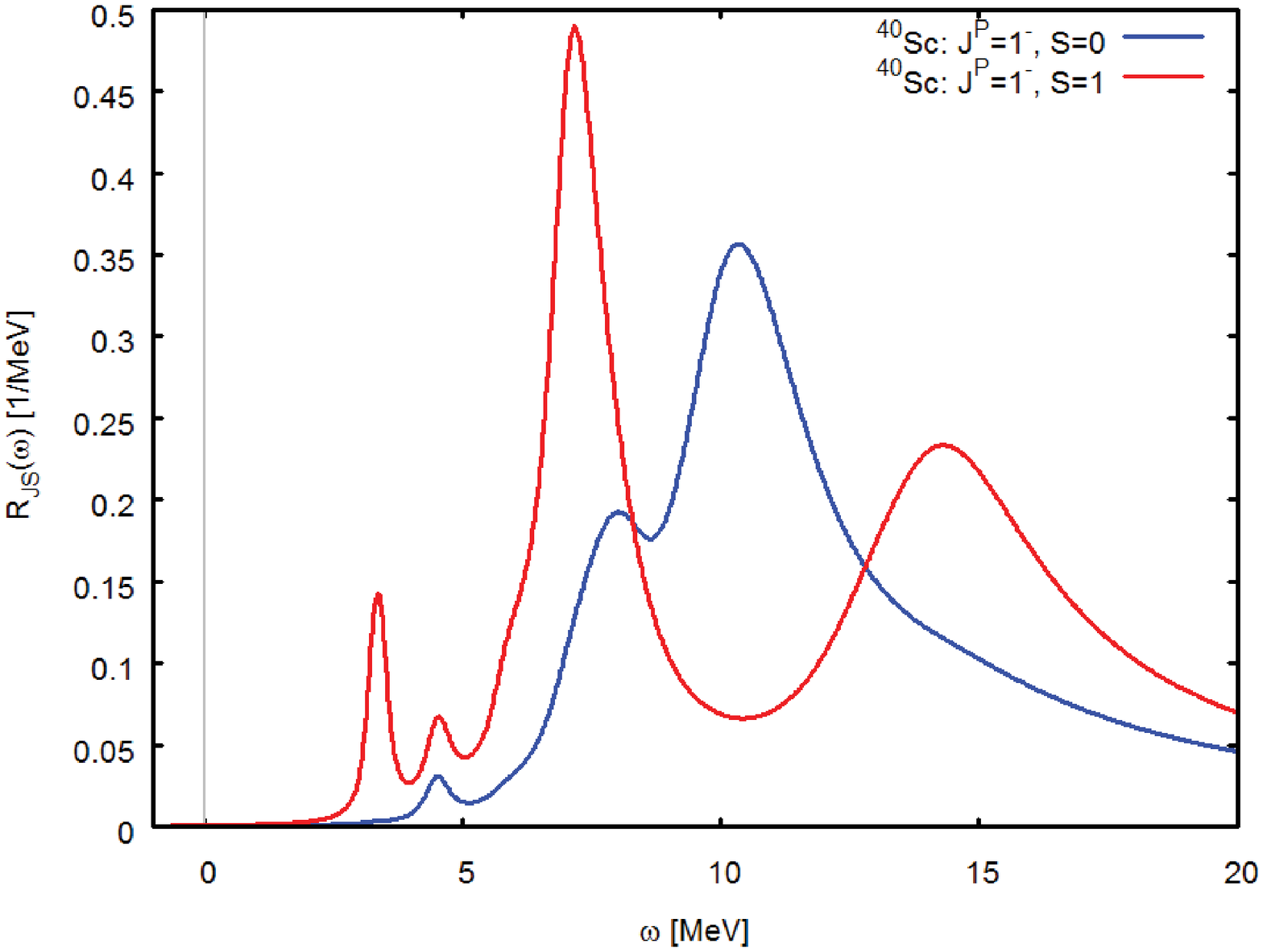}
\caption{(Color online) QRPA response functions for $^{40}Ca\to ^{40}Sc$ transitions. Results for the multipole transition operators $T_{LSJM}=\left(\frac{r}{R_d} \right)^L\left[ \bm{\sigma}^S\otimes Y_L\right]_{JM}\tau_{+}$ are shown where $R_d=3.72$~fm corresponds to the half-density radius of $^{40}Ca$.}
\label{fig:40Sc}
\end{center}
\end{figure*}

Using the same scheme as in the previous case, also here HFB single quasiparticle energies, pairing amplitudes, and wave functions for $^{40}Ca$ have been used to construct the polarization propagators. The QRPA spectra are shown for $^{40}K$ in Fig.\ref{fig:40K} and for $^{40}Sc$ in Fig.\ref{fig:40Sc}, respectively. The $A=40$ ground state multiplets are satisfactorily described: In both nuclei a $J^P=4^-$ ground state is obtained. In $^{40}K$ we obtain the level sequence $J^P=3^-,5^-,2-$ at $E_x=302,501,1008$~keV. As in the data,  $J^P=0^-,1^-$ states are found at higher energies, namely $E_x=3726$~keV and $E_x=3562$~keV. A very similar picture is emerging for $^{40}Sc$: There, we find again the ground state multiplet $J^P=3^-,5^-,2^-$ but at slightly different energies, $E_x=165,474,923$~keV, followed by a $J^P=0^-$ level at $E_x=3412$~keV and a $J^P=1^-$ state at $E_x=3355$~keV. In both nuclei, positive parity states occur at much high energy states, in fact beyond the continuum thresholds. The reason is that the $^{40}Ca$ HFB ground state is  given by a perfect double-magic shell closure. However, as discussed in \cite{Eckle:1989,Eckle:1990}, core polarization will modify that picture by dissolving the shell closures in $^{40}Ca$ on a level of about 10 to 15\% and intruder positive parity states may be present also at low energy.

We emphasize again that the same EDF was used as in the A=18 calculations, refraining from parameter adjustments. As typical for mean-field based theories, in this case the larger mass of the parent nucleus led to an even better agreement with data. Thus, we may conclude that the QRPA approach provides a quite reliable description of SCE spectra.

\subsection{Optical Potential and Elastic Scattering}

A key issue for understanding heavy ion reactions on a quantitative level is the proper treatment of ion-ion interactions. Their paramount role is evident by considering the huge total reaction cross sections which are reflecting the importance of absorption of the incoming flux into a multitude of reaction channels. These effects lead to self-energies with large imaginary parts. Because of the lack of elastic scattering data, empirical optical potentials are not available for the projectile-target systems under scrutiny. Thus, we calculate the optical potential fully microscopically in a folding approach. The HFB ground state densities discussed above are folded with the NN T-matrix interaction, including both the isoscalar and isovector components. Because heavy ion scattering is a strongly absorptive process, elastic scattering and peripheral inelastic reactions are mainly sensitive to the nuclear surface regions of the interacting nuclei. Thus, to a very good approximation in-medium modifications of interactions can be neglected in the elastic ion-ion interactions and we are allowed to use the free space NN T-matrix as the dominant leading order impulse  approximation. In the numerical calculations, the NN T-matrix derived  by Franey and Love \cite{Love:1981gb} was used, extrapolated down to the present energy region. The approach is used for calculating the real and  the  imaginary  part  of  the  optical  potentials  in  the  incident  and  the  exit  channels. The Pauli-principle is taken care of by the pseudo-potential approach in local momentum approximation \cite{Satchler:1983}. Distorted waves are obtained by solving the Schr\"odinger equation with these microscopically derived optical potentials as discussed in section \ref{sec:DisCoeff}.

\begin{figure}
\begin{center}
\epsfig{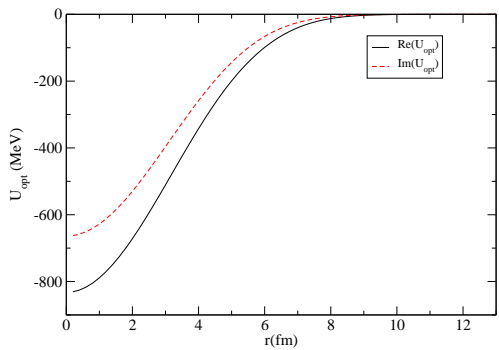}
\caption{(Color online) Double folding optical potential for $^{18}O+^{40}Ca$ at $T_{lab}=15$~AMeV: the real part (full line, blue) and the imaginary part (dashed line, red) are shown. }
\label{fig:OMP}
\end{center}
\end{figure}

In Fig. \ref{fig:OMP} the nuclear part of the optical potential for the $^{18}O+^{40}Ca$ incident channel is shown. Characterizing quantities like volume integrals (per nucleon), root-mean square radii, and the total reaction cross section are found in Tab. II. 

\begin{table}\label{tab:OMP}
\begin{center}
\begin{tabular}{|c|c|c|c|}
  \hline
  $U_{opt}$ & $I_0/N$ $[MeVfm^3]/N$ & $\sqrt{\lan r^2 \ran}$ $[fm]$ & $\sigma_{reac}$ $[b]$  \\ \hline
  Re$U_{opt}$ & -439.71 & 4.75 & --  \\ \hline
  Im$U_{opt}$ & -319.37 & 4.61 &  2.14  \\
  \hline
\end{tabular}
\caption{(Color online) Defining quantities of the double folding optical potential for the system $^{18}O+^{40}Ca$ at $T_{lab}=270$~MeV. Volume integrals per projectile and target nucleon numbers are denoted by $I_0/N$. HFB ground state densities and the free space NN T-matrix interaction obtained from Ref. \cite{Love:1981gb} were used.}
\end{center}
\end{table}

\begin{figure}
\begin{center}
\epsfig{file=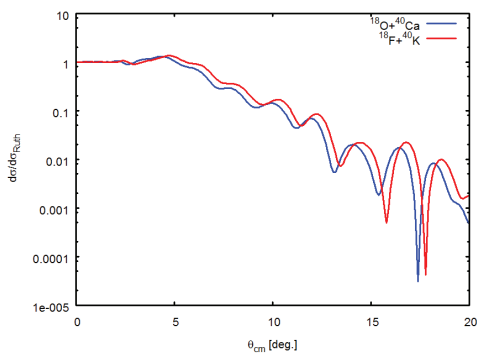, width=6.5cm}
\caption{(Color online) Elastic scattering angular distributions, normalized to the Rutherford cross section, are shown as a function of the center-of-mass scattering angle for $^{18}O+^{40}Ca$ (blue line) and $^{18}F+^{40}K$ (red line) at $T_{lab}=15$~AMeV .}
\label{fig:Elastic}
\end{center}
\end{figure}

Results for the elastic scattering cross sections, normalized to the Rutherford cross section are shown in Fig.\ref{fig:Elastic}. At extreme forward scattering angles it is dominated by pure Coulomb scattering but beyond $\theta \gtrsim 1.5$~degr. the short range nuclear parts are taking over.

\subsection{Comparison of PWBA and DWBA SCE Cross Sections }
Following the reaction and nuclear structure  formalism outlined above, numerical calculations
of single charge exchange cross section have been performed
employing the code HIDEX. Form factors are derived by folding the transition densities with the projectile-target residual charge exchange interaction where the momentum representation is used \cite{Franey:1985ye}. In order to maintain self-consistency as much as possible we use the same 2QP isovector interaction as in the nuclear structure calculations. The operator structure includes spin-dependent and spin-independent direct and exchange central interactions, together  with  second  rank  tensor terms. The  NN  spin-~orbit interactions have been neglected.
Then elastic scattering and SCE cross sections were obtained.
The procedure follows closely the approach used successfully already in our previous investigations of SCE reactions \cite{Lenske:1989zz,Cappuzzello:2004afa}.

\begin{figure}
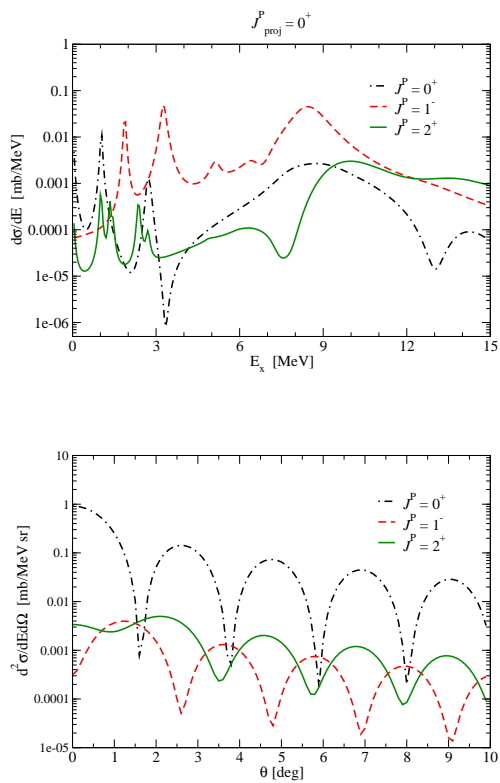

\begin{center}
\epsfig{file=fig10a_SCE.eps, width=6.5cm}
\vskip 1cm
\epsfig{file=fig10b_SCE.eps, width=6.5cm}
\caption{(Color online) DWBA cross section as a function of target excitation energy, integrated in the full angular range (top panel),
and angular distribution for $E_x=0\,MeV$ (bottom panel) for several multipoles, contributing to Fermi like transitions in the target.  Calculations are for the reaction $^{40}Ca\left(^{18}O,^{18}F\right)^{40}K$ reaction at $T_{lab}=270\,MeV$.
}
\label{fig:1_5Fermi}
\end{center}
\end{figure}

\begin{figure}[!hbt]
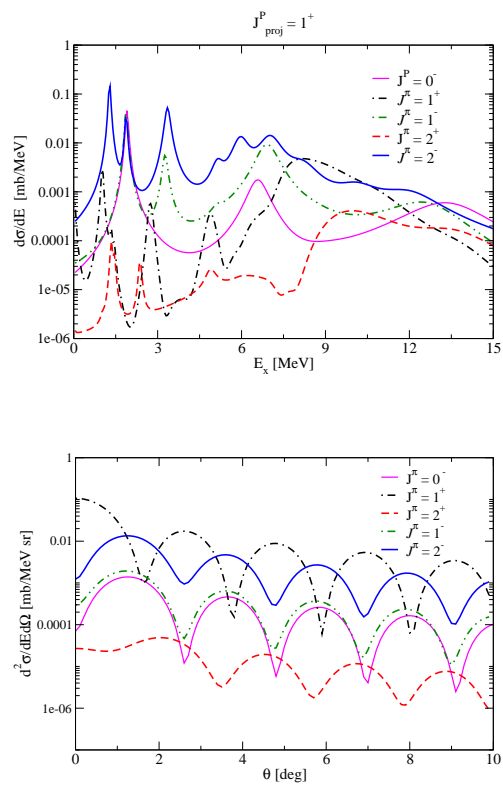

\begin{center}
\epsfig{file=fig11a_SCE.eps, width=6.5cm}
\vskip 1cm
\epsfig{file=fig11b_SCE.eps, width=6.5cm}
\caption{(Color online) DWBA cross section as a function of target excitation energy, integrated in the full angular range (top panel), and angular distribution for $E_x=0\,MeV$ (bottom panel) for several multipoles, contributing to Gamow - Teller like transitions
in the target. The system is the same as in Fig. \ref{fig:1_5Fermi}.}
\label{fig:1_5}
\end{center}
\end{figure}

The closest resemblance to nuclear beta decay processes is found in pure Gamow-Teller (spin-isospin flip with $J^P=1^+$) or pure Fermi (isospin flip with $J^P=0^+$) excitations, respectively. However, strong interaction processes are less selective on multipolarities than weak interactions. Moreover, due to the peripheral character of inclusive heavy ion reactions, very often transition of higher angular momentum transfer are favored. Thus, heavy ion SCE reactions enable to probe the whole spectrum of Gamow-Teller-like spin-isospin flip and Fermi-like isospin flip multipole transitions, discussed in the previous section, allowing to study multipolarities suppressed otherwise in weak decay processes.
From the theoretical discussion is it clear that distortion effects are playing a significant role in heavy ion SCE reactions. Results for SCE reaction cross sections and angular distributions in full DWBA are shown in Figs. \ref{fig:1_5Fermi} - \ref{fig:1_5}, for the reaction $^{18}O+^{40}Ca$ $\to$ $^{18}F+^{40}K$. The associated $Q$ value is $Q_{gs} = -2.97\, MeV$, whereas the alternative single charge changing process $^{18}O+^{40}Ca$ $\to$ $^{18}N+^{40}Sc$ would correspond to $Q_{gs} = -28.22\, MeV$ which is of a much larger magnitude. The strong kinematical mismatch will lead to a smaller cross section in this case.

For the Gamow - Teller (Fermi) case, we consider transitions leading to the
1$^+$ ground state (0$^+$ excited state) of $^{18}F$ and populating several $^{40}K$ excited states, identified by the spin $J$ and the
excitation energy $E_x$ . For the present purpose we neglect the small variations in excitation energy of the $^{18}F$ ground state multiplet, treating the states as energetically degenerate with vanishing excitation energy.  From Figs. \ref{fig:1_5Fermi} -
\ref{fig:1_5}, it is straightforward to note that $J^P=1^+$ and $J^{P}=0^+$ target transitions contribute significantly to the cross section at low excitation energies and dominate at small angles.

Having in mind in the first place illustrative purposes, we will focus thereafter on pure Gamow - Teller  excitations in both projectile and target. The results concerning distortion effects and the relation of the (physical) DWBA cross section to the plane wave counterpart and the beta-decay matrix elements is to a large extent independent of the multipolarity, at least at small momentum transfer. Thus, without loss of generality, it is sufficient to consider a single multipolarity.

\begin{figure}
\begin{center}
\epsfig{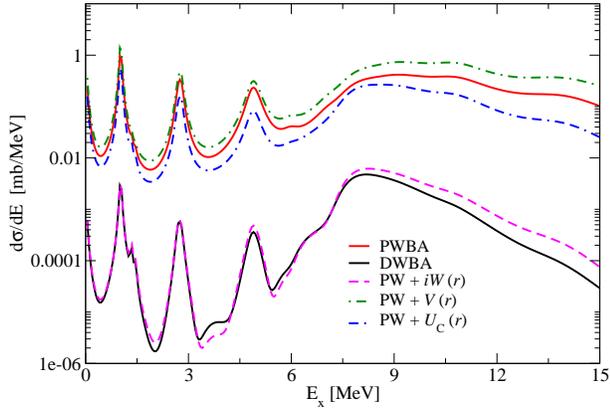}
\caption{(Color online) Cross sections as a function of the target excitation energy, $E_x$,
for the $J^P = 1^+$ transition,
for the SCE reaction $^{40}Ca\left(^{18}O,^{18}F\right)^{40}K$ reaction at $T_{lab}=270\,MeV$,
integrated over the full angular range. The different curves show the effect of Coulomb potential ($U_C(r)$), of real ($V(r)$) and imaginary ($W(r)$) components of the optical potential and of the full potential (DWBA), with respect to PWBA calculations. The system is the same as in the previous figures.}
\label{fig:2}
\end{center}
\end{figure}

In order to understand the influence of the elastic ion-ion interactions on SCE processes, we first disentangle the various contributions to the optical potentials. Fig. \ref{fig:2} displays the ($^{18}O,^{18}F(g.s.)$) total cross section $\sigma_{\alpha\beta}$ as a function of the target excitation energy, integrated over the full angular range. Calculations are performed in the Plane Wave Born Approximation (PWBA), as well as considering
separately the effects of Coulomb potential and of real and imaginary part of the nuclear optical potential, and, finally, combining all these potentials in the Distorted Wave Born Approximation (DWBA). Already at the PWBA level, one can appreciate the main excitation peaks contributing to $J^{P} = 1^+$ transitions in the target. With respect to the latter results, it is observed
that the cross section decreases when the effect of the Coulomb repulsion is taken into account or increases when considering
the contribution of the (attractive) real part of the nuclear optical potential. However, the most striking feature
is the strong suppression, by about a factor $500-600$, observed just taking into account the imaginary part of the optical potential,
which essentially brings the cross section down to the value associated with the full DWBA calculation.
This indicates that the DWBA result is mainly explained in terms of strong absorption effects, as expected
in heavy ion reactions, and justifies the strong absorption approach, underlying the black disk approximation to model the ion-ion initial and final state interactions (see Section \ref{ssec:BDA}).

\begin{figure}
\begin{center}
\epsfig{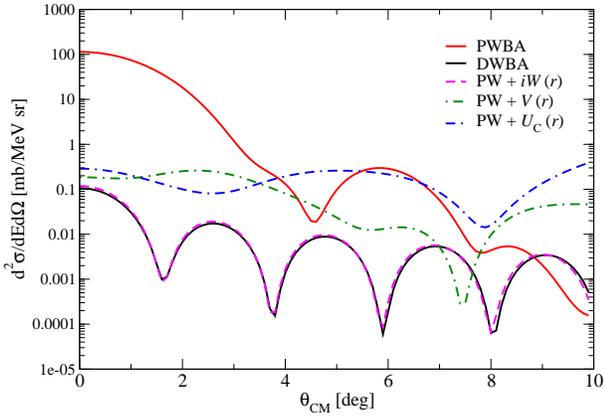}
\caption{(Color online) Angular distribution for the target state at $E_x=0\,MeV$ .
The different curves show the effect of each optical potential component and of the full DWBA case, with respect to
PWBA calculations. Same system as in the previous figures.}
\label{fig:3}
\end{center}
\end{figure}

For the state at the lowest excitation energy ($E_x=0\,MeV$, with respect to $^{40}K$ ground state), that as discussed before
is an intruder state for the $^{40}K$ ground state,
Fig.\ref{fig:3} represents the differential cross section,  $d^2\sigma_{\alpha\beta}/d\Omega dE$,
as a function of the angle $\theta$.  It appears that absorption effects
also lead to a different diffraction pattern (compare PWBA and DWBA results),
which reflects the size of the absorbing region.

\begin{figure}
\begin{center}
\epsfig{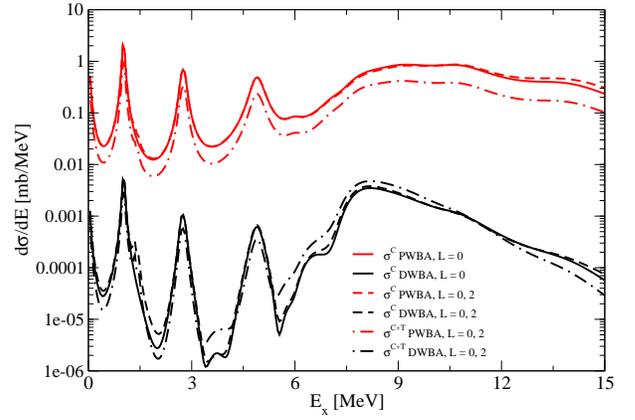}
\caption{(Color online) Cross section as a function of the target excitation energy, integrated over the full angular range. The plot shows the effects of $L=0,2$ multipolarities, involved in $J^{\pi}=1^+$ transition and of central and tensor components of nuclear interaction. Same system as in the previous figures.}
\label{fig:4}
\end{center}
\end{figure}

\begin{figure}
\begin{center}
\epsfig{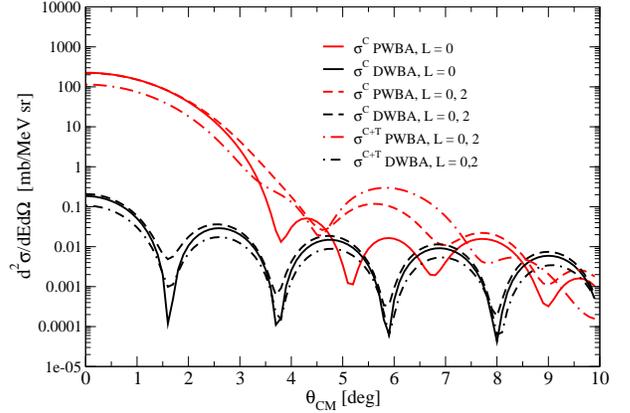}
\caption{(Color online) Angular distribution for the target state at $E_x$ = 0.
The plot shows the effects related to central and tensor components of the nuclear interaction, for the two multipolarities allowed by $J^{\pi}=1^+$ transitions: $L=0,2$. Same system as in the previous figures.}
\label{fig:5}
\end{center}
\end{figure}

The interplay between central and tensor terms of the effective interaction is investigated next.
Results are shown in Figs. \ref{fig:4} - \ref{fig:5}, together with the contributions associated with the two multipolarities ($L = 0,2$) leading to $J^{P} = 1^+$ transitions. One can see that the central interaction contribution to the
angle integrated cross section is fully dominated by $L=0$ transitions. The same conclusion holds for the differential cross section, as far as
the small angles shown on the figure are concerned.

\vskip 0.5cm

The tensor interaction is seen to slightly reduce the cross section in the
PWBA case and, in the full DWBA calculations, for the main excitation peaks.
Actually, as shown in Fig. \ref{fig:5}, the tensor contributions also shift
the cross section to larger angles, owing to the dominant role of $L=2$ transitions
in this case.
Guided by these results, in the following we will consider, for the sake of
simplicity, excitations corresponding to $L=0$ and we will neglect the tensor
part of the effective interaction.

\subsection{Cross section Factorization}\label{ssec:XsecFact}

\begin{figure}
\begin{center}
\epsfig{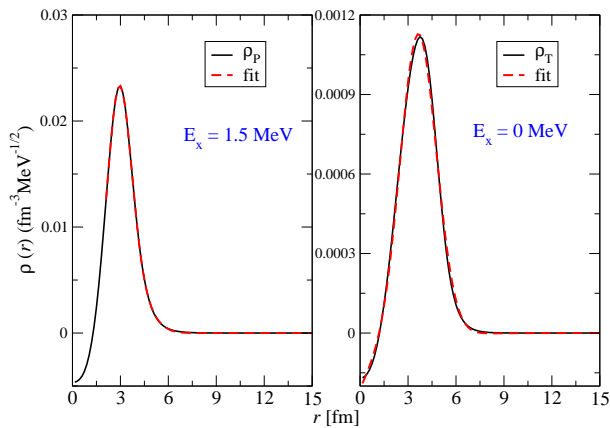}
\caption{(Color online) Projectile (left panel) and target (right panel) radial transition densities corresponding to a
selected transition (see text), fitted with a sum of two Gaussians.}
\label{fig:6}
\end{center}
\end{figure}

As stressed in Section \ref{sec:GaussFF}, the case when the transition form factors, Eqs.(\ref{eq:FprojM}),(\ref{eq:FtargM}), can be approximated by a Gaussian function is of a particular advantage for the separation of the distortion effects.  This implies that the spatial transition densities contained in Eq.(\ref{eq:fredXY}) correspond
to the multipole components of a Gaussian. Following the formalism outlined in Section \ref{ssec:GaussTr}, we perform a Gaussian fit of the
transition densities, as obtained from our QRPA calculations, for projectile and target. An example, corresponding to excitations leading to the ground state
of $^{18}F$ ($E_x=1.5\,MeV$) and to zero excitation energy  ($E_x=0$) for the target is shown in Fig. \ref{fig:6}. The fit is performed considering the superposition of two Gaussians. The Gaussian fit parameters $R$ and $\sigma$ are determined in the region of interest for direct reactions, i.e. the surface region.

\begin{figure}
\begin{center}
\epsfig{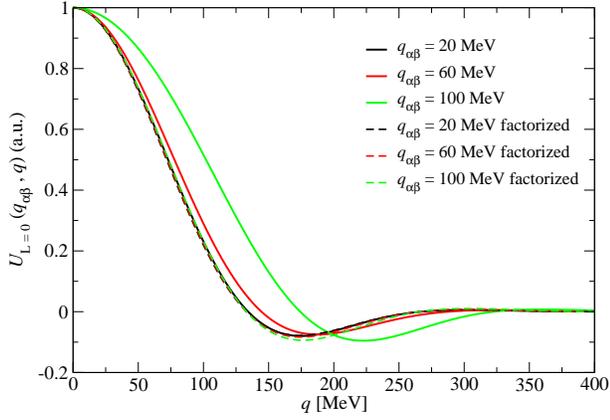}
\caption{(Color online) Comparison between the monopole component of the Gaussian reaction kernel, obtained in the separation ansatz,
with $h_{\alpha\beta}(q)$ given by Eq. (\ref{mono}), (dashed lines) and in the full BD approximation (solid lines). 
Different colors indicate different values of $q_{\alpha\beta}$.
The three dashed lines are difficult to distinguish because they are very
close to each other.} 
\label{fig:7}
\end{center}
\end{figure}

Combining the results of projectile and target Gaussian fits and neglecting the momentum dependence of
the interaction form factor $V_{ST}^C(p^2)$ , which is quite flat in the low momentum transfer
 range corresponding to  $\theta\in\left[0^{\circ}, 10^{\circ}\right]$,
one can finally extract the parameters ($R$ and $\sigma$) entering the expressions
(\ref{Ug_gaus}),(\ref{Ulm}) for  the full reaction amplitude in Born approximation,
$M^{(B)}_{\alpha\beta}(\mathbf{p})$.
It results: $R^2 = R_a^2 + R_A^2$ and $\sigma^2 = \sigma_a^2 + \sigma_A^2$,
being $R_a(R_A)$ and $\sigma_a(\sigma_A)$ the fit parameters referring to
the projectile (target) transition density.
We find R$\approx 5\,fm$ , $\sigma \approx 1.2\,fm$. 
Then it is possible to evaluate the quantity
$\bar{M}^{(B)}_{\alpha\beta}(\mathbf{q}_{\alpha\beta},q)$, i.e.  the Born amplitude
averaged over the orientation of the off-shell momentum $\mathbf{q}$,
which is particulary important for the calculation of the distortion effects.
Fig. \ref{fig:7} shows the results obtained,
employing the Gaussian fit described above, for the monopole term $U_{L=0}$,
according to the full expression Eqs.(\ref{M_2}),
(\ref{M_3}) or adopting the (partial) separation
ansatz, as in Eqs.(\ref{UG_eq})
,(\ref{mono}).
One can observe that, whereas the separation ansatz works quite well
for small values of $q_{\alpha\beta}$ (see for instance the results corresponding
to $q_{\alpha\beta} \approx 20 \,MeV/c$), important deviations from the exact results are seen for larger  $q_{\alpha\beta}$ values.

\begin{figure}
\begin{center}
\epsfig{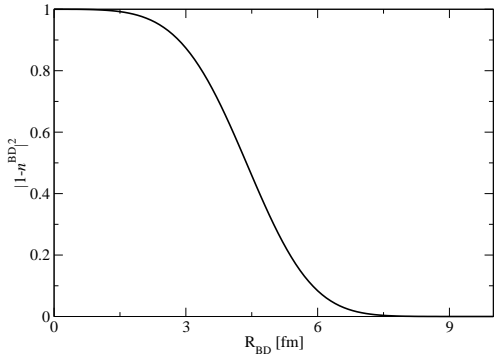}
\caption{Distortion factor as a function of $R_{abs}$, 
for the separation function $h_{\alpha\beta}(q)$ corresponding to Eq. (\ref{mono}) (see text).}
\label{fig:8}
\end{center}
\end{figure}

Let us first consider the case of small momentum transfer ($q_{\alpha\beta} = 20\, MeV/c$).
Using Eq.(\ref{scal}), the distortion factor $f_{BD} = |1-n_{\alpha\beta}|^2$
is readily obtained in the black disk approximation. This is shown in Fig.\ref{fig:8} as a function of the absorption radius $R_{abs}$.
Here, the results obtained with the full expression of $h(q)$, as given
in Appendix \ref{app:GaussBD}, practically coincide with the approximate expressions, Eq.(\ref{mono})
and Eq.(\ref{comp_fact}).
Guided by the total reaction cross section obtained numerically with the HIDEX code by the partial wave method ($\sigma_a\simeq 2.14\,b$), we adopt $R_{abs} =\sqrt(\sigma_a/\pi) \approx 8~fm$.
Correspondingly, the suppression factor is found to be
$f_{BD}(analytical)\vert_{\substack{E_x=0\\ \theta=0}}\simeq 8.14\cdot 10^{-4}$ , in good agreement with the HIDEX result, $f_{BD}(HIDEX)\vert_{\substack{E_x=0\\ \theta=0}}\simeq 8.35\cdot 10^{-4}$,
as it can be extracted from the ratio between DWBA and PWBA
calculations at zero angle, in Fig. \ref{fig:3}.
As already anticipated above, owing to the important effects associated with the imaginary
part of the optical potential, the black disk assumption represents quite well the distortion
effects predicted by the full DWBA calculations.

\begin{figure}
\begin{center}
\epsfig{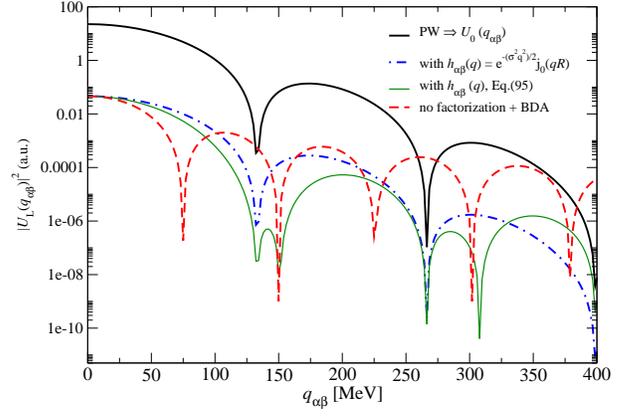}
\caption{(Color online) Square modulus of reaction kernel monopole component as a function of $q_{\alpha\beta}$, in plane wave (PW), full BD approximation and adopting the separation ansatz (for different choices
of $h_{\alpha\beta}(q)$, see text).}
\label{fig:9}
\end{center}
\end{figure}

To discuss the validity of the separation ansatz at finite momentum transfer,
we represent in Fig. \ref{fig:9}
the square modulus of the monopole component of the reaction amplitude $M_{\alpha\beta}$, evaluated considering the Gaussian fit of the form factors, 
as a function of $q_{\alpha\beta}$.  We note that this quantity
is closely linked to the reaction cross section.
Results have been  obtained in the full BD approximation, 
Eq.(\ref{M_1}), or adopting Eq.(\ref{eq:Mab_sep}), with several possibilities for the separation ansatz, Eqs.(\ref{mono}),(\ref{comp_fact}).
The square modulus of the Born reaction
amplitude is also represented in the figure (black line).
One can note that the black and red lines exhibit
interesting similarities with the results, presented in Fig.\ref{fig:3}, for PWBA and DWBA calculations, respectively.  This confirms
again  that the black disk approximation is indeed an appropriate
way to describe the absorption effects given by the full DWBA calculations.
Comparing black and red lines, one also observes that the
scaling factor generally depends on $q_{\alpha\beta}$, so that the full
separation
ansatz Eq.(\ref{comp_fact}) (blue line) can work well only up to $q_{\alpha\beta} \approx$ 50~MeV/c.
However, the green curve, corresponding to the
partial separation ansatz of Eq.(\ref{mono}), looks closer to the full BD results also at larger $q_{\alpha\beta}$ values.

The results discussed here for $E_x$ = 0  essentially depend on the momentum transfer, so they can be easily extended to  transitions leading to other excited states.
We conclude that the full cross section factorization is generally
valid for small momentum transfer, i.e. in the case of low-energy excitations and forward
angles.   Under these conditions, it is possible to isolate, in the total reaction
amplitude, the contribution of the Born amplitude, as done in
Eq.(\ref{eq:Mab_sep}).
This is particularly important because it would allow one
to access direct information on the nuclear transition densities,
which are linked, in turn, to $\beta$-decay strengths, as discussed in the following  section.

\subsection{Unit Cross Section and $\beta$-Decay Strengths}\label{F_section}
In the Born approximation, the reaction cross section, Eq. (\ref{eq:xsec_gen}), is simply
given by the product of a kinematical factor and the square modulus of the
reaction amplitude $\mathcal{U}_{\alpha,\beta}$ in Eq.(\ref{eq:TransPot}).

\vskip 0.5cm

As shown in the previous section, the distortion effects obtained in DWBA can be accounted for,
at small momentum transfer, by means of the scaling function:
$f_{BD}(R_{abs},R,\sigma) = |1-n_{\alpha\beta}|^2$.

Let us keep considering only L = 0 transitions, for both projectile and target,
and only the central part of the nuclear interaction. Then, the SCE cross section, for small momentum transfer, can be recast in the form
(see also Eqs.(\ref{cross_1}),(\ref{cross_2})):
\be
\begin{split}
&d\sigma_{\alpha\beta}=
K_f(T_{lab},\omega) (2S + 1) |V^{(C)}_{ST}(0)|^2
\left|b^{(ab)}_{0SS}\right|^2  \left|b^{(AB)}_{0SS}\right|^2 \\
&exp[-{1\over 3} q_{\alpha\beta}^2(<r^2>_a + <r^2>_A)]f_{BD}(R_{abs},R,\sigma)\\
\end{split}
\ee
where also the low-momentum expansion of the Bessel function 
in Eq.(\ref{eq:fredXY}) has been considered:
$j_0(q_{\alpha\beta}r)\approx 1 - 1/6~(q_{\alpha\beta}r)^2 \approx
exp(-1/6 ~(q_{\alpha\beta}r)^2)$.
Thus, in the above equation,
$<r^2>_a$ and  $<r^2>_A$ denote the mean square radius of proton
and neutron transition densities, respectively.
The kinematical factor $K_f(T_{lab},\omega)$ is given by:
\be
K_f(T_{lab},\omega) = \frac{m_\alpha m_\beta}{(2\pi\hbar^2)^2}\frac{k_\beta}{k_\alpha}.
\ee
It essentially depends on the energy loss
$\omega=E_{tot}-\left(M_A+M_a-M_B-M_b\right)=E_{tot}-Q_{g.s.}$,  
where $E_{tot} = E_x^A + E_x^a$  is the total excitation energy.

The cross section can be rewritten as:
\be\label{sigma_analyt}
d\sigma_{\alpha\beta}=
F(q_{\alpha\beta},\omega) \sigma_U  \left|b^{(ab)}_{0SS}\right|^2  \left|b^{(AB)}_{0SS}\right|^2
\ee
where we define a ``unit'' cross section, in analogy with what is usually done for SCE reactions involving light
projectiles \cite{Taddeucci:1987}, as:
\be
\sigma_U = K_f(T_{lab},0) |V^{(C)}_{ST}(0)|^2 f_{BD}(R_{abs},R,\sigma)
\ee
The function $F$, mainly determining the shape of the cross section,  is given by:
\be\label{F_shape}
F(q_{\alpha\beta},\omega) ={\frac{K_f(T_{lab},\omega)}{K_f(T_{lab},0)}}
exp[-{1\over 3} q_{\alpha\beta}^2(<r^2>_a + <r^2>_A)]
\ee
We note that the two equations above retrace the formalism developed in
Ref.\cite{Taddeucci:1987}.
From Eq.(\ref{F_shape}), it follows that $F(q_{\alpha\beta},\omega)\to 1$ for $(q_{\alpha\beta},\omega)\to (0,0)$, so that the proportionality coefficient between the SCE cross section and the product of the beta decay strengths relative to projectile and target reduces to $\sigma_U$. In the plane wave limit $\sigma_U$ becomes
\be
\sigma_U = K_f(T_{lab},0) |V^{(C)}_{ST}(0)|^2
\ee
so that it is characterized by a weak mass dependence \cite{Taddeucci:1987}.
On the other hand, the distortion factor $f_{BD}$ may vary significantly with the system mass. 

\section{Relation to Eikonal Theory}\label{sec:Eikonal}

\subsection{Kinematical Conditions}
As discussed in Appendix \ref{app:EDC}, the kinematical conditions of the reactions considered here are supporting in fact a description by eikonal theory. At first sight, this might be unexpected and surprising because eikonal theory is thought to be suited best for reactions at energies comparable to or exceeding the rest mass of the projectile. However, what really counts is not the energy but the wave length $\lambda_\alpha = 1/k_\alpha$ \cite{Joachain:1984,Lenske:2005nt}: A description of a reaction by eikonal theory becomes physically meaningful if $\lambda_\alpha$ is much shorter than the size of the interaction zone. In our case the scale is defined by the potential radius $R_{opt}\sim R_W$.  In other words, the decisive figure is the relation $\xi =k_\alpha R_W \gg 1$ which in our case is well fulfilled with $k_\alpha \sim 11fm^{-1}$ and $R_W\sim 3.5$~fm leading to $\xi \sim\mathcal{O}(40)$. Thus, eikonal theory will be a useful tool at least for qualitative investigations of heavy ion reactions like the present one 
\footnote{For proton- and $^3He$-induced reactions on $^{40}Ca$, the same conditions would be obtained only for incident energies $T^{(p)}_{Lab}\sim 1500$~MeV and $T^{(He)}_{Lab}\sim 900$~MeV!}. By Taddeucci et al. \cite{Taddeucci:1987} elements of eikonal theory have been applied to light ion-induced charge exchange reactions. However, for heavy ion reactions distortion and absorption effects have to be considered in more detail because of their strong influence on the selectivity of reaction channels and the magnitude of cross sections.

\begin{figure}
\begin{center}
\includegraphics[width = 6.5cm]{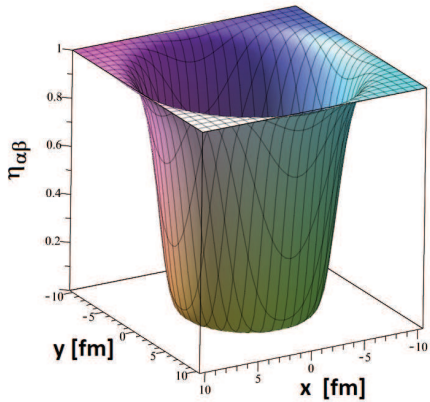}
\caption{The distorted wave density $\eta_{\alpha\alpha}$, Eq.(\ref{eq:DisAmp}) shown in the $(x,y)$-plane. The imaginary part of the double folding potential in Fig. \ref{fig:OMP} was parameterized in terms of a Gaussian form factor (see text) allowing to evaluate the distortion coefficient in closed form according to Appendix \ref{app:EDC}. The profile of the distribution resembles closely a step function in three dimensions.}.
\label{fig:DisAmp}
\end{center}
\end{figure}

In this section, we use eikonal theory to investigate the evolution of heavy ion charge exchange reactions with mass and incident energy. The primary goal is to understand the dependencies of the SCE cross sections on these external, physical parameters. Not to the least, this may serve to encircle favorable reaction scenarios on projectile-target combinations and energies. For the sake of analytical results, we continue to use the Gaussian approximation for nuclear transition form factors and the effective transition potentials. As shown in Appendix 
\ref{app:EDC}, the resulting Gauss-Eikonal-Approach (GEA) leads to analytical results for the quantities of interest.

\subsection{Mass and Energy Dependence of Absorption Effects}\label{ssec:MassEnergy}

An important conclusion from the foregoing discussion is the paramount role of absorption effects for which the absorption radius $R_{abs}$  is the key quantity. Moreover, according to Appendix \ref{app:EDC}, in the strong absorption limit the distortion coefficient is fixed once $R_{abs}$ is known together with the nuclear shape parameters. For our purpose, it is enough to consider the imaginary part $W(\mathbf{r})$. In the present context, $W(\mathbf{r})$ plays the role of an effective Eikonal-potential which has to be adjusted such that the quantal results are reproduced as close as possible. A spherical-symmetric potential of Gaussian shape is used
\be
W(r)=-W_0e^{-r^2/R^2_W}.
\ee
The radius parameter $R_W$ and the potential strength are fixed by comparison to quantum mechanical results, as given by the HIDEX code,  
for the two systems $^{18}O+{}^{40}Ca$ and $^{18}O+{}^{116}Sn$ \cite{Cappuzzello:2016mxt}, such that the total reaction cross sections 
are reproduced. Denoting the mass numbers of projectile and target by $A_{p,T}$, a proper description of the two systems is obtained with $R_W=r_0\sqrt{A^\frac{2}{3}_p + A^\frac{2}{3}_T}$, where $r_0=0.783$~fm, and
\be
W_0=w\left(A^\frac{2}{3}_p + A^\frac{2}{3}_T\right)^{-3/4}
\ee
with $w=5902.743$~MeV. Interestingly, the potential depth behaves according to the so-called $UR^\alpha$-law which was found in the early days of the nuclear optical model by Hodgson \cite{Hodgson:1962,Hodgson:1978} when studying ambiguities of optical potentials. In our case we have $\alpha=3/2$. 
{We also note that the eikonal approximation works rather well for shallow optical potentials, as given by our parametrization.}
For the system $^{18}O+{}^{40}Ca$ we find $R_W\simeq 3.375$~fm and strength $W_0\simeq 660$~MeV, resulting in $\sigma^{(\alpha)}_{abs}\simeq 2.14$~b and $R_{abs}\simeq 8.26$~fm. The transition potential is described by a surface-centered Gaussian,
\be
U_G(\mathbf{r},\mathbf{R}_G)\sim e^{-(\mathbf{r}-\mathbf{R}_G)^2/2\sigma^2_G}
\ee
with $R_G=r_G\sqrt{A^{\frac{2}{3}}_p+A^{\frac{2}{3}}_T}$, $r_G=1.2$~fm. 
The width parameter $\sigma_G\approx 1$~fm  
corresponds to the width obtained  by folding two Gaussian nuclear transition form factors 
with $\sigma_{tr}\sim 0.7$~fm.  

Thus, we have at hand all quantities necessary to evaluate by the formalism of Appendix \ref{app:EDC} the distortion amplitude $\eta_{\alpha\beta}$ and the total absorption cross section $\sigma^{(\alpha,\beta)}_{abs}$  as functions of mass and energy. 
{Then, from the absorption radius, we derive, within the black disk approximation,  
the distortion coefficient $n_{BD}$ and the absorption factor $f_{BD}=|1-n_{BD}|^2$.} 

In Fig. \ref{fig:DisAmp}, the (diagonal) distortion density $\eta_{\alpha\alpha}$ is displayed for the system $^{18}O+{}^{40}Ca$ at $T_{Lab}=270$~MeV. The strong suppression in the interaction zone resembles indeed a spherical symmetric Heaviside distribution in three dimensions $\Theta(R^2_{abs}-\mathbf{r}^2)$, thus confirming our previous conjecture. At the edges, a diffuse smoothing is found, which, however, will not affect the leading order behaviour and, in particular, leaves the overall conclusions unaltered.

\begin{figure}
\begin{center}
\epsfig{file=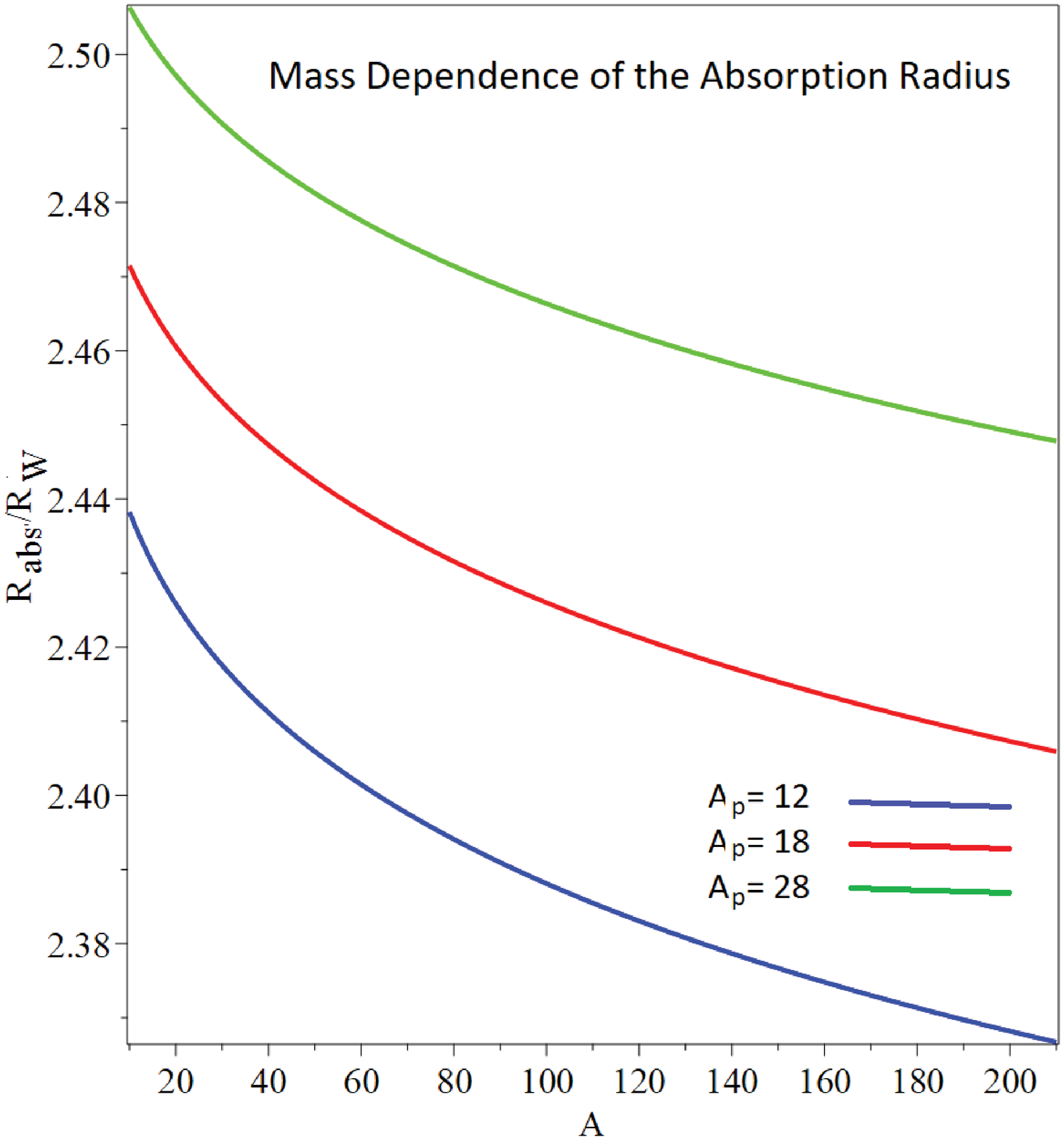, width=6.5cm}
\epsfig{file=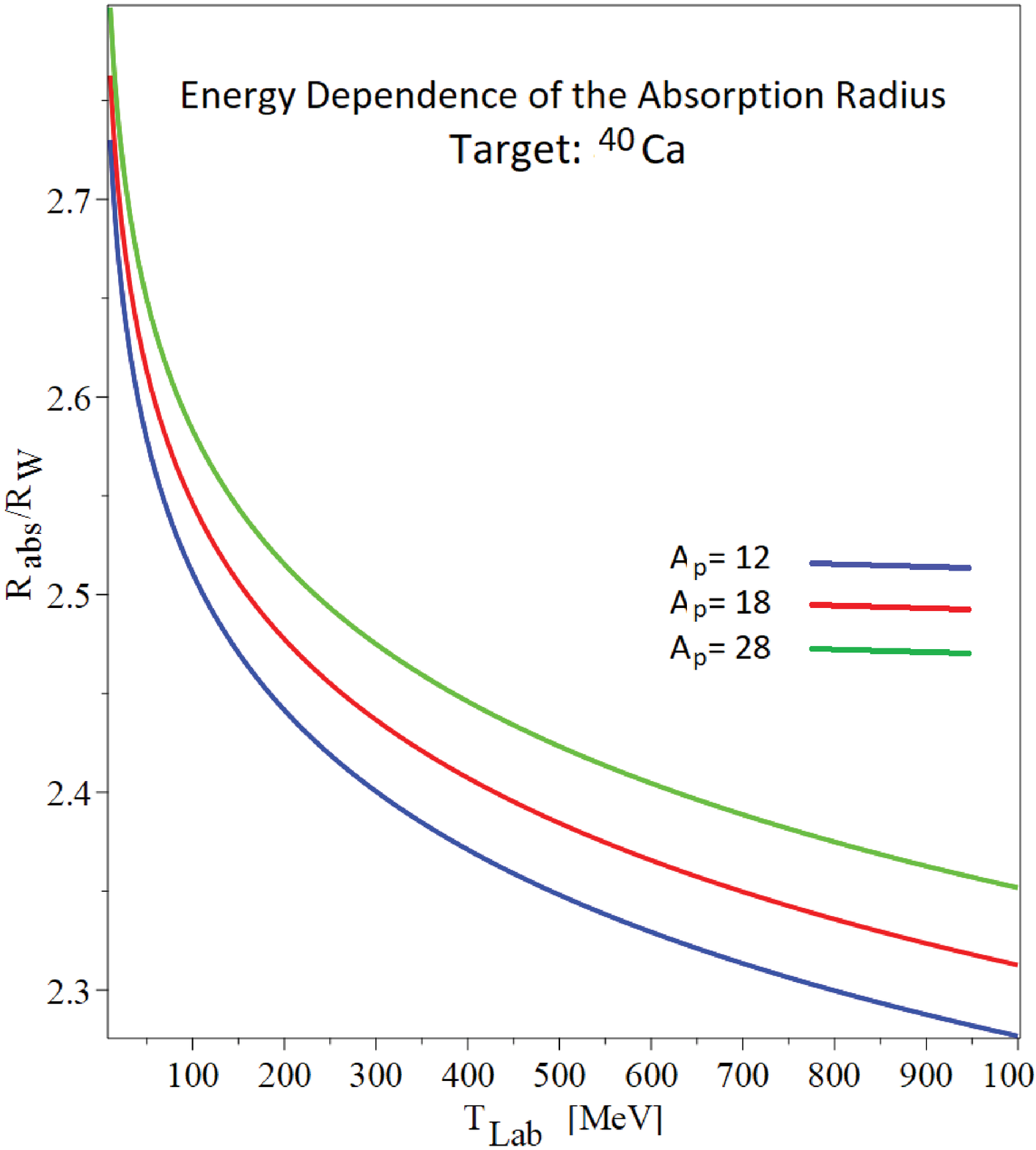, width=6.5cm}
\caption{(Color online) Upper panel: Variation  of the effective absorption radius $R_{abs}$ with projectile and target mass. GEA results are shown for the projectiles $^{12}C$ (lower curve, blue), $^{18}O$ (center curve, red), and $^{28}Si$ (upper curve, green), respectively, on targets with mass numbers $10\leq A_T \leq 210$. Lower panel: Variation of the effective absorption radius $R_{abs}$ with incident energy. GEA results are shown for reactions on $^{40}Ca$ with the projectiles $^{12}C$ (lower curve, blue), $^{18}O$ (center curve, red), and $^{28}Si$ (upper curve, green), respectively.}
\label{fig:RoRw}
\end{center}
\end{figure}

The dependence of $R_{abs}(A_P,A_T,T_{Lab})$ on the ion masses and the incident energy is illustrated in Fig. \ref{fig:RoRw}. The variation of the ratio $R_{abs}/R_W$ on the target mass number $A_T$ is displayed for three choices of projectiles, namely $^{12}C$, $^{18}O$, and $^{28}Si$ at fixed energy $T_{Lab}=270$~MeV. The ratio decrease mildly by a few percent with increasing $A_T$, implying a $A^{1/3}$-dependence for $R_{abs}$. A slight increase with $A_P$ is found, reflecting the slight increase of the strength of the absorptive potential with $A_p$.

In the lower panel of Fig.\ref{fig:RoRw}, the dependence of $R_{abs}/R_W$ on the incident energy is shown. The target is fixed to $^{40}Ca$. Here, one finds a behaviour similar to the mass-dependence: The absorption radii decrease continuously with increasing incident energy. From Eq.(\ref{eq:fshape}) and Eq.(\ref{eq:argShape}) one finds for small energies a logarithmically divergent dependence on $T_{Lab}$  which for large energies changes to a dependence on $1/k_\alpha\sim 1/\sqrt{T_{lab}}$.

\begin{figure}
\begin{center}
\epsfig{file=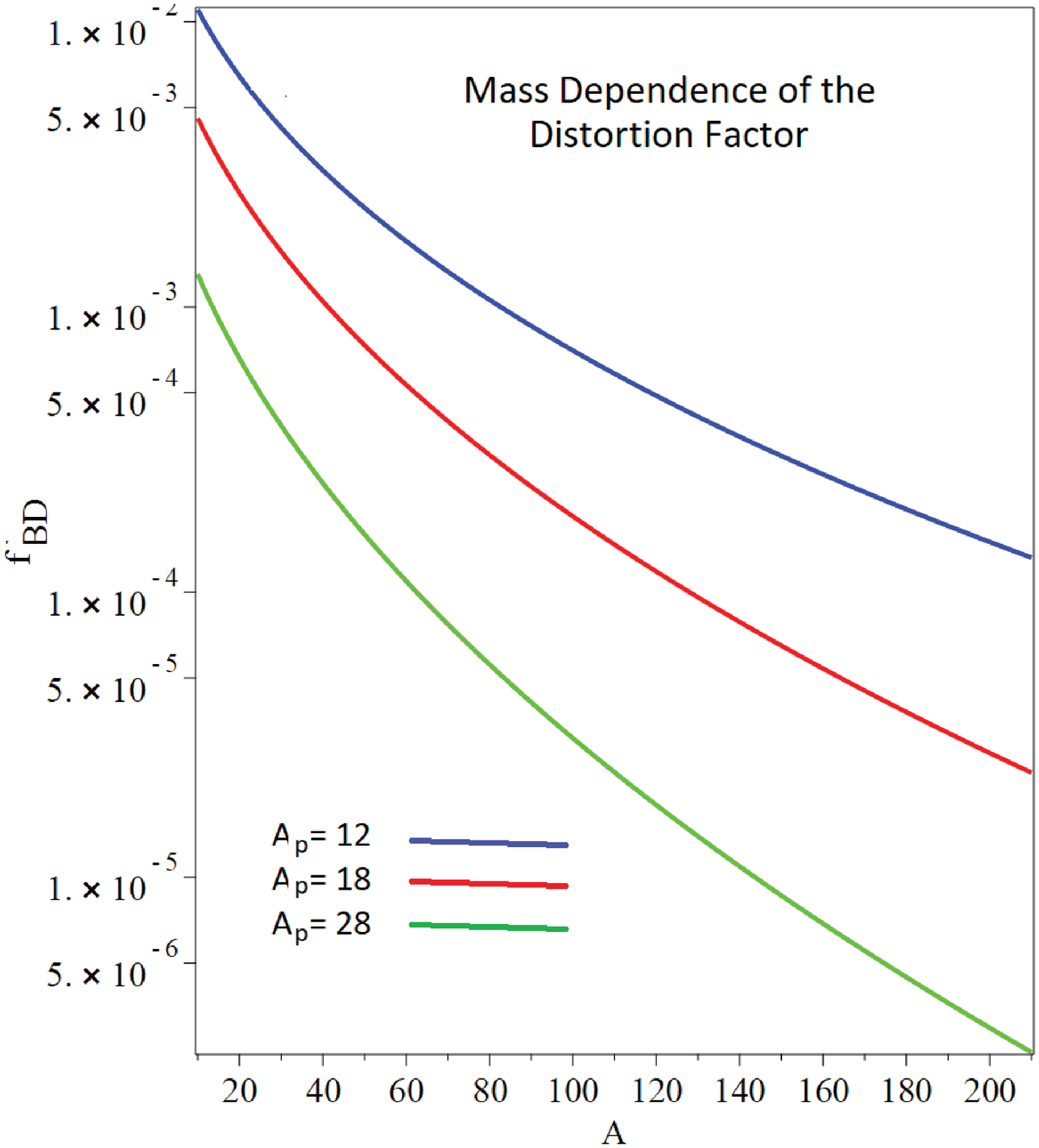, width=6.5cm}
\epsfig{file=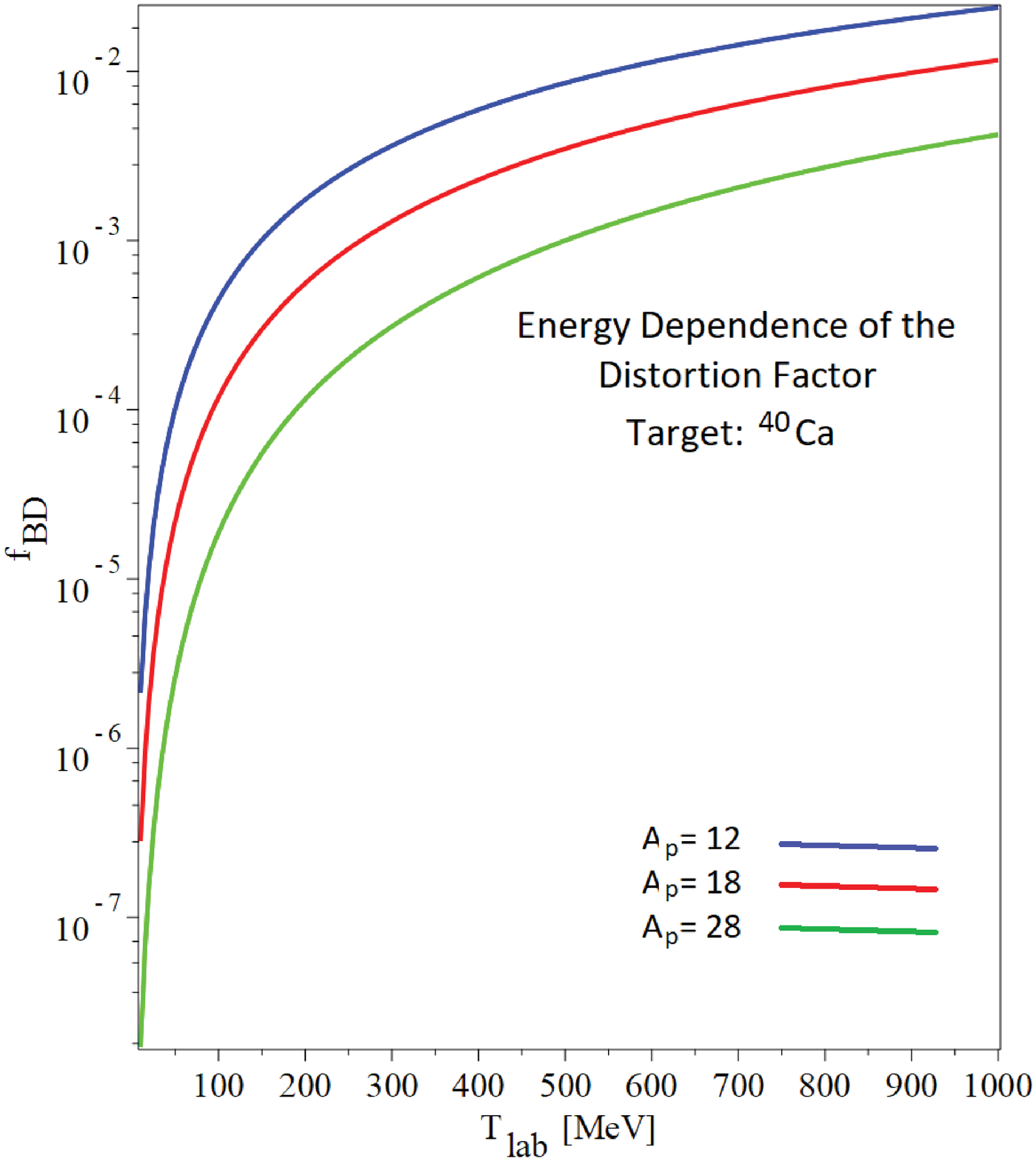, width=6.0cm}
\caption{(Color online) Upper panel: Variation  of the distortion factor $f_{BD}$ with projectile and target mass. GEA results are shown for the projectiles $^{12}C$ (lower curve, blue), $^{18}O$ (center curve, red), and $^{28}Si$ (upper curve, green), respectively, on targets with mass numbers $10\leq A_T \leq 210$. Lower panel: Variation of the distortion factor $f_{BD}$ with incident energy. GEA results are shown for reactions on $^{40}Ca$ with the projectiles $^{12}C$ (upper curve, blue), $^{18}O$ (center curve, red), and $^{28}Si$ (lower curve, green), respectively.}
\label{fig:fBD_MassEnergy}
\end{center}
\end{figure}

The mass and energy dependence of the absorption factor $f_{BD}$ is explored in Fig.\ref{fig:fBD_MassEnergy}. 
Over the shown mass range, a decrease by several 
orders of magnitude is found, indicating the smallness of cross sections to be expected for heavy targets and increasing projectile mass. The results indicate, on the other hand, that lighter projectiles are leading to a less extreme suppression.

As a function of energy, $f_{BD}$ increase rapidly with $T_{Lab}$ as seen in the lower panel of Fig.\ref{fig:fBD_MassEnergy}. Thus, combining these results with those on the mass dependence, we conclude that already a moderate increase of the incident energy will lead to considerably larger cross sections also for heavier target-projectile combinations.

\section{Summary and Outlook}\label{sec:SumLook}
Heavy ion reactions are of wide interest by their own because of the possibility to explore several excitation mechanisms by the same experiment. Charge changing reactions, in particular, open unprecedented perspectives for detailed nuclear structure investigations of the many-body dynamics underlying also $\beta$-decay processes. In this paper, we have presented a revised approach to the theoretical modeling of nuclear SCE reactions. In a strictly microscopic picture we have reformulated the reaction dynamical aspects in the framework of DWBA theory. Central and rank-2 tensor interactions were considered. In momentum representation the reaction amplitude was separated formally into projectile and target transition form factors and the distortion coefficient, accounting for ISI and FSI ion-ion interactions.

HFB theory is used to describe the projectile and target ground states. The charge changing nuclear excitations were described by correlated 2QP configurations including residual quasiparticle interactions. Effects beyond mean-field dynamics were treated by introducing dynamical self-energies of a universal character. Thus, an extended QRPA approach was obtained. The QRPA problem as solved by direct solution of the  Dyson-equation  which is closely connected to the polarization propagators formalism. Nuclear response functions were introduced as the appropriate method for investigating charge changing external fields which in the present context are provided by the isovector NN projectile-target interactions.

The nuclear structure approach has been illustrated in calculations for charge-changing excitations off $^{18}O$ and $^{40}Ca$. Response functions for multipole operators, intimately connected to weak interactions at low momentum transfer, have been discussed. As illustrative -- and typical -- examples results for both $\tau_\pm$ branches have been presented. The $^{18}N$ and the $^{18}F$ spectra, respectively, could be reproduced satisfactorily well. An even better agreement with spectral data was obtained for the heavier systems $^{40}K$ and $^{40}Sc$.

Our previous investigations of heavy ion SCE reactions have shown that fully quantum mechanical DWBA calculations with microscopic nuclear structure input describe observed cross sections quantitatively. Thus, from the theoretical side we have a powerful and successful toolbox at hand. However, for the direct extraction of spectroscopic quantities from measured cross sections, a separation of reaction and nuclear dynamics contributions is of great advantage. In this respect, a central goal of our investigation was to explore in detail the interplay of reaction and nuclear structure aspects in heavy ion SCE cross section, aiming at identifying the conditions under which the two parts factorise, thus giving access to nuclear matrix elements relevant for $\beta$-decay processes. We note that this point has been widely investigated for reactions involving light projectiles (such as protons or $^{3}$He) at energies of a few hundred MeV per nucleon, found to be a quite useful tool to extract direct information on the $\beta$-decay strength of the target. Here, we could show that heavy ion reactions of a few tens of MeV per particle are in principle as well suited for such explorations.

Performing explicit calculations for the reaction $^{18}O+{}^{40}Ca$ at $T_{lab}= 15$~AMeV, we have shown that in heavy ion reactions the distortion effects are strongly amplified, where the imaginary part of the nuclear optical potential is playing the major role. Indeed, SCE cross sections obtained with only the imaginary potential (i.e. neglecting real part and Coulomb interaction) in the initial and final state elastic ion-ion interactions coincide almost perfectly with results by full DWBA calculations. Thus, the distortion effects are in fact mainly absorption effects which are well reproduced in the strong absorption limit by the black disk approximation. Within such a scheme, we have worked out a factorization of the reaction cross section which is well suited for reactions characterized by small momentum transfer.

The investigations have shown that heavy ion SCE reactions indeed allow to extract from the forward angle cross section, i.e. at small momentum transfer, a direct information on the product of the $\beta$ decay strengths in projectile and target. Hence, also in the case of heavy ion reactions, we are led to define a ``unit'' cross section,  which allows to relate the SCE differential cross section to the $\beta$-decay strengths. For a given projectile, calculations performed in the black disk approximation suggest a $1/A$ dependence of the distortion coefficient on the target mass. The dominant role played by ISI and FSI demands for studies of elastic scattering. Angular distribution data in at least one of the involved channels would be of high importance as a counter-check the accuracy of the microscopically derived ion-ion potentials which are central parts of the presented approach.

Eikonal theory provides an elegant approach to investigate universal aspects of the mass and energy dependence of distortion effects. This kind of predictions are of interest especially for estimates of yields to be expected in future experiments. The intentions of the present eikonal studies is to encircle global trends and variations of cross sections. We emphasize agian that for a quantitative analysis of a given reaction a fully quantum mechanical calculation as indicated will be the preferred method. Here, we have used physically meaningful but schematic descriptions for the mass and energy dependence of the input quantities, as there are optical potentials and transition potnetials. Both quantities were approximated by Gaussian form factors.

Within the Gauss-Eikonal-Approximation the mass and energy dependence of the distortion factor was investigated in the strong absorption limit. An attractive feature of GEA approach is that it allows to connect the phenomenological concept of black disk (or black sphere) scattering to the microscopic background. The results for the absorption factor are clearly indicating certain constraints on heavy ion SCE reactions: The magnitude of the cross sections will depend critically on the projectile-target combination. At fixed energy, systems with low total mass are favored, meaning that e.g. an increase of the target mass will result in a steeply decreasing cross section. The same is true for a variation of the projectile mass. However, this mass effect can be compensated to a large extent by varying the energy because $f_BD$ increases rapidly with incident energy.

Broad space was given to a formulation covering reaction and nuclear structure aspects on the same footing. By practical considerations, the main weight was laid on reactions at energies well above the Coulomb barrier. Such reactions are of high interest for currently active experiments, e.g. within the NUMEN project at LNS Catania \cite{Cappuzzello:2015ixp}. The theoretical results, however, apply to other choices of projectile-target combinations and energy as well.These developments of the theory of heavy ion SCE reactions open new interesting perspectives for studies of double charge exchange (DCE) reactions. The extension to heavy ion DCE reactions will be the topic of a forthcoming paper. In fact, with appropriate extensions the methodology developed in this work is a suitably entry point for investigations of second order processes as the heavy ion DCE reactions, allowing to establish their relation to double $\beta$ decay processes.

\subsection*{Acknowledgement}
Part of this work was done during visits of H.L. at LNS Catania. The warm hospitality of the NUMEN group and the financial support by the NUMEN project are gratefully acknowledged. 
We wish to thank F.Cappuzzello and M.Cavallaro for their help with the use of the HIDEX code and for fruitful discussions.  
This project has received funding from the European Union's Horizon 2020 research and innovation programme under grant agreement N. 654002, and from the Spanish Ministerio de Economia y Competitividad and FEDER funds under Project FIS2017-88410-P.

\appendix
\section{Angular Momentum Decomposition of the Reaction Kernel}\label{app:Irreps}
The decomposition of the full reaction kernel and correspondingly the reaction amplitude relies on their decomposition into irreducible tensorial components. For that purpose, the product of the projectile and target form factors, Eq.(\ref{eq:Fproj}) and Eq.(\ref{eq:Ftarg}), must be recoupled to total angular momentum. We use the addition theorem of spherical harmonics \cite{Edmonds:1957}
\bea\label{eq:AddTheor}
&&i^{L_1}Y_{L_1m_1}(\mathbf{\hat{p}})i^{L_2}Y_{L_2m_2}(\mathbf{\hat{p}})= \nonumber \\
&&\sum_{LM_L}(-)^{\frac{L_1+L_2-L}{2}}\frac{\hat{L}_1\hat{L}_2}{\sqrt{4\pi}\hat{L}}\left(L_10L_20|L0 \right) \nonumber \\
&&\left(L_1m_2L_2m_2|LM \right)i^LY_{LM}(\mathbf{\hat{p}}).
\eea
Then, for a central interaction the product of nuclear form factors is obtained as
\bea\label{eq:Multi_C}
&&F^{(\alpha\beta)}_{ST}(\mathbf{p})=
\sum_{J_1,M_1,J_2,M_2,L,M}\left(J_aM_aJ_bM_b|J_1M_1 \right)\nonumber \\
&&\left(J_AM_AJ_BM_B|J_2M_2 \right)\left(J_1M_1J_2M_2|LM \right)\nonumber \\
&&i^LY_{LM}(\bm{\hat{p}})F^{J_1J_2}_{LS}(p^2),
\eea
with the reduced multipole form factors
\be\label{eq:Irep_C}
F^{J_1J_2}_{LS}(p^2)=\sum_{L_1L_2}{A_{LS}(L_1L_2,J_1J_2)f^{(ab)}_{L_1SJ_1}(p^2)f^{(AB)}_{L_2SJ_2}(p^2)}.
\ee
We have introduced the recoupling coefficients
\bea
&&A_{LS}(L_1L_2,J_1J_2)= \nonumber \\
&&(-)^{\frac{L_1+L_2-L}{2}}\frac{\hat{L}_1\hat{L}_2}{\sqrt{4\pi}\hat{L}}\left(L_10L_20|L0 \right)\nonumber \\
&&(-)^{L_2+J_1-L}W\left(L_1J_1L_2J_2;LS \right)\hat{J}_1\hat{J}_2
\eea
where $W(abcd;ef)$ is a Racah-coefficient \cite{Edmonds:1957}.

The rank-2 tensor component requires additional recoupling of the spin and orbital angular momentum operators in order to comply with quadrupole character of the vertex. The resulting form factor can be cast into from similar to Eq.(\ref{eq:Multi_C}):
\bea\label{eq:Multi_Tn}
&&H^{(\alpha\beta)}_{ST}(\mathbf{p})=\sum_{J_1,M_1,J_2,M_2,L,M}\left(J_aM_aJ_bM_b|J_1M_1 \right)\nonumber \\
&&\left(J_AM_AJ_BM_B|J_2M_2 \right)\left(J_1M_1J_2M_2|LM \right)\nonumber \\
&&i^LY_{LM}(\mathbf{\hat{p}})H^{J_1J_2}_{L1}(p^2)
\eea
The reduced form factors, however, is of a somewhat more involved structure
\bea
&&H^{J_1J_2}_{L1}(p^2)=\sum_{L_1,L_2,L'}B_{LL'}(L_1L_2,J_1J_2)\nonumber \\
&&f^{(ab)}_{L_11J_1}(p^2)f^{(AB)}_{L_21J_2}(p^2).
\eea
In this case, the recoupling coefficient is given by
\bea
&&B_{LL'}(L_1L_2,J_1J_2)=\sqrt{\frac{24\pi}{5}} \nonumber \\
&&(-)^{\frac{L_1+L_2-L}{2}}\frac{\hat{L}_1\hat{L}_2\sqrt{5}}{4\pi\hat{L}}
\left(L_10L_20|L'0 \right)\left(L'020|L0 \right)\nonumber \\
&&\hat{J}_1\hat{J}_2\hat{L}'\sqrt{5}
\left\{ {\begin{array}{*{20}{c}}
{{L_1}}&1&{{J_1}}\\
{{L_2}}&1&{{J_2}}\\
{L'}&2&L
\end{array}} \right\}
\eea
where the object in the last line is a 9-j symbol \cite{Edmonds:1957}.

\section{Gaussian Form Factors and Microscopic Nuclear Structure}\label{app:GFF}
The price paid for the advantage of the Gaussian approximation that the dependencies on the ion masses and sizes are directly accessible by closed form expressions is that the connection to microscopic nuclear structure seems to be lost. However, by re-interpretation of the parametrical dependence on the yet to be specified radius $R$, that connection can be restored under certain constraints. In leading order the transition potential Eq.(\ref{eq:FF_SCE}) is given by replacing the NN T-matrix by a contact interaction where the strength if given by the momentum space amplitude at $p=0$, i.e. the volume integral. Denoting the intrinsic projectile and target coordinates by $\mathbf{r}_{1,2}$, respectively, and the ion-ion relative coordinate by $\mathbf{r}$, the zero-range assumption implies $\mathbf{r}_1+\mathbf{r}-\mathbf{r}_2=0$. For a contact interaction, the folding integral defining the transition potential reduces to the folding of the nuclear transition form factors. For that purpose we assume that the intrinsic nuclear transitions $a\to b$ and $A\to B$ are described by Gaussian form factors
\be
F_{N}(\mathbf{r}_N)=C_N e^{-\frac{1}{2\sigma^2_N}|\mathbf{r}_N-\mathbf{R}_N|^2}
\ee
where $N\in \{ab,AB\}$ ($N = 1,2$) and the normalization constant is chosen as $C_N=(\sqrt{2\pi}\sigma_N)^{-3}$ such that $F_N$ has a volume integral equal to unity. In coordinate space have to evaluate a folding integral of the type
\be
F_{12}(\mathbf{r})=C_1C_2\int{d^3r_1 e^{-\frac{1}{2\sigma^2_1}\left(\mathbf{r}_1-\mathbf{R}_1 \right)^2} e^{-\frac{1}{2\sigma^2_2}\left(\mathbf{r}_1+\mathbf{r}-\mathbf{R}_2 \right)^2} }.
\ee
With the substitutions $\mathbf{x}=\mathbf{r}_1-\mathbf{R}_1$ and $\bm{\rho}=\mathbf{r}+\mathbf{R}_1-\mathbf{R}_2$ the integral becomes
\be
F_{12}(\mathbf{r})=C_1C_2\int{d^3x e^{-\frac{1}{2\sigma^2_1}x^2} e^{-\frac{1}{2\sigma^2_2}\left(\mathbf{x}+\bm{\rho} \right)^2} }.
\ee
The angle integrations lead to modified Bessel function of order $n=0$. The remaining integration can be performed in closed form with the final result
\be
F_{12}(\mathbf{r},\mathbf{R})=\frac{1}{(\sqrt{2\pi}\sigma)^3}e^{-\frac{1}{2\sigma^2}\left(\mathbf{r}-\mathbf{R}\right)^2}
\ee
The width is given by
\be
\sigma^2=\sigma^2_{1}+\sigma^2_{2},
\ee
and the centroid radius is found as
\be
\mathbf{R}=\mathbf{R}_{1}-\mathbf{R}_{2},
\ee
which plays the role of  a scale-defining quantity. Considered as classical mathematical objects, the vectors $\mathbf{R}_{1,2}$ are free parameters reflecting the nuclear scales. Thus, we use $R_{1,2}\sim A^\frac{1}{3}_{1,2}$. Since the relative orientation of the two centroid vectors is arbitrary, we use the averaging, resulting in $R^2=R^2_1+R^2_2\simeq A^{\frac{2}{3}}_1+A^{\frac{2}{3}}_2$.

Within the above \emph{zero-range} approximation, the transition potential, Eq.(\ref{eq:UGauss}) is given by the superposition of a spin-scalar (S=0) and a spin-vector (S=1) component
\be
U_G(\mathbf{r})=\sum_{S=0,1,T=1}{I_{ST}B^{(ab,AB)}_{ST}F^{(ab,AB)}_{12,ST}(\mathbf{r},\mathbf{R}_{ST})}
\ee
where $I_{ST}\equiv V^{(C)}_{ST}(p=0)$ denotes the volume integral of the interaction. The crucial point is how to incorporate the underlying microscopic nuclear structure physics. The simplest, but rather schematic approach is to use projectile and target spectral distributions averaged over multipolarities. Such a solution is indicated above: The nuclear charge-changing spectral transition strengths for projectile and target are contained in $B^{ab,AB}_{ST}$, obtained e.g. by the response function formalism, section \ref{sec:CC_Spectra}.

In a refined approach the multipole structure of the form factors and spectral distributions should be combined explicitly. On a formal level, this is achieved by identifying $Y_{LM}(\hat{\mathbf{R}})$ as a dynamical quantity
with an operator structure inducing intrinsic nuclear transitions. Formally, this is achieved by imposing the quantization conditions
\be
Y_{LM}(\hat{\mathbf{R}})\to \mathcal{Y}_{LM}(\Omega^\dag)=\sum_{\lambda \kappa}{b_\lambda(ab)b_\kappa(AB)\left[\Omega^\dag_\lambda(ab) \Omega^\dag_\kappa(AB)\right]_{LM} }
\ee
i.e. a representation by the state operators $\Omega^\dag_{\lambda,\kappa}$ of projectile and target, respectively. $\lambda$ and $\kappa$ include spin and orbital angular momenta. The coupling to good total angular momentum transfer $L$ is indicated. The expansion coefficients are given by nuclear multipole transition amplitudes. Thus, we have obtained a relation similar to the collective model approach to nuclear spectroscopy of Bohr and Mottelsen \cite{BM:1969}, widely used in the past for nuclear reactions. Thus, the essence of the Gaussian form factor is seen to separate the state dependent transition form factors into a state-independent spatial form factor $U_{LM}(\mathbf{r})$, Eq.(\ref{eq:ULM}), and state-dependent amplitudes $b_\mu$ giving rise to the multipole spectral distributions
\be
B^{(aA,bB)}_{L}=\lan bB|\mathcal{Y}_{LM}|aA\ran
\ee
and the multipole transition potentials
\be
U^{aA,bB}_{ST,LM}(\mathbf{r})=B^{(aA,bB)}_{L}I_{ST}U_{LM}(\mathbf{r})
\ee
where the spectral amplitudes and the reduced form factors will depend in general also on the spin transfer $S$.

\section{Distortion Coefficient in Eikonal Approximation}\label{app:EDC}

For wave lengths $\lambda\sim 1/k$ short against the scale $R_{pot}$ of the interaction zone, i.e. $\xi=k R_{pot}\gg 1$
semi-classical descriptions become an appropriate description for nuclear reactions. In the case considered here, we have $k_\alpha\simeq 11$~ fm$^{-1}$ and $R_{pot}\simeq 4$~fm (see Fig.\ref{fig:OMP}), leading to $\xi \simeq 40$. Thus, despite the rather low energy of $T_{lab}=270$~MeV the kinematical conditions allow to apply eikonal theory \cite{Joachain:1984}. Then, the distorted waves are given as
\bea
\chi^{(+)}_\alpha(\mathbf{k}_\alpha,\mathbf{r})&=&e^{iS^{(+)}_\alpha(\bm{\rho},z)}e^{+ i \mathbf{k}_\alpha\cdot \mathbf{r}}\\
\chi^{(-)*}_\beta(\mathbf{k}_\beta,\mathbf{r})&=&e^{iS^{(-)*}_\beta(\bm{\rho}',z')}e^{- i \mathbf{k}_\beta\cdot \mathbf{r}}
\eea
with the asymptotically in- and outgoing eikonals
\bea
S^{(+)}_\alpha(\bm{\rho},z)&=&\int^z_{-\infty}{d\zeta \left( Q_\alpha(\bm{\rho},\zeta)-k_\alpha \right)}\\
S^{(-)*}_\beta(\bm{\rho}',z')&=&\int^{z'}_{+\infty}{d\zeta \left(Q_\beta-k_\beta \right)}
\eea
with the local channel momenta
\be
Q_{\gamma}(\bm{\rho},\zeta)=\sqrt{k^2_\gamma-\frac{2m_\gamma}{\hbar^2}U_\gamma(\bm{\rho},\zeta)}
\ee
and
$\bm{\rho},z$ are oriented such $z$ coincides with the direction of $\mathbf{k}_\alpha$ and $\bm{\rho}',z'$ are taken accordingly with respect to $\mathbf{k}_\beta$.
Hence, we identify $u^{(\pm)}_{\alpha,\beta}=e^{iS^{(\pm)}_{\alpha,\beta}}$, leading to
\be
\eta_{\alpha\beta}=u^{(-)*}_{\beta}u^{(+)}_{\alpha}=e^{i\left(S^{(+)}_\alpha+S^{(-)*}_\beta\right)}= e^{i\phi_{\alpha\beta}-\kappa_{\alpha\beta}}
\ee
where the (real) phase shift $\phi_{\alpha\beta}$ and the attenuation exponent $\kappa_{\alpha\beta}$ are given by the sum of the real and imaginary parts of the eikonals, respectively. For small momentum and energy transfer, we may neglect the differences in the channel momenta and potentials. Under such conditions the distortion amplitude is given by the (diagonal) distortion phase shift and attenuation exponent
\bea
&&\phi_{\alpha\alpha}(\bm{\rho})\simeq \int^{+\infty}_{-\infty}{d\zeta Re\left(Q_\alpha(\rm{\rho},\zeta)\right)-k_\alpha }\\
&&\kappa_{\alpha\alpha}(\bm{\rho})\simeq \int^{+\infty}_{-\infty}{d\zeta Im\left(Q_\alpha(\rm{\rho},\zeta)\right)}
\eea
Neglecting terms equal to and higher than $\mathcal{O}(U_\alpha/k^2_\alpha)$, these expressions are given by
\bea
&&\phi_{\alpha\alpha}(\bm{\rho})\simeq -\frac{m_\alpha}{\hbar^2k_\alpha}\int^{+\infty}_{-\infty}{d\zeta Re\left(U_\alpha(\rm{\rho},\zeta)\right)}\\
&&\kappa_{\alpha\alpha}(\bm{\rho})\simeq \frac{m_\alpha}{\hbar^2k_\alpha}\int^{+\infty}_{-\infty}{d\zeta Im\left(U_\alpha(\rm{\rho},\zeta)\right)}.
\eea
For the present purpose it is sufficient to consider primarily the attenuation exponent. Assuming Gaussian form factors and spherical symmetry
\be
U_\alpha(r)=-U_0 e^{-r^2/R^2_U}-iW_0e^{-r^2/R^2_W}
\ee
with potential strengths $U_0>0$ and $W_0>0$, the leading order absorption exponent is obtained in closed form:
\be
\kappa_{\alpha\beta}(\rho)= \sqrt{\pi}\frac{m_\alpha W_0}{\hbar^2k_\alpha}R_We^{-\rho^2/R^2_W}
\ee
depending only on the modulus of $\bm{\rho}$. The resulting distortion density $\eta_{\alpha\alpha}(\rho)$ is displayed in Fig.\ref{fig:DisAmp}. In general, the phase shift $\phi_{\alpha\beta}$ should be considered as well but the modifications of the pure attenuation result will be small and not affecting the overall picture.

Within the Gaussian approximation the reaction cross section can be evaluated in closed form. As shown in \cite{Lenske:2005nt} the key point is to consider the continuity equation of the distorted waves from which one derives the relation
\bea
\sigma^{(\alpha)}_{abs}&=&\frac{2m_\alpha}{\hbar^2k_\alpha}\int{d^3r|\chi^{(+)}(\mathbf{k}_\alpha,\mathbf{r})|^2W(\mathbf{r})}\nonumber \\
&=&\frac{2m_\alpha}{\hbar^2k_\alpha}\int{d^3r e^{-2Im\left(S^{(+)}_\alpha(\bm{\rho},z) \right)} W(\mathbf{r})}.
\eea
For a Gaussian $W(r)$ the integration can be performed analytically. As anticipated before, the result may be expressed indeed in a form resembling in structure the black disk expression
\be
\sigma^{(\alpha)}_{abs}(\sqrt{s_\alpha})=\pi R^2_{abs}(\sqrt{s_\alpha}),
\ee
but where the effective absorption radius is related to the potential radius $R_W$ by the shape function $f(x)$
\be\label{eq:Rabs}
R^2_{abs}(\sqrt{s_\alpha})=f(\xi(W_\alpha,k_\alpha))R^2_W.
\ee
The shape function is given analytically by
\be\label{eq:fshape}
f(x)=\gamma+log(x)+Ei(1,x)
\ee
where $\gamma=0.5772...$ denotes Euler's constant and $Ei(1,x)$ is an exponential integral. $f(x)$ is increasing steadily with x, vanishing at $x=0$ and diverging logarithmically for $x\gg 1$.
The argument
\be\label{eq:argShape}
\xi(W,k)=\sqrt{\pi}k R_W \frac{W_0}{T_{cm}}
\ee
depends on the reduced mass, the energy, and the absorption potential.
$T_{cm}=(\hbar k)^2/2m$ is the kinetic energy in the rest frame and $kR_W\sim \ell_g$ corresponds to a grazing angular momentum with respect to the potential $W$.  Hence, the absorption and the potential radius are relate in a non-trivial manner, changing with mass and energy. Results have been shown in Fig.\ref{fig:RoRw}.

\section{Gaussian Approach to the Black Disk Distortion Factor}\label{app:GaussBD}
As discussed, the separation function $h(q)$ is well approximated by the modified Gaussian in Eq.(\ref{mono_sep}),
\be
h(q)=e^{-{1\over 2}q^2\sigma^2}j_0(q\rho).
\ee
The parameter $\sigma$ controls the slope of the momentum distribution around the momentum transfer $p=q_{\alpha\beta}$. The (off-shell) diffraction structure is determined by $\rho$. Thus, we have to evaluate the integral
\be
n^{BD}=\frac{2R_{abs}}{\pi}\int^\infty_0
{dqj_0(qR_{abs})\frac{\partial}{\partial q}(q h(q)) }
\ee
which is given explicitly by the 3-parameter form
\begin{widetext}
\be
n^{BD}={\frac {{2\it R_{abs}}}{\pi }\int^{\infty}_{0}\!{\frac {\sin \left( q{
\it R_{abs}} \right) }{q{\it R_{abs}}}{{\rm e}^{-{1\over 2}{\sigma}^{2}{q}^{2}}} \left( -{\frac {{\sigma}^{2}q }{\rho}}\sin \left( q{\rho} \right)+\cos \left( q{\rho} \right)
 \right) }\,{\rm d}q}
\ee
\end{widetext}
The absorption radius, $R_{abs}$, is fixed by the total reaction cross section. 
The integral can be performed in closed form, with the result

\bea
n^{BD}&=&\frac{1}{2}\left[erf\left(\frac{1}{\sqrt{2}\sigma '}(R'-\rho ')\right)+erf\left(\frac{1}{\sqrt{2}\sigma '}(R'+\rho ')\right) \right]\nonumber \\
&-&\sqrt{\frac{2}{\pi}}\frac{\sigma '}{2\rho '}\left[e^{-\frac{1}{2\sigma '^2}(R'-\rho ')^2}-e^{-\frac{1}{2\sigma '^2}(R'+\rho ')^2}  \right]
\label{eq_app}
\eea
expressed in terms of the scaled (dimensionless) quantities  $R'=R_{abs}/R$,
$\sigma' = \sigma / R$, $q'_{\alpha\beta} = q_{\alpha\beta}R$ and $\rho '=\sqrt{1-\sigma '^4q_{\alpha\beta}'^2+2i\sigma '^2q_{\alpha\beta}'\cos\gamma}$ .
$n^{BD}$ is a complex-valued function, because it depends on the complex pseudo-radius
$\rho'$.
Moreover, $n^{BD}$ contains the full set of multipoles in $\mathbf{q}_{\alpha\beta}$.
Typical results for $n^{BD}$ are displayed in Fig.\ref{fig:8}, for $ {q}_{\alpha\beta} \approx 0$.

\end{document}